\documentclass{article}
\usepackage{graphicx} 
\usepackage[utf8x]{inputenc}
\usepackage[margin=1in]{geometry}
\usepackage[T1]{fontenc}

\usepackage{caption}
\usepackage{subcaption}
\usepackage{amsmath}
\usepackage{amssymb}
\usepackage{wasysym}
\usepackage{xcolor}
\usepackage{float}
\usepackage{mathtools}
\usepackage[normalem]{ulem}

\usepackage[square,numbers,sort&compress]{natbib}
\usepackage[pdftex,linkcolor=black,pdfborder={0 0 0}]{hyperref}

\usepackage{calc} 
\usepackage{enumitem} 

\usepackage[all]{nowidow}
\usepackage[protrusion=true,expansion=true]{microtype}
\usepackage{tikz}

\usepackage{authblk}

\DeclareMathOperator*{\sumint}{\sum\mkern-25mu\int}
\newcommand{\smallsumint}{\sum\mkern-20mu\int}
\def \Kummer {{}_1F_1}

\def \re {\text{Re}}
\def \dm {\text{DM}}
\def \BH {\text{BH}}

\def \cc {\text{c.c.}}
\def \const {\text{const.}}
\def \years {\text{years}}
\def \GeV {\text{GeV}}
\def \eV {\text{eV}}
\def \cm {\text{cm}}
\def \rel {\text{rel}}
\def \db {\text{dB}}

\title{Capturing Ultralight Dark Matter Around Black Holes}
\author[1,2]{Salvatore Bottaro}
\author[1]{Viktor Firsov}
\author[1]{Gilad Perez}
\author[1]{Halil Steuer}
\author[3]{Ofri Telem}
\affil[1]{Department of Particle Physics and Astrophysics,\protect\\ Weizmann Institute of Science, Rehovot 7610001, Israel}
\affil[2]{Center for Theoretical Physics – a Leinweber Institute,\protect\\ Massachusetts Institute of Technology, Cambridge, MA 02139, USA}
\affil[3]{Racah Institute of Physics, Hebrew University of Jerusalem, 91904 Jerusalem, Israel}
\date{}

\begin{document}

\maketitle

\begin{abstract}
Ultralight dark matter (DM) can form hydrogen-like bound states around black holes (BHs), that can be perturbatively described at small gravitational fine-structure constant $\alpha=GM\mu\ll1$, where $M$ and $\mu$ are the masses of the BH and the DM scalar, respectively. We show that self-interactions lead to the formation of a $2p$-halo around a Schwarzschild BH, with the final cloud density as high as $10^5$ times the background density.
We also show that for the case of a Kerr BH with a superradiant cloud, the decay of the $211$ state can populate the ground state faster than it decays into the BH. Consequently, the $211$ cloud mass is bounded from above by $M_{211}^\text{crit}/M\simeq5\cdot10^{-3}\,\alpha^{-1/2}\left[f_a/10^{16}\,\GeV\right]^2$ up to an $O(1)$ factor depending on the spin of the BH. Crucially, for rapidly spinning BHs with $M\lesssim4\cdot10^6\,M_{\astrosun}$, $\alpha\lesssim0.03$, and $f_a\lesssim9\cdot10^{15}\,\GeV$ the resulting superradiant cloud mass is lower than previously assumed, which might have an impact on axion bounds from measurements of stellar BH spins.
\end{abstract}

\tableofcontents

\section{Introduction}
Ultralight dark matter (DM) is a compelling class of models in which DM is comprised of one or more scalars in the ultralight mass range, which could be lighter than $ 10^{-10}\ \eV$~\cite{ULDM_cosmology,fuzzy_DM,axion_cosmology_review,wave_DM_review,ULDM_review}. In this mass range, the de Broglie wavelength of the scalar is astrophysical in size, and it can form coherent bound halos around astrophysical objects~\cite{superradiance_review,Clough_2019,relaxion_stars,gilads_halo_paper}, also known as \textit{gravitational atoms}~\cite{superradiance_axiverse,Baumann_2019,gravitational_atom_energy}. In~\cite{gilads_halo_paper}, one of the present authors showed that ultralight DM scalars with a quartic self interaction $\propto \mu^2/f_a^2$, with $\mu$ the DM mass and $f_a$ its decay constant, can be captured efficiently by stellar objects. These halos can form on astrophysical timescales and reach a large occupation number in the ground state of the system. The condition for the halo to form is that the background DM waves have an average reduced de Broglie wavelength, $\lambda_\text{dB}$, larger than the system's Bohr radius, $r_b$. This condition can be written using the dimensionless focusing parameter\footnote{Note that this definition differs from the one in~\cite{gilads_halo_paper} by a factor of $2\pi$.}:
\begin{equation}
    \label{eq:xi definition}
    \xi\equiv \frac{\lambda_\text{dB}}{r_b}=\frac{\alpha}{v_\dm},
\end{equation}
with $v_\dm$ the background DM's average velocity relative to the stellar object, and $\alpha=GM\mu$ the gravitational fine structure constant, with $M,\mu$ the masses of the stellar object and DM, respectively.

When $\xi\gg1$, the background waves undergo gravitational focusing~\cite{gravitational_focusing}. Their density around the star increases, leading to an increase in the capture of the background DM into a bound state around the star. Due to Bose enhancement this capturing is exponential~\cite{gilads_halo_paper}. The system also captures the background DM into excited states of the gravitational atom (denoted by the same quantum numbers of the hydrogen atom system $nlm$), but due to the self-interactions, all excited states decay into the ground state. This decay is also Bose enhanced by the ground state occupation number, so a large halo in an excited state cannot form, and only the ground state halo is stable~\cite{gilads_halo_paper}.

Scalar clouds around black holes (BHs) have been studied extensively, mostly for Kerr BHs, where the superradiance process populates excited states of the halo~\cite{Detweiler_1980,superradiance_axiverse,superradiance_review}, and where self-interactions lead to various phenomenological implications including bosenovae and axion wave emission~\cite{superradiance_axiverse,QCD_axion_BHs,Baryakhtar_2021}. Bound clouds around BHs have also been numerically shown to grow by accreting the background DM~\cite{Clough_2019,Bamber_2021}.

In this work, we extend the analytic halo formation mechanism developed in~\cite{gilads_halo_paper} to the case of a Schwarzschild BH and show that unlike the case of a star, halos of excited states can form. In the limit $\alpha\ll1$, the BH's effect is an added complex correction to the bound state energies, making them quasi-bound. The imaginary part describes the decay width caused by the absorption of DM into the BH, and is proportional to $\alpha^{5+4l}$~\cite{Detweiler_1980,Leaver,Dolan_2007,gravitational_atom_energy}. For a Schwarzschild BH these widths are always positive, and there is no superradiant amplification. The halos we find form via self-interaction-induced capture of the background DM.

As the states with smaller angular momentum decay faster, for a large part of the relevant DM parameter space a ground state halo cannot form due to its fast decay. Because of the low population of the ground state in this region, excited states with non-zero angular momentum can form long-lived halos. We focus on the simplest case of the $2p$ ($21m$) states.

We show that the DM can be efficiently captured into $1s$ or $2p$-states around astrophysical objects. For the case of BHs the cloud can reach density $10^5$ times larger than the background. However, for both $1s$ and $2p$ clouds due to small $r_b$ the mass of the cloud does not exceed $10^{-12}$ of the BH's mass unless considering a BH close to the galactic center where $\rho_\dm$ is higher. 

In the context of superradiant clouds around Kerr BHs, we consider a decay process where two DM particles in the $211$ state scatter and produce a particle in the ground state and an unbound particle; the rate of this process was computed in~\cite{Baryakhtar_2021}. We show that for a $211$ cloud above a critical mass this decay is strong enough to drive growth of the ground state despite its decay into the BH. This process gives a new limit for the mass of the $211$ cloud for decay constants as high as $f_a\sim9\cdot10^{15}\ \GeV$.

The paper is organized as follows: In section \ref{sec:gravitational atom} we define the gravitational atom system and its properties, and give a brief overview of the capturing mechanism developed in~\cite{gilads_halo_paper}. In section \ref{sec:dynamics} we describe in detail the dynamics of the system and how the halos form from the background and evolve over time. Finally, in section \ref{sec:implications} we discuss the final size and density of the halos that form, the relevant DM parameter space for the system, and implications of these results. Sections 2–4.1 assume a Schwarzschild BH, while in subsection~\ref{subsec:superradiance} we turn to Kerr BHs and show that the $211\rightarrow1s$ transition can affect superradiant clouds.

In what follows, we use natural units, with $\hbar=c=1$.

\section{Qualitative description of DM halos around BHs}\label{sec:gravitational atom}
\subsection{The gravitational atom}
We consider a real scalar field $\phi$ of mass $\mu$ with potential $V(\phi)$ in the background of the Schwarzschild metric for a BH of mass $M$: $g_{\mu\nu}=\mathrm{diag}\left(\left(1-\tfrac{r_s}{r}\right),-\left(1-\tfrac{r_s}{r}\right)^{-1},-r^{2},-r^{2}\sin^{2}\theta\right)$, with $r_s=2GM$ the Schwarzschild radius of the BH. We assume that the effects of the field on the metric can be neglected. We take the scalar potential to consist of a quartic self-interaction: 
\begin{equation}
    V(\phi)=\frac{1}{2}\mu^2\phi^2+\frac{\lambda}{4!}\phi^4,
\end{equation}
which, in the case of an attractive self-interaction, could be considered an expansion of an axion-like potential: $V(\phi)=-\mu^2 f_a^2\cos(\phi/f_a)$, with $\lambda=-\mu^2/f_a^2$ and $f_a$ the decay constant of the particle. It is also useful to define the dimensionful coupling $g$:
\begin{equation}
    g\equiv\frac{\lambda}{8\mu^2}=-\frac{1}{8f_a^2}.
\end{equation}

In most of what follows, the processes are proportional to $g^2$, so the sign of the self-interaction does not matter, and we will use $f_a$ without necessarily assuming the type of interaction.

The Klein-Gordon equation for the field in the background of the metric is:
\begin{equation}
\left(g^{\mu\nu}D_{\mu}D_{\nu}+\mu^{2}\right)\phi=\frac{1}{\sqrt{-g}}\partial_{\sigma}\left(g^{\sigma\tau}\sqrt{-g}\partial_{\tau}\right)\phi+\mu^{2}\phi=-\frac{\lambda}{6}\phi^3,
\end{equation}
with $D_\mu$ a covariant derivative. In the non-relativistic limit we can decompose our field as:
\begin{equation}
    \phi\equiv \frac{1}{\sqrt{2\mu}}(\psi e^{-i\mu t}+\cc),\quad \text{assuming}\;\dot{\psi}\ll\mu\psi
\end{equation}

The metric can be expanded in powers of the gravitational fine structure constant:
\begin{equation}
    \alpha\equiv GM\mu \cong 7.48\cdot10^{-5} \left[\frac{\mu}{10^{-14}\,\eV}\right] \left[\frac{M}{M_{\astrosun}}\right],
\end{equation}
which gives to leading order in $\alpha$ a Gross-Pitaevskii equation for $\psi$:

\begin{equation}
    \label{eq:GP equation}
    \left( i\partial_t+\frac{\nabla^2}{2\mu}+\frac{\alpha}{r} \right)\psi = g|\psi|^2\psi.
\end{equation}

When neglecting the self-interactions, this is the Schrödinger equation for the hydrogen atom, which is why the system is usually referred to as a gravitational atom. The solutions have at leading order in $\alpha$ the same bound and unbound wavefunctions $\psi_{nlm},\;\psi_\mathbf{k}$ as the hydrogen atom, but differ from the regular hydrogen atom solutions due to the boundary conditions imposed at the BH's event horizon: an incoming wave describing particles traveling into the horizon leads to a complex correction to the bound states' energies, leading to the decay into the BH and a decrease of the states' occupation numbers:
\begin{equation}
    E_{nlm}=\omega_{nlm}-i\Gamma_{nlm\rightarrow \BH}/2
\end{equation}

In the limit $\alpha\ll1$ (we take $\alpha<0.2$), the real energy corrections and imaginary decay widths were computed analytically in~\cite{gravitational_atom_energy}, and give for the Schwarzschild case: 
\begin{equation}\label{eq:energy corrections}
    \omega_{nlm}=-\frac{\alpha^2\mu}{2n^2}+\alpha^4\mu\left(\frac{15}{8n^4}-\frac{6}{(2l+1)n^3}\right)+O(\alpha^5),
\end{equation}
\begin{equation}
    \label{eq:BH decay rate}
    \Gamma_{nlm\rightarrow \BH}/2=\mu\alpha^{4l+5}\frac{2^{4l+3}\left(n+l\right)!}{n^{2l+4}\left(n-l-1\right)!}\left(\frac{(l!)^{2}}{(2l)!\left(2l+1\right)!}\right)^{2}.
\end{equation}

The factor of $2$ in the definition of $\Gamma_{nlm\rightarrow\BH}$ is introduced since the decay rate for the occupation number of a particular state is twice that of its amplitude: the state amplitude's time evolution in the free system is $c_{nlm}(t)=c_{nlm}(0)e^{-i\omega_{nlm}t-\Gamma_{nlm\rightarrow\BH}t/2}$, so the occupation number evolves as $N_{nlm}(t)=|c_{nlm}(t)|^2=N_{nlm}(0)e^{-\Gamma_{nlm\rightarrow\BH} t}$. The proportionality $\Gamma_{nlm\rightarrow\BH}\propto \alpha^{4l+5}$ means that states with higher angular momentum have a slower decay into the BH. We add this decay by hand into the equations we develop in the next section.

In the non-relativistic limit, the number density of particles is given by $|\psi|^2$. Decomposing $\psi$ over the complete hydrogen atom basis:
\begin{equation}
    \psi=\sum_{nlm}c_{nlm}(t)e^{-i\omega_{nlm}t}\psi_{nlm}+\int\frac{d^3\mathbf{k}}{(2\pi)^3}c_\mathbf{k}(t)e^{-i\omega_\mathbf{k}t}\psi_{\mathbf{k}},
\end{equation}
means that $|c_{nlm}|^2$ is the total number of particles in the state $nlm$, while $|c_\mathbf{k}|^2$ describes the number density of unbound states in $k$-space.

\paragraph{Bound states}\mbox{}\\
The bound states are the usual hydrogen atom wavefunctions:
\begin{equation}
    \psi_{nlm}(\mathbf{r})=Y_l^m(\hat{r})R_{nl}(r),\quad R_{nl}(r)=\left(\frac{2}{nr_{b}}\right)^{3/2}\sqrt{\frac{(n-l-1)!}{2n(l+n)!}} \left(\frac{2r}{nr_b}\right)^{l}L_{n-l-1}^{2l+1}\left(\frac{2r}{nr_b}\right)e^{-\frac{r}{nr_b}},
\end{equation}
with $Y_l^m$ the spherical harmonics, $L_n^m$ associated Laguerre polynomials, and $r_b=(\mu\alpha)^{-1}$ the Bohr radius. The wavefunctions are normalized such that $\int d^3x\  \overline{\psi_{nlm}}\psi_{n'l'm'}=\delta^n_{n'}\delta^l_{l'}\delta^m_{m'}$. We will mainly focus on the $nlm=100,200,21m$ states, also known as $1s$, $2s$, $2p$ states, respectively. In cases where the $21m$ states can be viewed together, we will use those names interchangeably. For those states the decay rates are:
\begin{equation}
    \Gamma_{1s\rightarrow \BH}=16\mu \alpha^5,\quad \Gamma_{2s\rightarrow \BH}=2\mu\alpha^5,\quad\Gamma_{2p\rightarrow \BH}=\frac{\mu\alpha^{9}}{6},
\end{equation}
which can be converted into timescales with some benchmark parameters:
\begin{align}
    \tau_{1s\rightarrow \BH}&\equiv\frac{1}{\Gamma_{1s\rightarrow \BH}}=\frac{1}{16\mu\alpha^5}\cong 6\cdot10^{10}\ \years \left[\frac{10^{-14}\,\eV}{\mu}\right]^6 \left[\frac{M_{\astrosun}}{M}\right]^5\\
    \tau_{2p\rightarrow \BH}&\equiv\frac{1}{\Gamma_{2p\rightarrow \BH}}=\frac{6}{\mu\alpha^9}\cong 2\cdot10^{29}\ \years\left[\frac{10^{-14}\,\eV}{\mu}\right]^{10}\left[\frac{M_{\astrosun}}{M}\right]^{9}
\end{align}
At leading order the energies of the states are:
\begin{equation}
    \omega_{1s}=-\frac{\mu\alpha^2}{2}+O(\alpha^4),\quad \omega_{2s}\approx\omega_{2p}=-\frac{\mu\alpha^2}{8}+O(\alpha^4),
\end{equation}
and the fine splitting energy between the $2s$ and $2p$ states is:

\begin{equation}
    \omega_{2p}-\omega_{2s}=\frac{\mu\alpha^4}{2}=\frac{1}{4\alpha}\Gamma_{2s\rightarrow\BH}.
\end{equation}

As we are working in the limit $\alpha\ll1$, the fine splitting energy will always be much larger than the width of the $2s$ or $2p$-states, so their energy levels are clearly split\footnote{Another requirement is that the relevant processes are slow enough that $t\gg\left(\omega_{2p}-\omega_{2s}\right)^{-1}$, which is easily fulfilled for the relevant parameters.}. While for the multiplet of $2p$-states, due to the spherical symmetry of the Schwarzschild metric, there is no hyperfine splitting between their energies or decay widths.

\paragraph{Unbound states}\mbox{}\\
The unbound states are parametrized by a vector $\mathbf{k}$ denoting their asymptotic momentum~\cite{gravitational_focusing}:
\begin{equation}
    \psi_{\mathbf{k}}(\mathbf{r})=e^{i\mathbf{k}\cdot\mathbf{r}}\Gamma\left(1-\frac{i}{kr_{b}}\right)e^{\frac{\pi}{2kr_{b}}} \Kummer\left(\frac{i}{kr_{b}},1,ikr\left(1-\hat{k}\cdot\hat{r}\right)\right),
\end{equation}
with $\Kummer(a,b,z)$ the confluent hypergeometric function. Their energy is $\omega_k=k^2/2\mu$, and they have a $\delta$-function normalization $\int d^3x\ \overline{\psi_{\mathbf{k}}}\psi_{\mathbf{k'}}=(2\pi)^3\delta^{(3)}(\mathbf{k-k'})$. They are also orthogonal to the bound states: $\int d^3x\ \overline{\psi_\mathbf{k}}\psi_{nlm}=0$. The behavior of unbound states' wavefunctions can be parametrized by the dimensionless focusing parameter $\xi\left(k\right)=(kr_b)^{-1}$, defined in eq. \eqref{eq:xi definition}.

In the limit where $\xi\gg 1$ we can approximate $\psi_\mathbf{k}$ using a large parameter expansion for $\Kummer(a,b,z)$ and $\Gamma(z)$, giving us (up to order $O\left(1/\sqrt{\xi}\right)$):
\begin{equation}
    \psi_{\mathbf{k}}(\mathbf{r})\big|_{\xi\gg1}= e^{i\varphi(k)}\sqrt{\frac{2\pi}{kr_{b}}}J_{0}\left(2\sqrt{\frac{r}{r_{b}}(1-\hat{k}\cdot\hat{r}})\right) ,\quad \varphi(k)= \frac{1}{kr_{b}}\left(\ln\left(kr_{b}\right)+1\right)-\frac{\pi}{4}+kr\left(1+\hat{k}\cdot\hat{r}\right)/2.
\end{equation}

If we only evaluate $\psi_\mathbf{k}$ at $r\sim r_b$, the oscillations inside of $\varphi(k)$ can be neglected and we can view the phase as constant in space. In this regime, the stellar object leaves a wake behind it, where $\psi_\mathbf{k}\sim \sqrt{\xi(k)}\gg1$:
\begin{equation}
    \psi_{\mathbf{k}}(\hat{r}=\hat{k})|_{\xi(k)\gg1}=e^{i\varphi(k)}\sqrt{\frac{2\pi}{kr_b}}J_0(0)=e^{i\varphi(k)}\sqrt{2\pi\xi(k)}.
\end{equation}

This effect is known as gravitational focusing~\cite{gravitational_focusing}, and it increases overlap integrals with the bound states' wavefunctions in two ways: first because the wavefunction is large at the wake, and second because it does not oscillate.

We can describe the background of DM scattering states by expanding it in the hydrogen atom basis:
\begin{equation}
    \psi_\text{background}(\mathbf{r})=\int \frac{d^3\mathbf{k}}{(2\pi)^3} c_\mathbf{k} e^{-i\omega_kt}\psi_\mathbf{k}(\mathbf{r}),
\end{equation}
where $c_\mathbf{k}$ describes the statistical properties of the background. Over times longer than the coherence time $\tau_\text{coh}=(\mu \sigma_\dm^2)^{-1}$ --- with $\sigma_\dm$ the velocity dispersion of the galactic DM --- we can treat $c_\mathbf{k}$ as a random variable, with norm describing the number density of states with momentum $\mathbf{k}$, and with a uniformly distributed random phase. If we average $c_\mathbf{k}$ over times longer than $\tau_\text{coh}$ (referred to as applying "coherence time averaging") we expect to get $\langle c_\mathbf{k} \rangle=0$ due to the random phases canceling, while for a pair:
\begin{equation}
       \langle c_\mathbf{k} c_\mathbf{k'}\rangle=0,\quad\langle c_\mathbf{k} \overline{c_\mathbf{k'}}\rangle=(2\pi)^3 f(\mathbf{k})\delta^{(3)}(\mathbf{k-k'}),
\end{equation}
where Wick's theorem can be used for a general average of $c_{\mathbf{k}_i},\overline{c_{\mathbf{k}'_j}}$. The function $f(\mathbf{k})$ describes the DM occupation number at momentum $\mathbf{k}$, which can be approximated with a Maxwell-Boltzmann distribution~\cite{gravitational_focusing}:
\begin{equation}
    f(\mathbf{k})=(2\pi)^{3/2}\frac{\rho_\dm}{\sigma_\dm^3\mu^4}\exp\left( -\frac{(\mathbf{k}-\mu\mathbf{v}_\dm)^2}{2\mu^2\sigma_\dm^2}\right),
\end{equation}
with $\mathbf{v}_\dm,\rho_\dm$ the DM's average velocity in the BH's reference frame and background density, respectively.  In the virialized halo the dispersion $\sigma_\dm$ is taken as $\sigma_\dm=v_\dm/\sqrt{2}$~\cite{galactic_DM_halo}, which gives for the coherence time:
\begin{equation}
    \tau_{\text{coh}}=\frac{1}{\frac{1}{2}\mu v_\dm^{2}}\cong7\cdot 10^{-3}\ \years\left[\frac{10^{-14}\,\eV}{\mu}\right]\left[\frac{240\,\mathrm{km/s}}{v_{\dm}}\right]^{2},
\end{equation}
where we chose as a fiducial velocity $v_{\rm fid}\equiv240\,\mathrm{km/s}$, the velocity of the Sun around the center of the Milky Way. Since the Milky Way rotation curve is approximately flat in the region $1$--$20$ kpc away from the galactic center~\cite{weird_rotation_curves,rotation_curves} we can in fact use $v_\dm\approx v_{\rm fid}$ throughout most of the galaxy. By changing $v_\dm, \rho_\dm$, our formalism can be applied to other galaxies.

The background density is taken from the common NFW density profile~\cite{NFW_profile_paper}, describing the density as a function of distance from the galactic center and defined by:
\begin{equation}\label{eq:NFW}
    \rho_\dm(r)=\frac{\rho_\text{NFW}}{\frac{r}{r_{\text{NFW}}}\left(1+\frac{r}{r_{\text{NFW}}}\right)^{2}},
\end{equation}
with $\rho_\text{NFW},r_\text{NFW}$ the parameters of the profile, and for the Milky Way $r_{\text{NFW}}=18.6\ \text{kpc},\ \rho_\text{NFW}=0.38\ \GeV/\cm^3$~\cite{DM_density_profile}. We will use as a benchmark value the DM density around the Sun, which is $\rho_\dm\approx0.4\ \GeV/\cm^3$~\cite{local_DM_density}.

The timescale for our processes is the DM relaxation time, defined as the typical time a particle in a gas with density $\rho_\dm$ and average square velocity $v_\dm^2$ takes to change its velocity by order one via the self-interactions mediated by $g$, in the absence of external gravitational potentials:
\begin{equation}
    \tau_{\text{rel}}\equiv\frac{\mu^3v_\dm^2}{g^2\rho_\dm^2} \cong 9\cdot10^{9}\ \years\left[\frac{f_a}{10^{8}\,\GeV}\right]^{4}\left[\frac{\mu}{10^{-14}\,\eV}\right]^{3}\left[\frac{0.4\,\GeV/\cm^{3}}{\mathrm{\rho_{\dm}}}\right]^{2}\left[\frac{v_{\dm}}{240\,\mathrm{km/s}}\right]^{2}.
\end{equation}

From this timescale we can define the typical rate of processes relating to self-interactions: $\Gamma_\rel\equiv1/\tau_\rel$.
\subsection{DM capture}
Due to the nonlinear term in eq. \eqref{eq:GP equation}, background particles will be captured into bound states over time and change their occupation numbers. This capture will be described in detail in the next section using perturbation theory, but for a single bound state it has a simple result. For a state ${nlm}$ interacting with the continuum, which we denote with "$\infty$", there are three relevant processes: capture of the background DM into the state at a constant positive rate $S_{\infty\rightarrow nlm}$, Bose-enhanced capture at a rate $\Gamma_{\infty\rightarrow nlm}$, and a similar depletion of the state back into the continuum at $\Gamma_{nlm\rightarrow \infty}$. 

These terms give a differential equation for the occupation number of the state $N_{nlm}$\footnote{It is worth mentioning that the Bose enhancement factor for the capture should be $N_{nlm}+1$ and not simply $N_{nlm}$ (unlike the decay, which is indeed proportional to $N_{nlm}$), but since we are dealing with large occupation numbers of the order of $O(10^{60})$ this $+1$ can be safely neglected.}:
\begin{equation}
    \dot{N}_{nlm}= S_{\infty\rightarrow nlm} + (\Gamma_{\infty\rightarrow nlm}-\Gamma_{nlm\rightarrow \infty})N_{nlm}.
\end{equation}
When focusing is in effect, $\xi\gg1$, the decay process is exponentially suppressed, $\Gamma_{nlm\rightarrow \infty}\ll \Gamma_{\infty\rightarrow nlm}$, and the cloud will grow exponentially at a rate $\Gamma_{\infty\rightarrow nlm}\propto \Gamma_\rel$. This growth will continue until the cloud reaches a density where the  perturbative treatment of the problem breaks down, as described in Appendix \ref{app:cloud size}.

Besides capture and decay into the BH, the evolution equations for the scalar halo also involve quartic-driven transitions between bound states with different quantum numbers. For every $nlm\ne100$ state there is a transition $nlm\rightarrow 1s$ proportional to $N_{nlm}^2N_{1s}$. There are also $1s\rightarrow nlm$ transitions, but none of them are proportional to $N_{1s}^2$, so the processes do not balance each other. This means that once a $1s$-halo forms, excited states tend to decay into it very fast, and only a large $1s$-halo will be present.

In the case of a star, such as the Sun, when capturing is applicable a $1s$-halo will usually form so no excited state halos are expected. On the other hand, for a BH the decay $\Gamma_{1s\rightarrow\BH}$ can be bigger than the $\Gamma_{\infty\rightarrow1s}$ exponential growth and stop a $1s$-halo from forming. In this case, most decays of excited states are suppressed by factors of $N_{1s}$, which is small compared to other occupation numbers in the system, so excited states can reach a large occupation number. Due to the fact that $\Gamma_{{nlm}\rightarrow\BH}\propto \alpha^{4l+5}$, all $s$-states have comparable decay widths. Therefore if the $1s$ is unstable, it usually means that all $s$-states do not capture, and states with $l\ne0$ are the ones that are expected to form a stable halo. The lowest energy $l\ne0$ states are the $2p$ states, which we will focus on in this paper.

\section{Dynamics}\label{sec:dynamics}
For convenience, we will restate the Gross-Pitaevskii equation from eq. \eqref{eq:GP equation}:
\begin{equation}
    \left(i\partial_{t}+\frac{\nabla^{2}}{2\mu}+\frac{\alpha}{r}\right)\psi=g\left|\psi\right|^{2}\psi.
\end{equation}

We will solve eq. \eqref{eq:GP equation} perturbatively in orders of the self-interaction parameter $g$. Expanding $\psi = \psi^{(0)}+\psi^{(1)}+\psi^{(2)}+O\left(g^3\right)$, with the superscript number in brackets denoting the order of the term, we get the following system of equations:
\begin{equation}
    \label{eq:GP system}
    \begin{cases}
       \left(i\partial_t+\frac{\nabla^2}{2\mu}+\frac{\alpha}{r}\right)\psi^{(0)}=0,\\
       \left(i\partial_t+\frac{\nabla^2}{2\mu}+\frac{\alpha}{r}\right)\psi^{(1)}
       =g{\psi^{(0)}}^2\overline{\psi^{(0)}},\\
       \left(i\partial_t+\frac{\nabla^2}{2\mu}+\frac{\alpha}{r}\right)\psi^{(2)}
       =g\left({\psi^{(0)}}^2\overline{\psi^{(1)}}
       +2\psi^{(0)}\overline{\psi^{(0)}}\psi^{(1)}\right).
    \end{cases}
\end{equation}

The full solution $\psi$ can be expanded as a superposition of bound and unbound hydrogen atom states
\begin{equation}
    \label{eq:psi expansion}
    \psi=\sum_{nlm}\tilde{c}_{nlm}(t)e^{-i\tilde{\omega}_{nlm}t}\psi_{nlm}+\int\frac{d^3\mathbf{k}}{(2\pi)^3}\tilde{c}_\mathbf{k}(t)e^{-i\tilde{\omega}_\mathbf{k}t}\psi_{\mathbf{k}}
    \equiv \sumint_{s}\tilde{c}_{s}(t)e^{-i\tilde{\omega}_{s}t}\psi_{s},
\end{equation}
where $\smallsumint$  denotes a sum over all states $[s\equiv(nlm)\,\,{\rm or}\,\,\mathbf{k}]$, both bound and unbound; $\tilde{c}_s$ and $\tilde{\omega}_s$ are the amplitude and energy of a given state to all orders in perturbation theory. We will be interested in the amplitudes of different states, expanding them up to second order $\tilde c_s= c_s^{(0)}+c_s^{(1)}+c_s^{(2)}+O\left(g^3\right)$. The total occupation number in a state is given by:
\begin{multline}
    \label{eq:number of particles expansion}
    N_s(t)=|\tilde{c}_s(t)|^2\approx\left|c_s^{(0)}+c_s^{(1)}(t)+c_s^{(2)}(t)\right|^2\\
    \approx |c_s^{(0)}|^2+\left(\overline{c_s^{(0)}}c_s^{(1)}(t)+c_s^{(0)}\overline{c_s^{(1)}(t)}\right)
    +\left\{\left|c_s^{(1)}(t)\right|^2+\left(\overline{c_s^{(0)}}c_s^{(2)}(t)+c_s^{(0)}\overline{c_s^{(2)}(t)}\right)\right\}
    \equiv N_s^{(0)}+N_s^{(1)}+N_s^{(2)}.
\end{multline}

The first equation of \eqref{eq:GP system} is trivially satisfied by the hydrogen atom wavefunctions that we chose to expand $\psi$ over, so it just yields $c_s^{(0)}=\text{const}$. The second and third equations allow us to find first- and second-order corrections, $c_s^{(1)}$ and $c_s^{(2)}$, to the amplitude $\tilde c_s$ and express them using the unperturbed zeroth order $c_s^{(0)}$, in which from now on we will omit the superscript for brevity.

The first order amplitude correction $c_s^{(1)}$ comes from the second equation in \eqref{eq:GP system}, which has three entries of $\psi^{(0)}$ in it, so a contribution to $c_s^{(1)}$ will contain amplitudes of three states $s_i$ and depend on the matrix element of these states with $s$: 
\begin{equation}
    M_{s_1+s_2\rightarrow s_3+s}\equiv \int d^3x\ \psi_{s_1}\psi_{s_2}\overline{\psi_{s_3}\psi_s}.
\end{equation}

This gives rise to the following correction to the number of particles in the state $s$:
\begin{equation}
    \label{eq:1st correction to number of particles}
    N_s^{(1)}=\overline{c_s}c_s^{(1)}(t)+\cc\supset
    -ig\sumint_{s_i}c_{s_1}c_{s_2}\overline{c_{s_3}c_s}
    M_{s_1+s_2\rightarrow s_3+s}
    \int_0^td\tau e^{-i\tau\left(\omega_{s_1}+\omega_{s_2}-\omega_{s_3}-\omega_s\right)}+\cc.
\end{equation}

It will be convenient to describe these expressions diagrammatically, with each diagram describing a transition rate between different states (bound or unbound).
We can treat this contribution as arising from $2\rightarrow2$ scattering process $s_1+s_2\rightarrow s_3+s$. In first-order (one vertex) diagrams we will use a dashed line to denote the state $s$ being produced.

\begin{figure}[H]
    \centering
    $
    N_s^{(1)}\supset
    \displaystyle
    \begin{tikzpicture}[scale=0.55, baseline=(current bounding box.center)]
        \coordinate (v1) at (0, 0);

        \draw[thick] (-1, 1) node[left] {$s_1$} -- (v1);
        \draw[thick] (-1, -1) node[left] {$s_2$} -- (v1);
        \draw[thick] (v1) -- (1, 1) node[right] {$s_3$};
        \draw[thick, dashed] (v1) -- (1,-1) node[right] {$s$} ;
        \filldraw (v1) circle (1.5pt);    
    \end{tikzpicture}
    \quad
    N_s^{(2)}\supset
    \begin{tikzpicture}[scale=0.55, baseline=(current bounding box.center)]
        \coordinate (v1) at (0, 0);
        \coordinate (v2) at (1.5, -1.5);

        \draw[thick] (-1, 1) node[left] {$s_1$} -- (v1);
        \draw[thick] (-1, -1) node[left] {$s_2$} -- (v1);
        \draw[thick] (v1) -- (1, 1) node[right] {$q_3$};

        \draw[thick] (0.5, -2.5) node[left] {$s_3$} -- (v2);
        \draw[thick] (v2) -- (2.5, -0.5) node[right] {$q_2$};

        \draw[thick, dashed] (v1) -- node[above right] {$s$} (v2);
        \draw[thick] (v2) -- (2.5, -2.5) node[right] {$q_1$};

        \filldraw (v1) circle (1.5pt);
        \filldraw (v2) circle (1.5pt);
    \end{tikzpicture}
    -
    \begin{tikzpicture}[scale=0.55, baseline=(current bounding box.center)]
        \coordinate (v1) at (0, 0);
        \coordinate (v2) at (1.5, -1.5);

        \draw[thick] (-1, 1) node[left] {$s_1$} -- (v1);
        \draw[thick] (-1, -1) node[left] {$s_2$} -- (v1);
        \draw[thick] (v1) -- (1, 1) node[right] {$q_3$};

        \draw[thick] (0.5, -2.5) node[left] {$s_3$} -- (v2);
        \draw[thick] (v2) -- (2.5, -0.5) node[right] {$q_2$};

        \draw[thick, dashed] (v1) -- node[above right] {$d$} (v2);
        \draw[thick, dashed] (v2) -- (2.5, -2.5) node[right] {$s$};

        \filldraw (v1) circle (1.5pt);
        \filldraw (v2) circle (1.5pt);
    \end{tikzpicture}
    +
    \begin{tikzpicture}[scale=0.55, baseline=(current bounding box.center)]
        \coordinate (v1) at (0, 0);
        \coordinate (v2) at (1.5, -1.5);

        \draw[thick] (-1, 1) node[left] {$s_1$} -- (v1);
        \draw[thick] (-1, -1) node[left] {$s_2$} -- (v1);
        \draw[thick, dashed] (v1) -- (1, 1) node[right] {$s$};

        \draw[thick] (0.5, -2.5) node[left] {$s_3$} -- (v2);
        \draw[thick] (v2) -- (2.5, -0.5) node[right] {$q_2$};

        \draw[thick, dashed] (v1) -- node[above right] {$d$} (v2);
        \draw[thick] (v2) -- (2.5, -2.5) node[right] {$q_1$};

        \filldraw (v1) circle (1.5pt);
        \filldraw (v2) circle (1.5pt);
    \end{tikzpicture}$
    \caption{Classes of diagrams contributing to first- and second-order corrections to the numbers of particles in a given state $s$. The diagram in $N_s^{(1)}$ corresponds to the expression in eq. \eqref{eq:1st correction to number of particles}. The first diagram in $N_s^{(2)}$ comes from the $|c_s^{(1)}|^2$ term in eq.\eqref{eq:number of particles expansion}, while the others arise from the terms involving $c_s^{(2)}$.}
    \label{fig:classes of diagrams}
\end{figure}
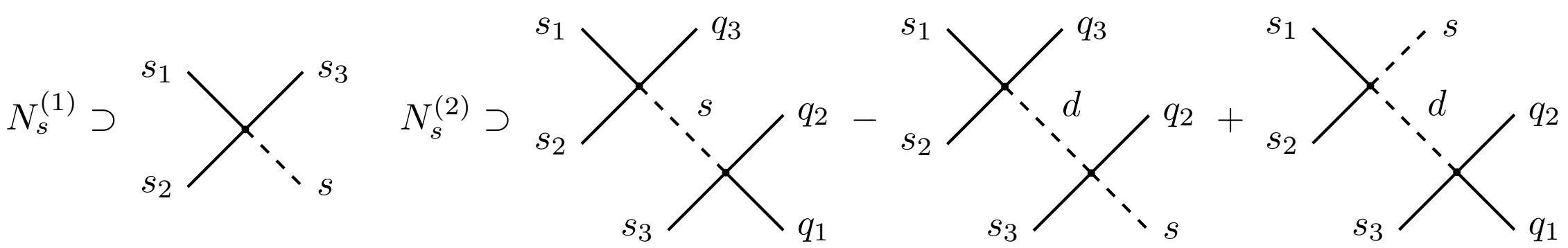

Similarly, the second order contributions will be represented with 2 vertex diagrams. The contribution coming from the $|c_s^{(1)}|^2$ term will be represented by two first-order diagrams "glued" together by the dashed line of the produced state, which will also be the intermediate state for these diagrams. The diagrams coming from the $\overline{c_s^{(0)}}c_s^{(2)}(t)$ term will be represented by diagrams with two separate dashed lines: one for the intermediate state $d$ and one for the produced state $s$. Examples of these diagrams appear in fig. \ref{fig:classes of diagrams}. 

It is important to state that despite some similarities, these are \underline{not} Feynman diagrams, and properties of Feynman diagrams do not apply here. These diagrams describe rates, not amplitudes.

To convert a diagram to an analytical expression, we take a product of amplitudes of all incoming states and conjugated amplitudes of all outgoing states (the diagram should be read from left to right, with incoming states defined as those coming into a vertex, and outgoing as those going out of a vertex); the intermediate state's amplitude does not enter the expression. For every vertex we get a factor of $g$, a matrix element, and a time integral over $e^{i\Delta\omega t}$ with $\Delta\omega$ the energy difference of the incoming and outgoing states in a vertex; and lastly a symmetry factor coming from permutations of the outer legs of the diagram. The exact expressions are derived in Appendix \ref{app:Derivation of diagrams}, see fig. \ref{fig:three second-order diagrams}.   
Due to the integrals over factors of $e^{i\Delta\omega t}$, the energy difference at every vertex must be small compared to the widths of the states involved, otherwise the contribution oscillates and remains small. 

Due to the finite width of the states, transitions that don't conserve energy are not prohibited, but they are suppressed by a factor of $\Gamma/\Delta\omega$, with $\Delta\omega$ the energy difference and $\Gamma$ the largest width of the involved states. They are neglected in the main text, but can be relevant for $2p\leftrightarrow2s$ processes due to both large widths of $s$-states and small fine splitting energy, giving $\Gamma/\Delta\omega\sim\alpha$.

Since coherence time averaging is done over all expressions, the external unbound states come in incoming-outgoing pairs with the same momentum, each giving a factor of $f(\mathbf k)$ (except for an unbound intermediate state). The total energy of the incoming bound states must therefore equal that of the outgoing ones, and since the levels are resolved (even the fine splitting exceeds the widths of the $2s$ and $2p$ states), the bound states also come in incoming-outgoing pairs with the same $nl$. Due to the lack of hyperfine splitting, they can have different $m$.

Another important property of the diagrams is the conservation of angular momentum: Assuming we align the $z$ axis with $\mathbf{v}_\dm$, the axial symmetry of the unbound particle distribution $f(\mathbf{k})$ means that in total, they do not carry a magnetic quantum number $m$. Therefore, the total magnetic number of incoming and outgoing bound states must be the same. E.g., the first order process  $21m+k\leftrightarrow21m'+k$ can only be fulfilled for $m=m'$. A detailed derivation is shown in Appendix \ref{app:m conservation}.

In this work, we carry out a full analysis of dynamics of $1s, 2s$ and $2p$ populations. We assume that effects coming from the occupation numbers of states with $n\ge3$ are negligible. This assumption is justified, since higher energy states have a strong decay mode proportional to the population of a lower energy state they decay into. Meaning that, when $1s/2s/2p$-clouds are present, populations of all higher $n$ states should be negligible. Therefore, we do not consider diagrams that involve occupation numbers of states with $n\ge3$. Note, however, that they still can be the intermediate state of a second-order diagram.

Most of the processes we consider will be second-order ones, as due to coherence time averaging, energy and magnetic number conservation, all first order processes have the same incoming and outgoing states, with a notable exception of $210+210\leftrightarrow211+21\!-\!\!1$ discussed in Appendix \ref{app:Mixing}. A first-order process $a+b\rightarrow a+b$ will not give rise to a change of the occupation number:
\begin{equation}
    N_a^{(1)}=
    -igc_ac_b\overline{c_ac_b}
    M_{a+b\rightarrow a+b}
    \int_0^td\tau e^{-i\tau\left(\omega_a+\omega_b-\omega_a-\omega_b\right)}+\cc
    =-igN_aN_bt\int d^3x|\psi_a\psi_b|^2+\cc=0.
\end{equation}

In Appendix \ref{app:Mixing} we show that interactions between $2p$-states drive a process redistributing particles between them, this $2p$-mixing will turn out to be faster than all other relevant processes, so we can average these oscillations out and assume $N_{21-\!1},\;N_{210},\;N_{211}\approx N_{2p}$, so the total population is $3N_{2p}$. 

A full enumeration of possible first- and second-order contributions can be found in Appendix \ref{app:Classification of diagrams}, while in the following subsections we will only discuss the most relevant terms appearing in this simplified differential equation for the population of the $1s$ and $2p$-states:
\begin{equation}
    \begin{cases}
            \dot N_{1s}\supset\left(\Gamma_{\infty\rightarrow1s}-\Gamma_{1s\rightarrow\BH}\right)N_{1s}
            +\frac{3}{2}C_{2p\rightarrow1s}^{1s}N_{2p}^2N_{1s}+\frac{3}{2}C_{2p\rightarrow1s}^{k}N_{2p}^2\\
            \dot N_{2p}\supset\left(\Gamma_{\infty\rightarrow2p}-\Gamma_{2p\rightarrow2s}-\Gamma_{2p\rightarrow\BH}\right)N_{2p}
            -C_{2p\rightarrow1s}^{1s}N_{2p}^2N_{1s}-C_{2p\rightarrow1s}^{k}N_{2p}^2
    \end{cases},
\end{equation}
where $\Gamma_{1s/2p\rightarrow\BH}$ are the inherent decay rates of the states around a BH from eq. \eqref{eq:BH decay rate}. All other terms describe processes caused by the self-interactions and are discussed in the following subsections.

In subsections \ref{subsec:Capture} and \ref{subsec:Decay 2p->2s} we discuss exponential processes, whose rates are proportional to the numbers of particles in the state $\dot N_s\sim N_s$. In the former we show Bose-enhanced capture processes of unbound DM driving exponential growth of the bound state clouds and find $\Gamma_{\infty\rightarrow1s/2p}$. In the latter, we find $\Gamma_{2p\rightarrow2s}$ describing a competing process driving exponential decay of the $2p$-cloud into the $2s$-cloud, which then promptly decays into the BH. 

In subsections \ref{subsec:Decay 2p->1s} and \ref{subsec:Second decay 2p->1s} we consider decays of $2p$ straight into $1s$, whose rates will be proportional to $\dot N_{2p}\propto -N_{2p}^2$, providing an equilibrium point against the exponential processes with $\dot N_{2p}\propto + N_{2p}$. In the former we find $C_{2p\rightarrow1s}^{1s}$ and in the latter $C_{2p\rightarrow1s}^{k}$.

\subsection{Capture into $1s/2p$ $\left(\Gamma_{\infty\rightarrow1s/2p}\right)$}
\label{subsec:Capture}
In this subsection we reproduce the generic halo formation mechanism described in~\cite{gilads_halo_paper} for the $1s$ state, and extend the analysis to $2p$-states. The largest contributions arise from the first two diagrams of fig. \ref{fig:capture diagrams}:
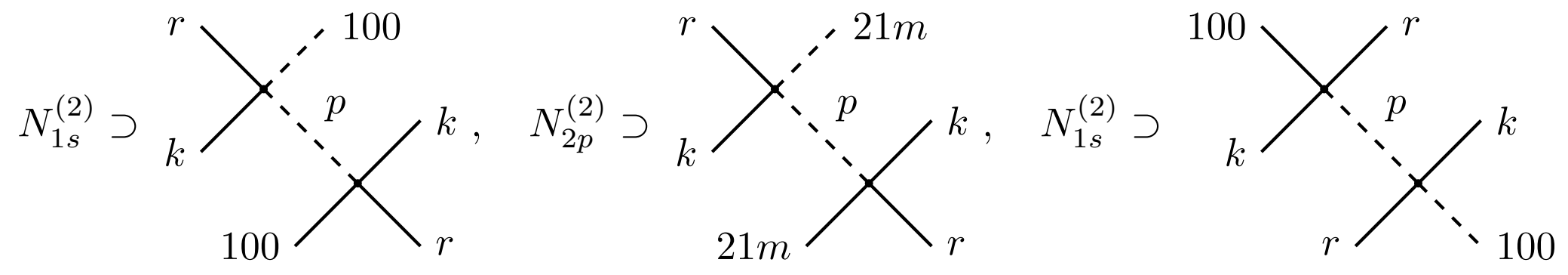
\begin{figure}[H]
    \centering
    $\displaystyle
    N_{1s}^{(2)}\supset
    \begin{tikzpicture}[scale=0.55, baseline=(current bounding box.center)]
        \coordinate (v1) at (0, 0);
        \coordinate (v2) at (1.5, -1.5);

        \draw[thick] (-1, 1) node[left] {$r$} -- (v1);
        \draw[thick] (-1, -1) node[left] {$k$} -- (v1);
        \draw[thick, dashed] (v1) -- (1, 1) node[right] {$100$};

        \draw[thick] (0.5, -2.5) node[left] {$100$} -- (v2);
        \draw[thick] (v2) -- (2.5, -0.5) node[right] {$k$};

        \draw[thick, dashed] (v1) -- node[above right] {$p$} (v2);
        \draw[thick] (v2) -- (2.5, -2.5) node[right] {$r$};

        \filldraw (v1) circle (1.5pt);
        \filldraw (v2) circle (1.5pt);
    \end{tikzpicture}
    ,\quad
    N_{2p}^{(2)}\supset
    \begin{tikzpicture}[scale=0.55, baseline=(current bounding box.center)]
        \coordinate (v1) at (0, 0);
        \coordinate (v2) at (1.5, -1.5);

        \draw[thick] (-1, 1) node[left] {$r$} -- (v1);
        \draw[thick] (-1, -1) node[left] {$k$} -- (v1);
        \draw[thick, dashed] (v1) -- (1, 1) node[right] {$21m$};

        \draw[thick] (0.5, -2.5) node[left] {$21m$} -- (v2);
        \draw[thick] (v2) -- (2.5, -0.5) node[right] {$k$};

        \draw[thick, dashed] (v1) -- node[above right] {$p$} (v2);
        \draw[thick] (v2) -- (2.5, -2.5) node[right] {$r$};

        \filldraw (v1) circle (1.5pt);
        \filldraw (v2) circle (1.5pt);
    \end{tikzpicture}
    ,\quad
    N_{1s}^{(2)}\supset
    \begin{tikzpicture}[scale=0.55, baseline=(current bounding box.center)]
        \coordinate (v1) at (0, 0);
        \coordinate (v2) at (1.5, -1.5);

        \draw[thick] (-1, 1) node[left] {$100$} -- (v1);
        \draw[thick] (-1, -1) node[left] {$k$} -- (v1);
        \draw[thick] (v1) -- (1, 1) node[right] {$r$};

        \draw[thick] (0.5, -2.5) node[left] {$r$} -- (v2);
        \draw[thick] (v2) -- (2.5, -0.5) node[right] {$k$};

        \draw[thick, dashed] (v1) -- node[above right] {$p$} (v2);
        \draw[thick, dashed] (v2) -- (2.5, -2.5) node[right] {$100$};

        \filldraw (v1) circle (1.5pt);
        \filldraw (v2) circle (1.5pt);
    \end{tikzpicture}
    $
    \caption{Some of the diagrams contributing to the populations of $1s/2p$-states at second order. $k/p/r$ denote unbound states characterized by the corresponding momentum. The first two diagrams lead to exponential growth of $1s/2p$-halos, the third diagram's contribution is suppressed.}
    \label{fig:capture diagrams}
\end{figure}

The third diagram in fig. \ref{fig:capture diagrams} is suppressed, since conservation of energy gives $\omega_{1s}+\omega_k=\omega_r+\omega_p$:
\begin{equation}
    \frac{k^2}{2\mu}=\frac{p^2}{2\mu}+\frac{r^2}{2\mu}+\frac{\mu\alpha^2}{2}\Rightarrow|k|\ge\mu\alpha=\xi\mu v_\dm\gg\mu v_\dm.
\end{equation}
So for focusing parameter $\xi\gg1$, energy conservation forces $k$ to be much higher than the typical momentum of oncoming dark matter, suppressing the contribution, as $N_{1s}^{(2)}\sim f(k)$. The first two diagrams do not experience this suppression, as their high-energy momentum is the intermediate state, whose occupation number does not enter the expression.

From now on, we will limit ourselves to the two leading diagrams. We will consider them together denoting $s=1s/2p$. The symmetry factor for these diagrams is 2, as out of 4 possible swaps of $k/r$ legs, only 2 are different when we account for relabeling $k\leftrightarrow r$. Denoting $\Delta\omega=\omega_r+\omega_k-\omega_s-\omega_p$, we get:
\begin{multline}
    N_s^{(2)}\supset2g^2N_s\int\frac{d^3\mathbf k\ d^3\mathbf p\ d^3\mathbf r}{(2\pi)^9}f(\mathbf k)f(\mathbf r)
    \left|M_{\mathbf r+\mathbf k\rightarrow s+\mathbf p}\right|^2
    \int_0^tdt'\int_0^{t'}dt''e^{i\left(t'-t''\right)\Delta\omega}+\cc\\
    =2g^2N_s \int\frac{d^3\mathbf k\ d^3\mathbf p\ d^3\mathbf r}{(2\pi)^9}f(\mathbf k)f(\mathbf r)
    \left|\int d^3x\ \psi_{\mathbf r}\psi_{\mathbf k}\overline{\psi_s\psi_{\mathbf p}}\right|^2\cdot2\pi t\delta(\Delta\omega).
\end{multline}
Now we first take the time derivative of both sides and expand $\delta(\Delta\omega)=\frac{\mu}{p_0}\delta\left(p-p_0\right)$, with $p_0$ being the modulus of $\mathbf{p}$ needed to satisfy energy conservation. To get the scaling of the expression, we change variables to a more convenient set $\tilde k=\frac{\mathbf k}{\mu v_\dm}$, $\tilde\psi_{nlm}=r_b^{3/2}\psi_{nlm}$, $\tilde\psi_k=\sqrt{kr_b}\psi_k$, $\tilde x=\frac{x}{r_b}$:
\begin{multline}
    \label{eq:growth rate}
    \dot N_s\supset4\pi g^2N_s\int\frac{d^3\mathbf k\ d^3\mathbf p\ d^3\mathbf r}{(2\pi)^9}f(\mathbf k)f(\mathbf r)
    \frac{\mu}{p}\delta\left(p-p_0\right)\left|\int d^3x\ \psi_{\mathbf r}\psi_{\mathbf k}\overline{\psi_s\psi_{\mathbf p}}\right|^2\\
    =\frac{N_s}{2\pi^5}\frac{g^2\rho_\dm^2}{v_\dm^2\mu^3}\int d^3\tilde k\ d^3\tilde r\ dS_{\hat p}
    \frac{e^{-(\tilde k-\hat v_\dm)^2}}{\tilde k}\frac{e^{-(\tilde r-\hat v_\dm)^2}}{\tilde r}
    \left|\int d^3\tilde x\ \tilde\psi_{\mathbf r}\tilde\psi_{\mathbf k}\overline{\tilde\psi_s\tilde\psi_{\mathbf p}}\right|^2
    =\frac{N_s}{\tau_\rel}\frac{1}{2\pi^5}\mathcal I_{\infty\rightarrow s}\left(\xi\right)
    =N_s\Gamma_{\infty\rightarrow s}.
\end{multline}
At leading order the remaining integral in \eqref{eq:growth rate} denoted with $\mathcal I$ depends only on the chosen process and the focusing parameter $\xi$. It was evaluated numerically as described in Appendix \ref{app:Numerical matrix element}. For numerical results and comparison against the $2p\rightarrow2s$ decay from subsection \ref{subsec:Decay 2p->2s} see fig. \ref{fig:capture vs decay}.

\subsection{Decay $2p\rightarrow2s$ $\left(\Gamma_{2p\rightarrow2s}\right)$}
\label{subsec:Decay 2p->2s}
In this subsection we discuss a process competing with the exponential growth of the $2p$-states discussed above: the $2p\rightarrow2s$ decay. As the $2s$ state has a similar decay time into the BH as the $1s$ state, its population will almost always be low, which is also confirmed by the numerical simulation, so the main contribution comes from the diagram that has $2s$ as the intermediate state:

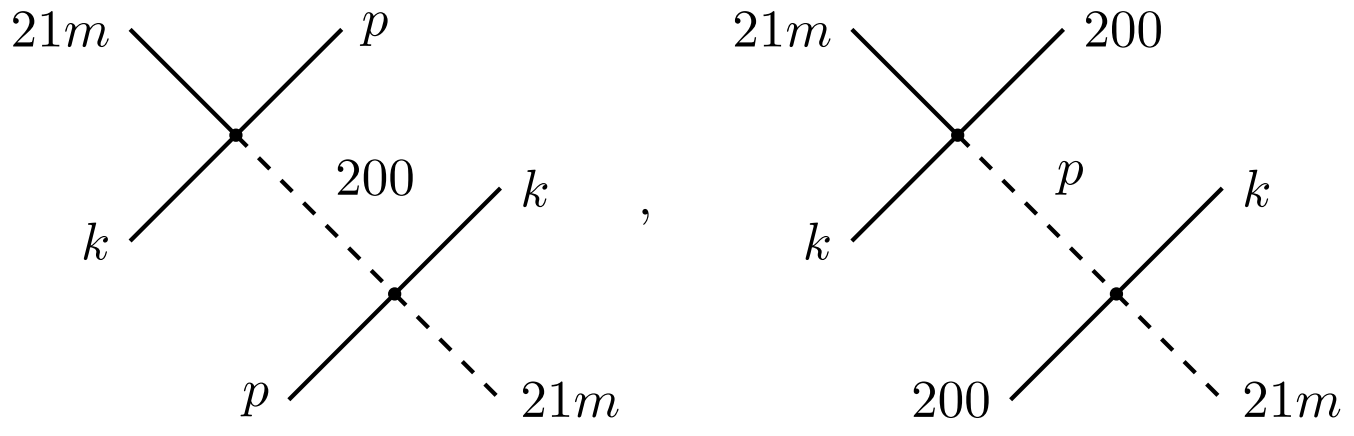
\begin{figure}[H]
    \centering
    $\displaystyle
    N_{2p}^{(2)}\supset
    \begin{tikzpicture}[scale=0.7, baseline=(current bounding box.center)]
        \coordinate (v1) at (0, 0);
        \coordinate (v2) at (1.5, -1.5);

        \draw[thick] (-1, 1) node[left] {$21m$} -- (v1);
        \draw[thick] (-1, -1) node[left] {$k$} -- (v1);
        \draw[thick] (v1) -- (1, 1) node[right] {$p$};

        \draw[thick] (0.5, -2.5) node[left] {$p$} -- (v2);
        \draw[thick] (v2) -- (2.5, -0.5) node[right] {$k$};

        \draw[thick, dashed] (v1) -- node[above right] {$200$} (v2);
        \draw[thick, dashed] (v2) -- (2.5, -2.5) node[right] {$21m$};

        \filldraw (v1) circle (1.5pt);
        \filldraw (v2) circle (1.5pt);
    \end{tikzpicture}
    ,\quad
    \begin{tikzpicture}[scale=0.7, baseline=(current bounding box.center)]
        \coordinate (v1) at (0, 0);
        \coordinate (v2) at (1.5, -1.5);

        \draw[thick] (-1, 1) node[left] {$21m$} -- (v1);
        \draw[thick] (-1, -1) node[left] {$k$} -- (v1);
        \draw[thick] (v1) -- (1, 1) node[right] {$200$};

        \draw[thick] (0.5, -2.5) node[left] {$200$} -- (v2);
        \draw[thick] (v2) -- (2.5, -0.5) node[right] {$k$};

        \draw[thick, dashed] (v1) -- node[above right] {$p$} (v2);
        \draw[thick, dashed] (v2) -- (2.5, -2.5) node[right] {$21m$};

        \filldraw (v1) circle (1.5pt);
        \filldraw (v2) circle (1.5pt);
    \end{tikzpicture}
    $
    \caption{Some of the diagrams contributing to the population in the $2p$-states coming from $2p\rightarrow2s$. The first diagram leads to exponential depletion of the $2p$-halo, the second diagram's contribution is suppressed by the number of particles in the $2s$ state $N_{2s}$.}
    \label{fig:2p->2s decay}
\end{figure}

It is important to note that unlike in subsection \ref{subsec:Capture}, the first diagram is not necessarily suppressed by $f(p)$, as the fine splitting energy difference is much smaller: $\omega_{2p}-\omega_{2s}=\mu\alpha^4/2$, so the energy conservation becomes:
\begin{equation}
    \frac{p^2}{2\mu}=\frac{k^2}{2\mu}+\omega_{2p}-\omega_{2s}\Rightarrow|p|\ge\alpha\xi\cdot\mu v_\dm,
\end{equation}
so for this decay to be suppressed we would need $\xi\ge 1/\alpha$, which is a much stronger requirement.

The symmetry factor for the diagrams in fig. \ref{fig:2p->2s decay} is 4. Performing the same evaluation as in subsection \ref{subsec:Capture}:
\begin{equation}
    \label{eq:2p->2s decay rate}
    \begin{aligned}
        &\dot N_{21m}\supset-N_{21m}\Gamma_\rel \frac{8\xi^2}{\pi^2}\mathcal{I}_{21m\rightarrow2s}(v_\dm,\xi),\quad
        \Gamma_{2p\rightarrow2s}=\Gamma_\rel \frac{8\xi^2}{\pi^2}
        \frac{\mathcal{I}_{211\rightarrow2s}+\mathcal{I}_{210\rightarrow2s}+\mathcal{I}_{21-\!1\rightarrow2s}}{3},\\
        &\mathcal{I}_{21m\rightarrow2s}(v_\dm,\xi)
        =\int d^3\tilde k\ dS_{\hat p}
        \frac{e^{-(\tilde k-\hat v_\dm)^2}}{\tilde k}e^{-(\tilde p-\hat v_\dm)^2}
        \left|\int d^3\tilde x\ \tilde\psi_{21m}\tilde\psi_{\mathbf k}\overline{\tilde\psi_{2s}\tilde\psi_{\mathbf p}}\right|^2.
    \end{aligned}
\end{equation}

Evaluating the integrals numerically, we see that for low $\xi$ the $\infty\rightarrow2p$ capture is faster than the $2p\rightarrow2s$ decay, for moderate $\xi$ the $\xi^2$ in the decay rate makes it faster than the capture, and for high $\xi$ the decay starts being suppressed by $f(p)$ and the capture overpowers it again. This pattern can be seen in fig. \ref{fig:capture vs decay}:  

\begin{figure}[H]
    \centering
    \includegraphics[width=0.95\textwidth]{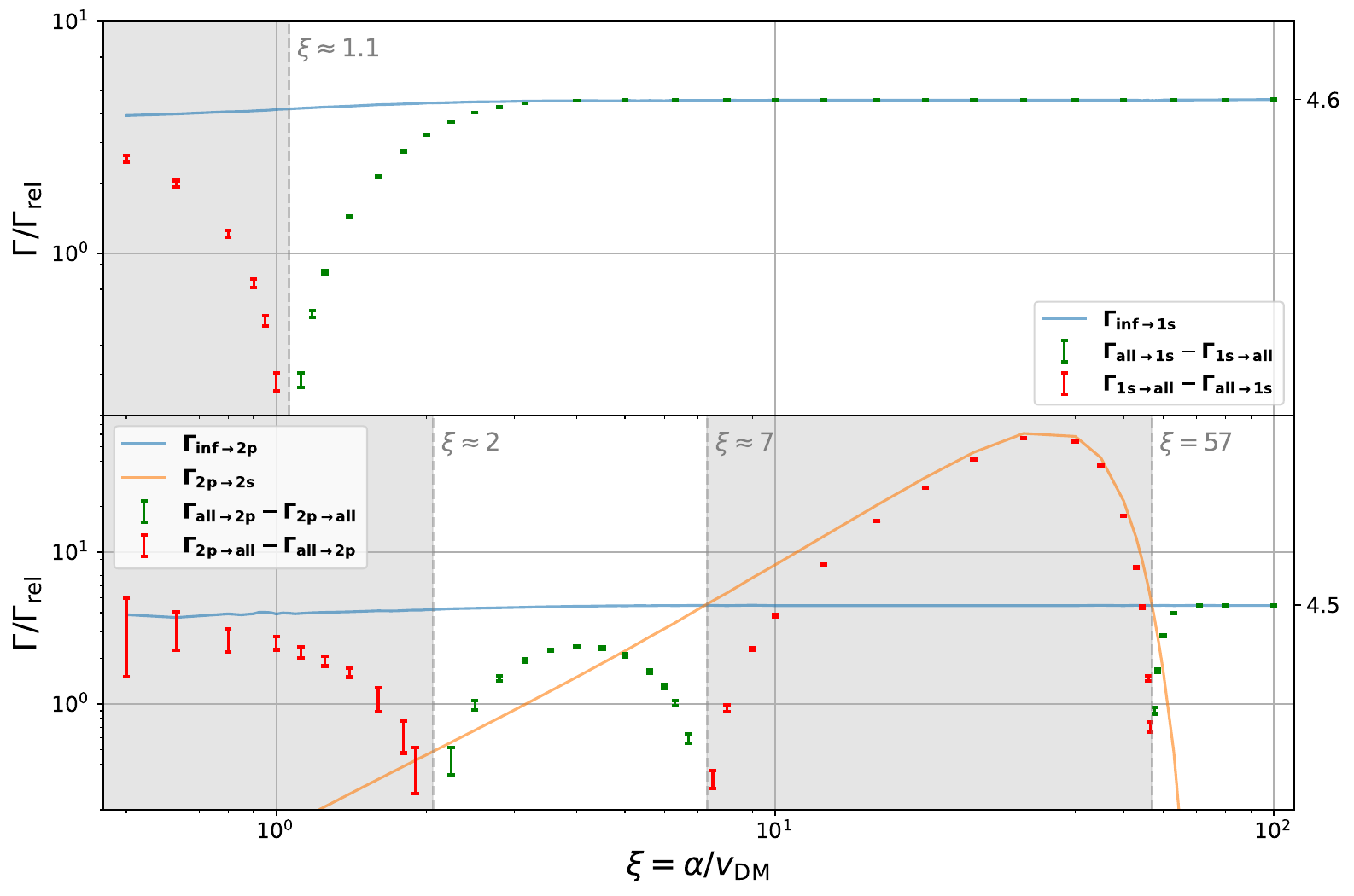}
    \caption{Growth and decay rates for $1s$ (top) and $2p$ (bottom) states. All data in both plots are numerical results for $v_\dm=240\ \text{km/s}$. Blue lines show the capture rate $\Gamma_{\infty\rightarrow1s/2p}$. The orange line in the bottom plot is the decay rate $\Gamma_{2p\rightarrow2s}$, error bars on these lines are omitted. The green and red error bar markers show the total rate of growth and decay respectively, accounting for all processes. \textbf{Top:} In the shaded region $\xi<1.1$ the diagrams like diagram 3 in fig. \ref{fig:capture diagrams} are not negligible, leading to an overall decay. For $\xi\gtrsim3$, the $1s$ rate approaches a constant value $\approx4.6\;\Gamma_\rel$. \textbf{Bottom:} For the shaded region $\xi<2$, we also get an overall decay. In the intermediate region $2<\xi<7$ capture prevails with the peak rate $\approx2.5\;\Gamma_\rel$. In the second shaded region $7<\xi<57$ the $2p\rightarrow2s$ decay is faster than the $\infty\rightarrow2p$ capture, so we get an overall decay. For $\xi>57$ the $2p\rightarrow2s$ decay becomes suppressed and the capture rate rises to $\approx4.5\;\Gamma_\rel$.}
    \label{fig:capture vs decay}
\end{figure}

\subsection{Decay $2p\rightarrow1s$ for $\xi\gg1$ $\left(C_{2p\rightarrow1s}^{1s}\right)$}
\label{subsec:Decay 2p->1s}
In this subsection we consider the $2p\rightarrow1s$ decay for large $\xi$. Similar to subsection \ref{subsec:Capture}, as we take $\xi\gg1$, the energy difference between $2p$ and $1s$ states is so high, that the high-momentum unbound state needs to be intermediate, giving us the following diagrams:
\begin{figure}[H]
    \centering
    $\displaystyle
    N_{211}^{(2)}\supset
    \begin{tikzpicture}[scale=0.5, baseline=(current bounding box.center)]
        \coordinate (v1) at (0, 0);
        \coordinate (v2) at (1.5, -1.5);

        \draw[thick] (-1, 1) node[left] {$211$} -- (v1);
        \draw[thick] (-1, -1) node[left] {$211$} -- (v1);
        \draw[thick] (v1) -- (1, 1) node[right] {$100$};

        \draw[thick] (0.5, -2.5) node[left] {$100$} -- (v2);
        \draw[thick] (v2) -- (2.5, -0.5) node[right] {$211$};

        \draw[thick, dashed] (v1) -- node[above right] {$k$} (v2);
        \draw[thick, dashed] (v2) -- (2.5, -2.5) node[right] {$211$};

        \filldraw (v1) circle (1.5pt);
        \filldraw (v2) circle (1.5pt);
    \end{tikzpicture}
    ,\ 
    \begin{tikzpicture}[scale=0.5, baseline=(current bounding box.center)]
        \coordinate (v1) at (0, 0);
        \coordinate (v2) at (1.5, -1.5);

        \draw[thick] (-1, 1) node[left] {$211$} -- (v1);
        \draw[thick] (-1, -1) node[left] {$210$} -- (v1);
        \draw[thick] (v1) -- (1, 1) node[right] {$100$};

        \draw[thick] (0.5, -2.5) node[left] {$100$} -- (v2);
        \draw[thick] (v2) -- (2.5, -0.5) node[right] {$210$};

        \draw[thick, dashed] (v1) -- node[above right] {$k$} (v2);
        \draw[thick, dashed] (v2) -- (2.5, -2.5) node[right] {$211$};

        \filldraw (v1) circle (1.5pt);
        \filldraw (v2) circle (1.5pt);
    \end{tikzpicture}
    ,\ 
        \begin{tikzpicture}[scale=0.5, baseline=(current bounding box.center)]
        \coordinate (v1) at (0, 0);
        \coordinate (v2) at (1.5, -1.5);

        \draw[thick] (-1, 1) node[left] {$211$} -- (v1);
        \draw[thick] (-1, -1) node[left] {$21\!-\!\!1$} -- (v1);
        \draw[thick] (v1) -- (1, 1) node[right] {$100$};

        \draw[thick] (0.5, -2.5) node[left] {$100$} -- (v2);
        \draw[thick] (v2) -- (2.5, -0.5) node[right] {$21\!-\!\!1$};

        \draw[thick, dashed] (v1) -- node[above right] {$k$} (v2);
        \draw[thick, dashed] (v2) -- (2.5, -2.5) node[right] {$211$};

        \filldraw (v1) circle (1.5pt);
        \filldraw (v2) circle (1.5pt);
    \end{tikzpicture}
    ,\ 
    \begin{tikzpicture}[scale=0.5, baseline=(current bounding box.center)]
        \coordinate (v1) at (0, 0);
        \coordinate (v2) at (1.5, -1.5);

        \draw[thick] (-1, 1) node[left] {$210$} -- (v1);
        \draw[thick] (-1, -1) node[left] {$210$} -- (v1);
        \draw[thick] (v1) -- (1, 1) node[right] {$100$};

        \draw[thick] (0.5, -2.5) node[left] {$100$} -- (v2);
        \draw[thick] (v2) -- (2.5, -0.5) node[right] {$21\!-\!\!1$};

        \draw[thick, dashed] (v1) -- node[above right] {$k$} (v2);
        \draw[thick, dashed] (v2) -- (2.5, -2.5) node[right] {$211$};

        \filldraw (v1) circle (1.5pt);
        \filldraw (v2) circle (1.5pt);
    \end{tikzpicture}
    ,
    $

    $\displaystyle
    N_{210}^{(2)}\supset
    \begin{tikzpicture}[scale=0.5, baseline=(current bounding box.center)]
        \coordinate (v1) at (0, 0);
        \coordinate (v2) at (1.5, -1.5);

        \draw[thick] (-1, 1) node[left] {$210$} -- (v1);
        \draw[thick] (-1, -1) node[left] {$210$} -- (v1);
        \draw[thick] (v1) -- (1, 1) node[right] {$100$};

        \draw[thick] (0.5, -2.5) node[left] {$100$} -- (v2);
        \draw[thick] (v2) -- (2.5, -0.5) node[right] {$210$};

        \draw[thick, dashed] (v1) -- node[above right] {$k$} (v2);
        \draw[thick, dashed] (v2) -- (2.5, -2.5) node[right] {$210$};

        \filldraw (v1) circle (1.5pt);
        \filldraw (v2) circle (1.5pt);
    \end{tikzpicture}
    ,\ 
    \begin{tikzpicture}[scale=0.5, baseline=(current bounding box.center)]
        \coordinate (v1) at (0, 0);
        \coordinate (v2) at (1.5, -1.5);

        \draw[thick] (-1, 1) node[left] {$210$} -- (v1);
        \draw[thick] (-1, -1) node[left] {$211$} -- (v1);
        \draw[thick] (v1) -- (1, 1) node[right] {$100$};

        \draw[thick] (0.5, -2.5) node[left] {$100$} -- (v2);
        \draw[thick] (v2) -- (2.5, -0.5) node[right] {$211$};

        \draw[thick, dashed] (v1) -- node[above right] {$k$} (v2);
        \draw[thick, dashed] (v2) -- (2.5, -2.5) node[right] {$210$};

        \filldraw (v1) circle (1.5pt);
        \filldraw (v2) circle (1.5pt);
    \end{tikzpicture}
    ,\ 
        \begin{tikzpicture}[scale=0.5, baseline=(current bounding box.center)]
        \coordinate (v1) at (0, 0);
        \coordinate (v2) at (1.5, -1.5);

        \draw[thick] (-1, 1) node[left] {$210$} -- (v1);
        \draw[thick] (-1, -1) node[left] {$21\!-\!\!1$} -- (v1);
        \draw[thick] (v1) -- (1, 1) node[right] {$100$};

        \draw[thick] (0.5, -2.5) node[left] {$100$} -- (v2);
        \draw[thick] (v2) -- (2.5, -0.5) node[right] {$21\!-\!\!1$};

        \draw[thick, dashed] (v1) -- node[above right] {$k$} (v2);
        \draw[thick, dashed] (v2) -- (2.5, -2.5) node[right] {$210$};

        \filldraw (v1) circle (1.5pt);
        \filldraw (v2) circle (1.5pt);
    \end{tikzpicture}
    ,\ 
    \begin{tikzpicture}[scale=0.5, baseline=(current bounding box.center)]
        \coordinate (v1) at (0, 0);
        \coordinate (v2) at (1.5, -1.5);

        \draw[thick] (-1, 1) node[left] {$211$} -- (v1);
        \draw[thick] (-1, -1) node[left] {$21\!-\!\!1$} -- (v1);
        \draw[thick] (v1) -- (1, 1) node[right] {$100$};

        \draw[thick] (0.5, -2.5) node[left] {$100$} -- (v2);
        \draw[thick] (v2) -- (2.5, -0.5) node[right] {$210$};

        \draw[thick, dashed] (v1) -- node[above right] {$k$} (v2);
        \draw[thick, dashed] (v2) -- (2.5, -2.5) node[right] {$210$};

        \filldraw (v1) circle (1.5pt);
        \filldraw (v2) circle (1.5pt);
    \end{tikzpicture}
    .
    $
    \caption{Some of the diagrams contributing to the populations in the $2p$-states coming from $2p\rightarrow1s$. The diagrams contributing to $N_{21-\!1}^{(2)}$ are similar to ones for $N_{211}^{(2)}$.}
    \label{fig:2p->1s decay}
\end{figure}
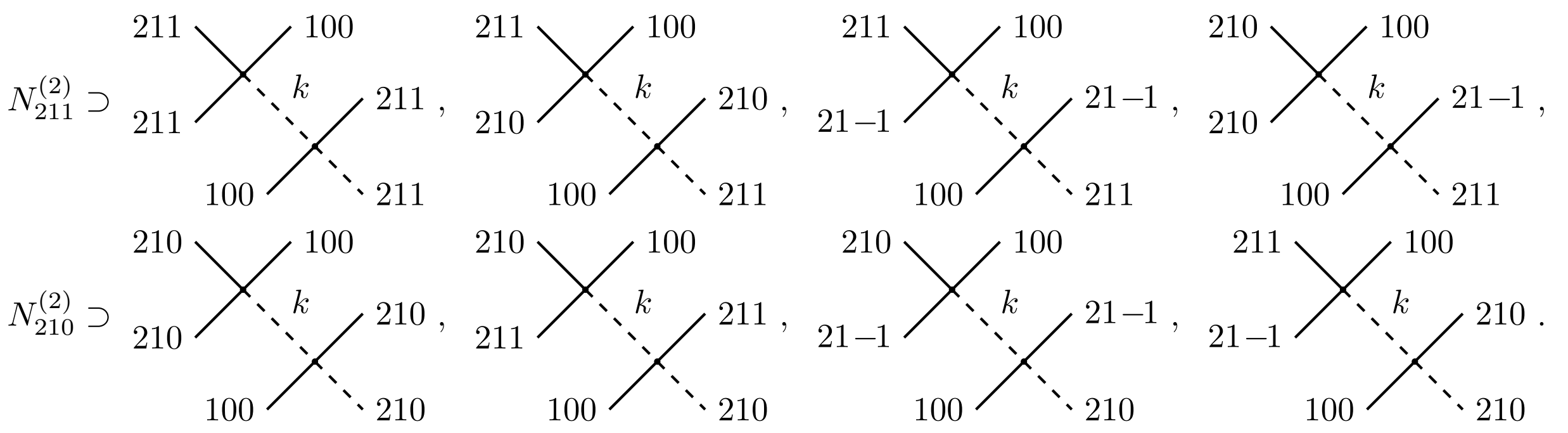

This process is notable because it has a rate proportional to the square of the current number of particles: $\dot N_{2p}\propto -N_{2p}^2$, creating an upper bound on the $2p$-halo population, so for a large enough $N_{2p}$ the $2p\rightarrow1s$ decay is faster than the capture terms.

The symmetry factor for diagrams in the first column is 2, as the dashed line is counted as distinguishable from a solid one; in the second and third columns it is 4; in the fourth column for diagrams contributing to $N_{21\pm1}$ it is 2, and to $N_{210}$ the symmetry factor is 4. It is important to note that half of the decaying $2p$-particles goes to unbound states and half to the $1s$-state. However, decrease of $N_{2p}$ by one corresponds to losing three particles --- one for every $m$, so for this process the rate of growth of the $1s$-state is $3/2$ the rate of the decay of $N_{2p}$.

The diagrams in the first three columns, which have the same incoming and outgoing states, are called symmetric. They will be straightforward to evaluate, denoting the number of unbound particles in a cube with a side of the reduced de Broglie wavelength $N_k\equiv\rho_\dm \lambda_\db^3/\mu=\rho_\dm/(\mu^4 v_\dm^3)$, we get:
\begin{equation}
    \label{eq:2p->1s decay rate}
    \begin{aligned}
        &\dot N_{21m}\supset-\Gamma_\rel \frac{\left(2-\delta^m_{m'}\right)\xi^4}{2\pi^2}
        N_{1s}\frac{N_{21m}}{N_k}\frac{N_{21m'}}{N_k}\mathcal{I}_{21m+21m'\rightarrow1s},\\
        &\mathcal{I}_{21m+21m'\rightarrow1s}
        =\int dS_{\hat k}
        \left|\int d^3\tilde x\ \tilde\psi_{21m}\tilde\psi_{21m'}\overline{\tilde\psi_{1s}\tilde\psi_{\mathbf k}}\right|^2.
    \end{aligned}
\end{equation}

It is interesting to note that this integral is independent of $v_\dm$ and is practically independent of $\xi$. There is only one unbound state in the diagram, so its magnitude is defined by the other states, and as it is an intermediate state, its occupation number does not enter the expression, so $v_\dm$ is not involved in any way. The dependence on $\xi$ (or, equivalently, on $\alpha$) does not appear at leading order: 

\begin{equation}
    k=\sqrt{2\mu E}\approx\sqrt{2\mu\left(\frac{\mu\alpha^2}{2}-2\frac{\mu\alpha^2}{8}\right)}=\frac{1}{\sqrt{2}}\mu\alpha=\frac{1}{\sqrt{2}}r_b^{-1}.
\end{equation}

That means that the momentum scales with the Bohr radius, so whenever we change $\xi$, the bound and unbound wavefunctions scale in the same way. However, there is still some dependence left, for example, coming from $\alpha^4$ corrections to the energies of bound states.

The diagrams from the fourth column are a bit trickier, as they depend not on the real population numbers, but on a complex combination of their amplitudes $c_{210}^2\overline{c_{211}c_{21-\!1}}$. We can consider these contributions together and show that the overall outcome is still the decay of $2p$-states:
\begin{equation}
    \label{eq:2p->1s mixing rate}
    \begin{aligned}
        &\sum_m\dot N_{21m}\supset-\Gamma_\rel \frac{2\xi^4}{\pi^2}N_{1s}\frac{\re\left(c_{210}^2\overline{c_{211}c_{21-\!1}}\right)}{N_k^2}
        \mathcal{I}_{\text{mix}\rightarrow1s},\\
        &\mathcal{I}_{\text{mix}\rightarrow1s}
        =\int dS_{\hat k}
        \int d^3\tilde x\ \tilde\psi_{210}\tilde\psi_{210}\overline{\tilde\psi_{1s}\tilde\psi_{\mathbf k}}
        \int d^3\tilde y\ \tilde\psi_{1s}\tilde\psi_{\mathbf k}\overline{\tilde\psi_{211}\tilde\psi_{21-\!1}}.
    \end{aligned}
\end{equation}

The results of \eqref{eq:2p->1s mixing rate} are purely real, and the integral is still independent of $\xi$, but the $\re\left(c_{210}^2\overline{c_{211}c_{21-\!1}}\right)$ can be negative, leading to a positive contribution to the $2p$-state population. However, as we have numerically verified, this potentially positive contribution is never greater than the negative contributions of the diagrams in the first three columns. Denoting $\arg\left(c_{210}^2\overline{c_{211}c_{21-\!1}}\right)=\beta$, we get:
\begin{multline}
    \label{eq:2p->1s with phases}
    \sum_m\dot N_{21m}\supset-\Gamma_\rel\frac{\xi^4}{2\pi^2}\frac{N_{1s}}{N_k^2}
    \left(\sum_{m,m'}\left(2-\delta^m_{m'}\right)N_{21m}N_{21m'}
    \mathcal{I}_{21m+21m'\rightarrow1s}
    +4\re\left(c_{210}^2\overline{c_{211}c_{21-\!1}}\right)
    \mathcal{I}_{\text{mix}\rightarrow1s}\right)\\
    \begin{aligned}
        =-\Gamma_\rel\frac{\xi^4}{2\pi^2}\frac{N_{1s}}{N_k^2}\cdot10^{-4}
        \Big(6.58\;N_{211}N_{210}
        +&3.51\;N_{210}\sqrt{N_{211}N_{21-\!1}}\cos\beta\\
        +3.29\;N_{211}^2+2.41\;N_{210}^2
        +&3.29\;N_{21-\!1}^2
        +3.07\;N_{211}N_{21-\!1}+6.58\;N_{210}N_{21-\!1}\Big),
    \end{aligned}
\end{multline} 
where we approximated the dependence $\mathcal{I}(\xi)$ by a constant, which has the relative error of less than $1\%$. The phase $\beta$ of the mixing term changes with time and will increase or decrease the rate of this decay. This leads to the biggest relative change for $N_{211}=N_{21-\!1}\approx N_{210}/2$, where it leads to up to $\approx15\%$ change in the rate of the process.

Taking $N_{211}, N_{210}, N_{21-\!1}\approx N_{2p}$, we get this simplified expression:
\begin{equation}\label{eq:2p->1s decay high xi}
    \dot N_{2p}\approx\frac{1}{3}\sum_m\dot N_{21m}
    \supset-\Gamma_\rel \xi^4\frac{N_{1s}N_{2p}^2}{ N_k^2}\cdot\left(4.3\pm0.6\right)\cdot10^{-5}\equiv-\Gamma_\rel \xi^4\frac{N_{1s}N_{2p}^2}{ N_k^2}\gamma^{1s}_{2p \rightarrow1s}
    \equiv-C_{2p\rightarrow1s}^{1s}N_{2p}^2N_{1s},
\end{equation}
where the $\pm0.6$ uncertainty mostly comes from the unknown phase $\beta$, and not from the numerical error.

\subsection{Decay $2p\rightarrow1s$ for $\xi\sim1$ $\left(C_{2p\rightarrow1s}^{k}\right)$}
\label{subsec:Second decay 2p->1s}
In this subsection we consider the $2p\rightarrow1s$ decay for $\xi\sim1$. In addition to the decay discussed in subsection \ref{subsec:Decay 2p->1s} where the unbound state was intermediate, for the case of $\xi\sim1$ we cannot outright neglect diagrams where the unbound state is not intermediate. 

The diagrams contributing to this process are exactly the same as the ones in fig. \ref{fig:2p->1s decay}, but with $100$ and $k$ swapped, so the decay is now proportional to $N_{2p}^2$ instead of $N_{2p}^2N_{1s}$. The symmetry factor for the diagrams is also the same, and again the rate of growth of the $1s$-state is $3/2$ the rate of the decay of $N_{2p}$. Performing the same analysis, we get:
\begin{equation}
    \begin{aligned}
        &\sum_m\dot N_{21m}\supset-\Gamma_\rel\frac{4\xi^4}{\sqrt{\pi}}\frac{1}{ N_k}
        \left(\sum_{m,m'}\left(2-\delta^m_{m'}\right)N_{21m}N_{21m'}
        \mathcal{I}_{21m+21m'\rightarrow k}
        +4\re\left(c_{210}^2\overline{c_{211}c_{21-\!1}}
        \mathcal{I}_{\text{mix}\rightarrow k}\right)\right),\\
        &\mathcal{I}_{21m+21m'\rightarrow k}
        =\int dS_{\hat k}\; e^{-\left(\tilde k - \hat v_\dm\right)^2}
        \left|\int d^3\tilde x\ \tilde\psi_{21m}\tilde\psi_{21m'}\overline{\tilde\psi_{1s}\tilde\psi_{\mathbf k}}\right|^2,\\
        &\mathcal{I}_{\text{mix}\rightarrow k}
        =\int dS_{\hat k}\; e^{-\left(\tilde k - \hat v_\dm\right)^2}
        \int d^3\tilde x\ \tilde\psi_{210}\tilde\psi_{210}\overline{\tilde\psi_{1s}\tilde\psi_{\mathbf k}}
        \int d^3\tilde y\ \tilde\psi_{1s}\tilde\psi_{\mathbf k}\overline{\tilde\psi_{211}\tilde\psi_{21-\!1}}.
    \end{aligned}
\end{equation}

Unlike the $2p\rightarrow1s$ decay in subsection \ref{subsec:Decay 2p->1s}, here $\mathcal{I}$ is dependent on $\xi$. From energy conservation we have:
\begin{equation}
    \frac{k^2}{2\mu}=2\omega_{2p}-\omega_{1s}=\frac{\mu\alpha^2}{4}\Rightarrow\left|\tilde k\right|=\frac{\left|k\right|}{\mu v_\dm}=\frac{\xi}{\sqrt{2}}.
\end{equation}
As in eq. \eqref{eq:2p->1s with phases} the mixing term can increase or decrease the rate of the decay. For $\xi<7$, where this decay can be noticeable, this change can be as high as $\approx50\%$. 

To order of magnitude, the $\xi$-dependence of the integrals can be approximated by $f(k)$: $\mathcal{I}\left(\xi\right)\sim e^{-\left(\xi/\sqrt{2}-1\right)^2}$. However, for the purpose of finding the fixed point, we parametrically fit this decay in the range $\xi\in(2, 7)$:
\begin{multline}
    \label{eq:2p->1s decay low xi}
    \dot N_{2p}\approx\frac{1}{3}\sum_m\dot N_{21m}
    \supset-\Gamma_\rel\xi^4\frac{N_{2p}^2}{ N_k}\cdot\big[1\pm\left(0.06\xi+0.15\right)\big]e^{-0.475\xi^2+0.94\xi+3.4}\cdot10^{-5}\\
    \equiv-\Gamma_\rel\xi^4\frac{N_{2p}^2}{ N_k}\gamma^{k}_{2p \rightarrow1s}
    \equiv-C_{2p\rightarrow1s}^{k}N_{2p}^2,
    \quad\xi\in(2, 7).
\end{multline}
The stated error includes both the contribution of the mixing diagram and numerical errors for the matrix elements. The fit for the central value is accurate to $2\%$.

\begin{figure}[H]
    \centering
    \includegraphics[width=\textwidth]{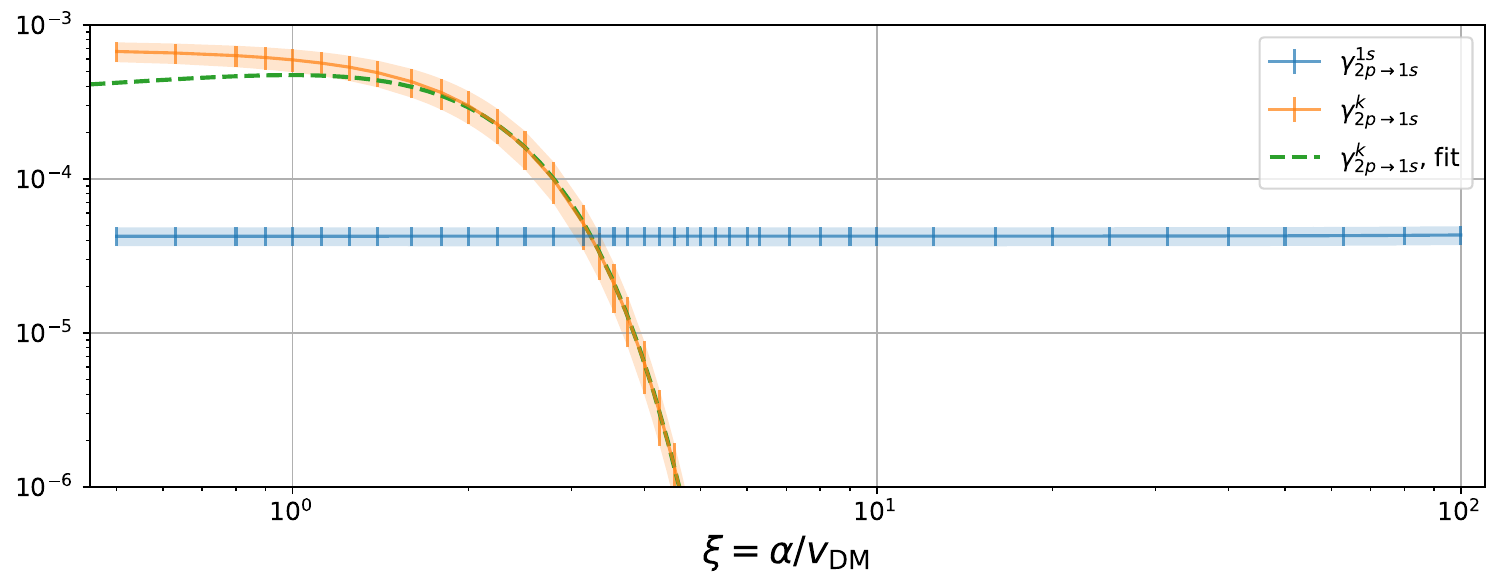}
    \caption{Comparison between the two decay modes with $\dot N\sim N_{2p}^2$. The blue line is $\gamma_{2p\rightarrow1s}^{1s}$ that gives rise to $\dot N\sim N_{2p}^2N_{1s}$ and the orange line is $\gamma_{2p\rightarrow1s}^{k}$ that gives rise to $\dot N\sim N_{2p}^2$. The green dashed line is the fit for $\gamma_{2p\rightarrow1s}^{k}$ in the region of $\xi\in(2, 7)$, where this decay mode affects halo dynamics. For $\xi<2$ the $2p$-halo does not form and for $\xi\gtrsim7$ the exponential suppression takes over and $\gamma_{2p\rightarrow1s}^{k}$ is negligible.}
    \label{fig:2p2p decays}
\end{figure}

\section{Implications}
\label{sec:implications}
As seen in the discussion of processes in section \ref{sec:dynamics} and its results in fig. \ref{fig:capture vs decay}, there are a few distinct regimes of the formation of a halo around a Schwarzschild BH. In this section we will go through every regime, establish final halo configurations and timescales of the halo formation. We will also cover some possible effects of this work on superradiance around Kerr BHs.

\subsection{DM halo parameter space for Schwarzschild BHs}
\begin{figure}[H]
    \centering
    \includegraphics[width=\textwidth]{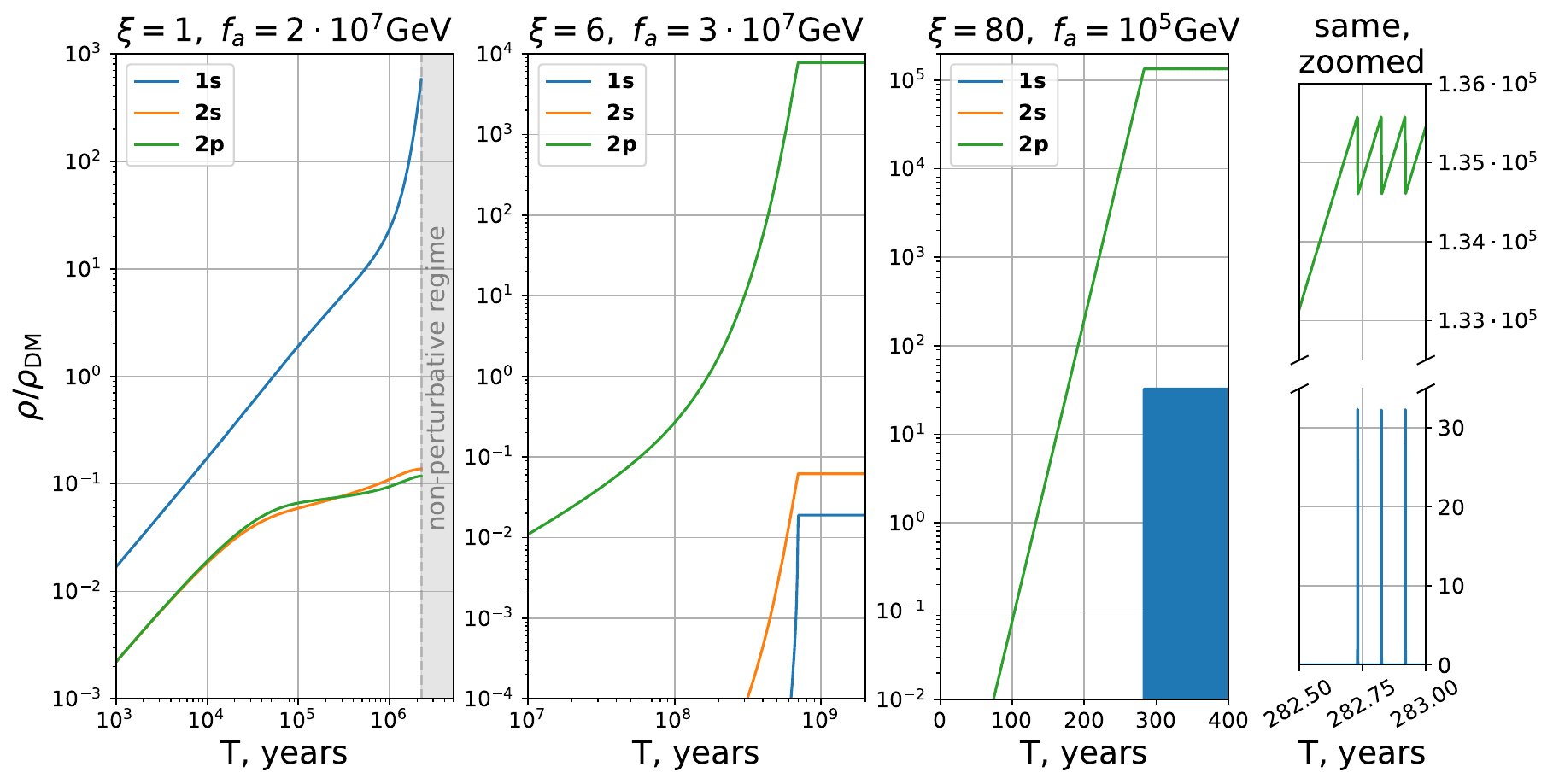}
    \caption{Evolution of the $1s$-, $2s$- and $2p$-halo densities at $r=r_b$ over time for selected points in the $\left(\xi,f_a\right)$ plane representing different regimes discussed below. From left to right: the $1s$-halo regime, dilute $2p$-halo regime and dense $2p$-halo regime. The fourth plot is a zoom in on the oscillations in the third plot. All plots are for a Schwarzschild BH with a mass of $10\ M_{\astrosun}$ and a distance of $r=1\ \text{kpc}$ from the galactic center (meaning a background density of $\rho_\dm\approx6\ \GeV/\cm^3$).}
    \label{fig:density over time}
\end{figure}

The ODE with full dynamics of the $1s/2s/2p$-states accounting for all their diagrams and the decay into the BH was solved numerically using the Euler method with adaptive step~\cite{code}. The simulation only tracks the occupation numbers; for the processes dependent on phases of amplitudes we always assume the phase giving a maximal decay. Results of the simulation for interesting points in the parameter space can be seen in fig. \ref{fig:density over time}, while the full list of possible regimes will be discussed below. 

\paragraph{No halo regime}\mbox{}\\
For $\xi\lesssim1$ both $1s$ and $2p$ states experience exponential decay from ionization even without accounting for the BH's absorption, so no large halo can be formed based on the described ODEs. It is possible that processes involving states with higher $n$ or third- or higher-order processes for lower $f_a$ might lead to an overall growth of the $1s$-halo, which might explain the results of the full GP equation simulations shown in fig. 11 of~\cite{gilads_halo_paper}. 

Similarly, for $1.1<\xi<2$ if $\Gamma_{1s\rightarrow\BH}+\Gamma_{1s\rightarrow\text{all}}>\Gamma_{\text{all}\rightarrow1s}$ the decay of $1s$ into the BH is faster than the capture process and no large halo is formed.

\paragraph{1s-halo regime}\mbox{}\\
For $\xi>1.1$ and $\Gamma_{\text{all}\rightarrow1s}>\Gamma_{1s\rightarrow\BH}+\Gamma_{1s\rightarrow\text{all}}$ the capture into the $1s$-state is faster than the decay and a $1s$-halo can form\footnote{Interestingly, the $1s$-halo can still grow exponentially for some $1\lesssim\xi<1.1$. For example, for $\xi=1$, $f_a=2\cdot10^7\ \GeV$ the remaining $2s$ and $2p$ populations are large enough to cause decays like $2p/2s+k\rightarrow1s+q$ and the $1s$ reaches the non-perturbative regime before these populations become exhausted.}. Any higher bound state rapidly decays into $1s$ for $N_{1s}\gtrsim N_k$. On the other hand, the $1s$-state does not have any decays limiting its growth, so the population of the halo will increase exponentially. An illustration of this can be seen in the first plot of fig. \ref{fig:density over time}. 

The growth continues until the halo's density reaches a point where self-interactions become strong and our perturbative expansion fails. At this point the cloud reaches saturation for repulsive self-interactions or may undergo a bosenova for attractive self-interactions. The approximate cloud mass where this happens can be evaluated by looking at the energy density of the field, as shown in Appendix \ref{app:cloud size}. This gives the mass and density:
\begin{equation}
    M_{1s}^{\max}=\mu N_{1s}^{\max}\cong  6\cdot10^{-13}\ M_{\astrosun}\left[\frac{f_{a}}{10^{8}\ \GeV}\right]^{2}\left[\frac{10^{-14}\ \eV}{\mu}\right]^{2}\left[\frac{M_{\astrosun}}{M}\right],
\end{equation}
and at this mass, the cloud reaches $\rho_\text{crit}$ at $r=0$:
\begin{equation}
    \rho_\text{crit}\cong 3\cdot 10^4 \rho_\dm \left[\frac{f_{a}}{10^{8}\ \GeV}\right]^{2}\left[\frac{\mu}{10^{-14}\ \eV}\right]^{4}\left[\frac{M}{M_{\astrosun}}\right]^{2}\left[\frac{0.4\ \GeV/\cm^{3}}{\rho_{\dm}}\right].
\end{equation}

\paragraph{Dilute $2p$-halo regime}\mbox{}\\
For $2<\xi<7$, when $\Gamma_{1s\rightarrow\BH}+\Gamma_{1s\rightarrow\text{all}}>\Gamma_{\text{all}\rightarrow1s}$ and $\Gamma_{2p\rightarrow\BH}+\Gamma_{2p\rightarrow\text{all}}<\Gamma_{\text{all}\rightarrow2p}$ the $2p$-halo grows exponentially until it reaches a fixed point where its exponential growth is in equilibrium with the $2p\rightarrow1s$ decays. An illustration of this can be seen in the second plot of fig. \ref{fig:density over time}. Due to the non-trivial $\xi$-dependence of the decays, the equilibrium point does not have a simple expression, although a semi-analytic expression can be derived (see Appendix \ref{app:static point}).

\paragraph{Intermediate-$\xi$ regime}\mbox{}\\
For $7<\xi<57$ and $\Gamma_{1s\rightarrow\BH}+\Gamma_{1s\rightarrow\text{all}}>\Gamma_{\text{all}\rightarrow1s}$ the $2p$-halo does not grow as it is dominated by the $2p\rightarrow2s$ decay and the growth of $1s$-halo is slower than its decay into the BH. It is possible that a halo of some higher $nlm$ state can be formed, although it too might decay into the BH via $nlm\rightarrow n00\rightarrow\BH$.

\paragraph{Dense $2p$-halo regime}\mbox{}\\
For $\xi>57$, $\Gamma_{1s\rightarrow\BH}+\Gamma_{1s\rightarrow\text{all}}>\Gamma_{\text{all}\rightarrow1s}$ and  $\Gamma_{2p\rightarrow\BH}+\Gamma_{2p\rightarrow\text{all}}<\Gamma_{\text{all}\rightarrow2p}$, similar to the dilute $2p$-halo regime, the $2p$-halo grows until a point where capture and $2p\rightarrow1s$ decay are in equilibrium. An illustration of this can be seen in the third plot of fig. \ref{fig:density over time}.

Distinct from the dilute $2p$-halo regime, here the $2p\rightarrow2s$ decay is suppressed due to energy conservation requiring the momentum of the produced unbound particle to be much larger than $\mu v_\dm$. As mentioned before, the processes that do not conserve energy are suppressed by the ratio of the widths to the energy difference $\Gamma/\Delta\omega\sim\alpha$. In the dense $2p$-halo regime $\alpha\sim0.1$, so this decay may affect the dynamics of the halo.  

We can estimate the fixed point semi-analytically as the point where the $1s$ exponential growth due to the $N_{2p}$ decay cancels with the exponential decay into the BH:
\begin{multline}
    \dot N_{1s}\approx+\Gamma_\rel \xi^4\frac{N_{1s}N_{2p}^2}{ N_k^2}\cdot\left(6.5\pm0.9\right)\cdot10^{-5}-\Gamma_{1s\rightarrow\BH}N_{1s}=0\\
    \Rightarrow N_{2p}(t=\infty)\approx (4\pm 0.3)\cdot 10^{3}\frac{f_a^2\sqrt{\alpha}}{\mu^2}\cong3.5\cdot 10^{63}\left[\frac{f_{a}}{10^{8}\ \GeV}\right]^{2}\left[\frac{10^{-14}\ \eV}{\mu}\right]^{3/2}\left[\frac{M}{M_{\astrosun}}\right]^{1/2},
\end{multline}
and the final cloud mass and density at $r_b$ will be (a detailed derivation for evaluating the density is given in Appendix \ref{app:cloud size}):
\begin{equation}
    M_{2p}=3\mu N_{2p}\cong (9.3\pm 0.7)\cdot10^{-17} M_{\astrosun} \left[\frac{f_{a}}{10^{8}\ \GeV}\right]^{2}\left[\frac{10^{-14}\ \eV}{\mu}\right]^{1/2}\left[\frac{M}{M_{\astrosun}}\right]^{1/2},
\end{equation}
\begin{equation}
    \rho_{2p}= (1.7\pm0.1)\cdot 10^{-2}\ \rho_\dm \left[\frac{f_{a}}{10^{8}\ \GeV}\right]^{2}\left[\frac{\mu}{10^{-14}\ \eV}\right]^{11/2}\left[\frac{M}{M_{\astrosun}}\right]^{7/2}\left[\frac{0.4\ \GeV/\cm^{3}}{\rho_{\dm}}\right].
\end{equation}

The numerical results assume maximal decay phases, so they should lie at the lower end of the quoted range.

\paragraph{Higher $n$ halos regime}\mbox{}\\
In the regime where $\Gamma_{2p\rightarrow \BH}>\Gamma_{\infty\rightarrow 2p}$ a $2p$-halo cannot form. Since $\Gamma_{3d\rightarrow \BH}\ll \Gamma_{2p\rightarrow \BH}$, there is a region in the parameter space for which the $3d$ (meaning $32m$) states are the lowest $n$ stable states, and we expect that they will form a stable halo in much the same way as the $2p$-halo. One can further estimate that there are regions where a $4f$ ($43m$ states) halo will be present, and so on (for large enough $l$ there are states that are stable everywhere in the relevant parameter space, so for an arbitrarily large $l$ state a halo cannot form). The dynamics of such high $l$ state halos are more complicated due to the larger number of states and diagrams involved, but using the formalism presented here it can be done systematically.

\begin{figure}[H]
    \centering
    \includegraphics[width=0.6\textwidth]{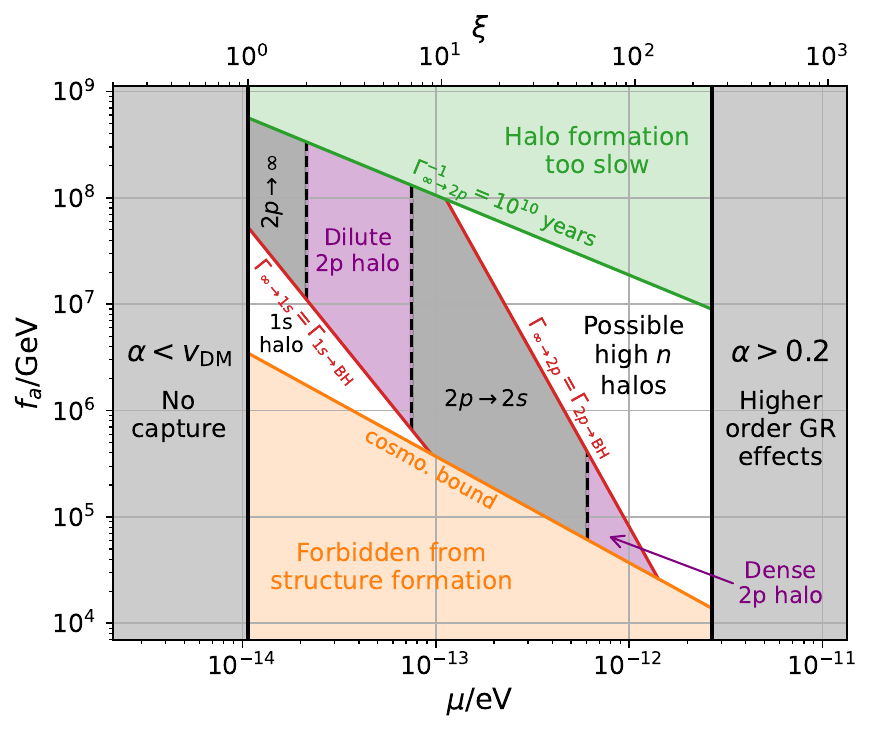}
    \caption{DM parameter space plot of $\mu$ vs. $f_a$, for a Schwarzschild BH of mass $M=10M_{\astrosun}$ and $\rho_\dm=6\ \GeV/\cm^3$, (the approximate density at 1 kpc from the galactic center). The shaded orange region is defined by eq. \eqref{eq:cosmo bound}. The purple shaded regions where a $2p$-halo forms coincide with positive-rate regions in fig. \eqref{fig:capture vs decay}. The top axis is parameterized by the focusing parameter $\xi$.}
    \label{fig:parameter space}
\end{figure}

With all of those regions defined, we can plot them schematically on the DM parameter space (fig. \ref{fig:parameter space}). Two additional restrictions are presented: that the formation time of the halo is smaller than $10^{10}\ \years$ for the halo to have enough time to form, and a cosmological upper bound on the strength of the self-interactions for ultralight DM of this mass range, described in~\cite{gilads_halo_paper}. This bound relates to possible effects of the self-interactions on the early universe evolution of DM excluded by CMB measurements, and is given by:
\begin{equation}\label{eq:cosmo bound}
    27\left[\frac{\mu}{10^{-14}\ \eV}\right]\left[\frac{f_a}{10^8\ \GeV}\right]\gtrsim1.
\end{equation}

The $2p$ cloud density is larger for large $\xi$ (mainly due to the Bohr radius scaling as $r_b\sim1/\xi^2$, meaning that the cloud mass is concentrated in a smaller volume for higher $\xi$), hence the naming of the $2p$-halo regimes. A more detailed description is given in Appendix \ref{app:static point}.

Numerical simulations of the ODEs and their resulting cloud densities for the $1s,2s,2p$ states are shown in fig. \ref{fig:density parameter space}. The density quoted is evaluated at $r=r_b$, for details see Appendix \ref{app:cloud size}. A plot showing the cloud masses, and describing the scaling of the numerical results with $M,\rho_\dm$ is given in Appendix \ref{app:mass plot and scaling}.

We can see that accounting for all the diagrams pushes the $1s$-halo region to $\xi=1$, making this region slightly larger, compared to fig. \ref{fig:capture vs decay}. For the $2p$-halo we can see that the numerical results match very well with the predicted regions and are clearly separated by the lines $\xi=2,7,57$. Energy non-conserving $2p\rightarrow 2s$ decay, which was not considered in this work, might affect the results of the dense $2p$-halo regime.

\begin{figure}[H]
    \centering
    \includegraphics[width=0.5\textwidth]{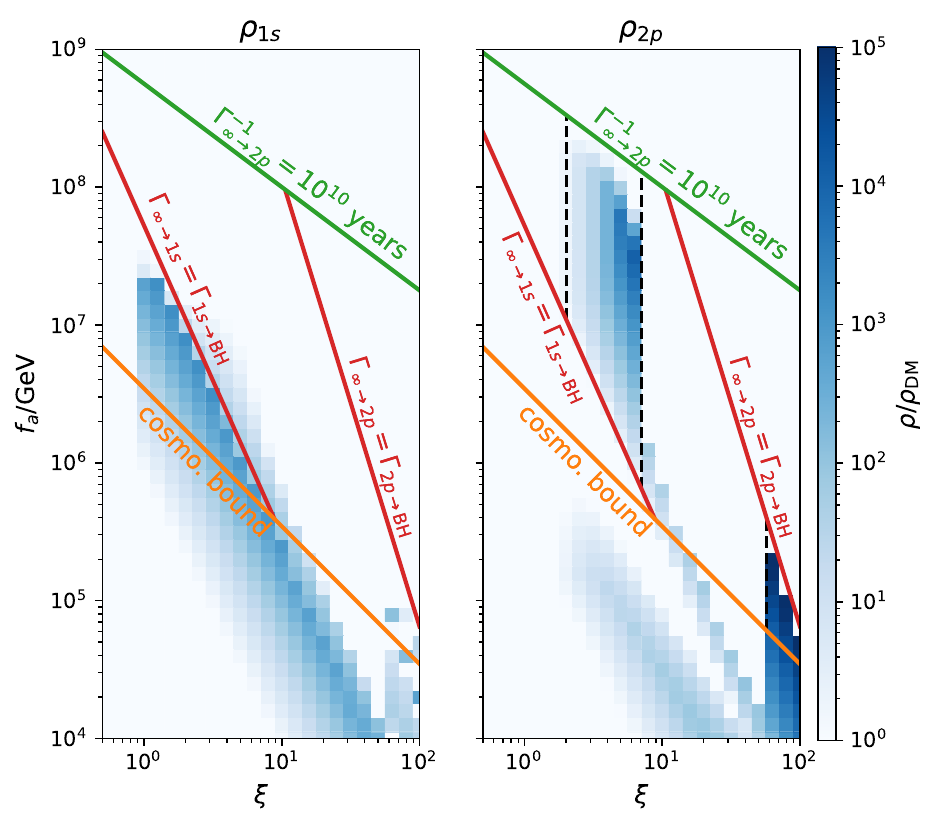}
    \caption{Halo densities at $r=r_b$ after $1.5\cdot10^{10}$ years for a Schwarzschild BH of mass $M=10 M_{\astrosun}$ at  1 kpc from the galactic center (meaning $\rho_\dm\approx6\ \GeV/\cm^3$). To help compare with the schematic parameter space plot fig. \ref{fig:parameter space}, the region boundaries are drawn on top of the numerical results.}
    \label{fig:density parameter space}
\end{figure}

\subsection{The Kerr case: superradiance}
\label{subsec:superradiance}
A different physical setup which leads to large excited state halos is the superradiance of Kerr BHs~\cite{Detweiler_1980,superradiance_review}. As in this subsection the halo forms through a purely gravitational process and not through a capture mechanism, we do not assume a background DM population. 

Although the Kerr metric is notably different from Schwarzschild, for small $\alpha$ the hydrogenic approximation still holds, and levels can be considered as the same $nlm$ states of the hydrogen atom~\cite{Detweiler_1980,gravitational_atom_energy}.

For a Kerr BH decay widths $\Gamma_{s\rightarrow\BH}$ are different from the Schwarzschild case. For the $1s$ state around a BH with dimensionless spin $\tilde a=J/(GM^2)$, with $J$ the BH's angular momentum, we have~\cite{Detweiler_1980,gravitational_atom_energy}:
\begin{equation}
    \Gamma_{1s\rightarrow\BH}^\text{Kerr}=\frac{1+\sqrt{1-\tilde a^2}}{2}\Gamma_{1s\rightarrow\BH}.
\end{equation}

While a detailed analysis of the self-interacting halo evolution is warranted, here we only try to point out a possible incompleteness of the description available in the literature. The interactions with unbound dark matter will be negligible, so we will use the hydrogen atom basis aligned with the direction of the spin of the BH. As a starting condition, we will assume that there is a large halo with $N\gg N_k$ wholly in the $211$ state and that it either still grows or that its decay into the BH is negligible. If other states are present, there will be additional decay modes that would contribute to the growth of the $1s$ state.

Unlike with the Bose-enhanced capture mechanism, the superradiance process only populates the $211$ state, so the populations of $m=0,-1$ states do not grow to become large enough to trigger transitions between the $2p$ states. Compared with the rest of this work, here we assume small self-interactions with $f_a>10^{10}\ \GeV$. The only non-negligible processes will be proportional to $N_{211}^2$, where the large $211$ population compensates for the transitions being $\sim f_a^{-4}$, and the decay of states to the BH, which is $f_a$-independent. These terms give us the following differential equation for the population of the $1s$ state and $211$ states:
\begin{equation}
    \label{eq:1s population superradiance}
    \begin{cases}
    \begin{aligned}
    \dot N_{1s}&=N_{1s}\left( \Gamma_{211\rightarrow 1s}N_{211}^2-\Gamma_{1s\rightarrow\BH}^\text{Kerr}\right)\\
    \dot N_{211}&=N_{211}\left(-2\Gamma_{211\rightarrow 1s}N_{211}N_{1s}+
    \Gamma_{\text{SR}}\right)
    \end{aligned}
    \end{cases},
\end{equation}
\begin{equation}
    \Gamma_{211\rightarrow 1s}\cong 2 \cdot 10^{-139}\ \years^{-1}\left[\frac{\mu}{10^{-14}\ \eV}\right]^{9}\left[\frac{M}{M_{\astrosun}}\right]^{4}\left[\frac{10^{8}\ \GeV}{f_{a}}\right]^{4},
\end{equation}
where $\Gamma_\text{SR}$ is the superradiance rate for the $211$ state, and $\Gamma_{211\rightarrow 1s}$ comes from the $3.29N_{211}^2$ term in eq. \eqref{eq:2p->1s with phases} which coincides with $\Gamma^{100\times\infty}_{211\times211}$ in table VI of~\cite{Baryakhtar_2021}. However, both there and in a recent analysis~\cite{Minimal_states_SR} the equation for the population of the $1s$-state was not considered, as it is rapidly absorbed by the BH. We point out that the ground state cannot be neglected, as for the $211$ occupation numbers reached during superradiant growth, the growth of $1s$ from depletion of the $211$-halo can overcome the ground state's decay into the BH. The growth of the ground state's population further Bose-enhances the depletion by a factor of $N_{1s}$ and limits the population of the $211$ cloud.

The $1s$ cloud grows if the expression in the brackets is positive, which happens for $M_{211}>M_{211}^{\text{crit}}=\mu N_{211}^{\text{crit}}$:
\begin{equation}
    \frac{M_{211}^{\text{crit}}}{M}=\frac{\mu}{M}\sqrt{\frac{\Gamma_{1s\rightarrow\BH}^\text{Kerr}}{\Gamma_{211\rightarrow1s}}}
    \cong\frac{5.25\cdot10^{-19}}{\sqrt\alpha}\sqrt{1+\sqrt{1-\tilde a^2}}\left[\frac{f_a}{10^8\ \GeV}\right]^2.
\end{equation}
This result is naturally independent of the properties of unbound DM $v_\dm$ and $\rho_\dm$, and the DM and BH masses enter only through $\alpha$.

While the first equation in \eqref{eq:1s population superradiance} is enough to find the fixed point, we need both to show that it is attractive, assuming, for simplicity, $\Gamma_{\text{SR}}=\text{const.}>0$. If $N_{211}<N_{211}^{\text{crit}}$, the $1s$-state decays into the BH, so its population is very small. Then, the rate of the decay becomes negligible and $N_{211}$ grows from the superradiant instability. On the other hand, if $N_{211}>N_{211}^{\text{crit}}$, $N_{1s}$ grows exponentially, increasing the rate of the decay until it matches $\Gamma_{\text{SR}}$, when the $211$-state starts to decay and its population falls.

We will limit ourselves to the regime where $\Gamma_{\text{SR}}>\text{Gyr}^{-1}$. For $\alpha\ll1$ we have~\cite{Detweiler_1980,gravitational_atom_energy}:
\begin{equation}
    \Gamma_{\text{SR}}\approx\frac{\alpha^8\tilde a\mu}{24}
    =2.7\cdot10^{20}\ \text{Gyr}^{-1}\alpha^9\tilde a\left[\frac{M_{\astrosun}}{M}\right]
    \quad\Rightarrow\quad\alpha>5.4\cdot10^{-3}\left(\frac{M}{\tilde aM_{\astrosun}}\right)^{1/9}.
\end{equation}

The upper limit for $\alpha$ comes from the condition that the mass of the superradiant cloud $M_{211}^\text{SR}$ in the "Large coupling" regime from eq. (55a) of~\cite{Baryakhtar_2021} is larger than $M_{211}^{\text{crit}}$:
\begin{equation}
    \frac{M_{211}^\text{SR}}{M}\approx\frac{2.6\cdot10^{-21}}{\alpha^2}\sqrt{\tilde a}\left[\frac{f_a}{10^8\ \GeV}\right]^2
    \quad\Rightarrow\quad\alpha<2.9\cdot10^{-2}\left(\frac{\tilde a}{1+\sqrt{1-\tilde a^2}}\right)^{1/3}.
\end{equation}

This limit is independent of $f_a$. However, for large enough $f_a$ the system transitions to the "Moderate coupling" regime, where the maximal size of the superradiant cloud is given by (see table III in~\cite{Baryakhtar_2021}):
\begin{equation}
    \frac{M_{211}^\text{SR,mod.}}{M}=\alpha\left(\tilde a-\frac{4\alpha}{1+4\alpha^2}\right),
\end{equation}
and requiring $M_{211}^\text{crit}<M_{211}^\text{SR,mod.}$ gives us an upper limit for $f_a$:
\begin{equation}
    M_{211}^\text{crit}<M_{211}^\text{SR,mod.}\quad\Rightarrow\quad f_a<1.4\cdot10^{17}\ \GeV\frac{\alpha^{3/4}\sqrt{\tilde a-\frac{4\alpha}{1+4\alpha^2}}}{\left(1+\sqrt{1-\tilde a^2}\right)^{1/4}}
\end{equation}
These conditions can be simultaneously fulfilled for BHs with masses up to $M\lesssim4\cdot10^6\ M_{\astrosun}$.

\begin{figure}[H]
    \centering
    \includegraphics[width=0.75\textwidth]{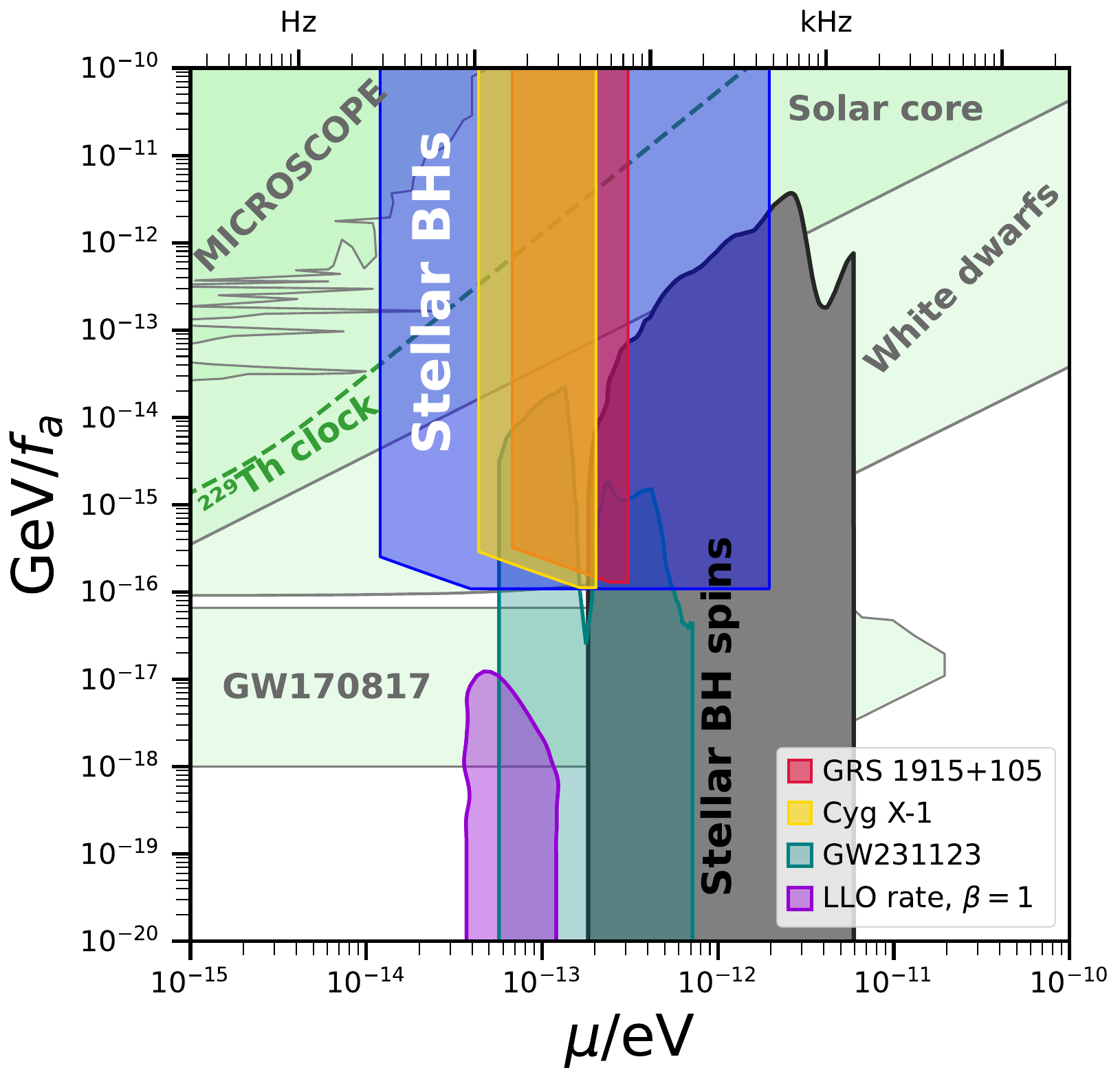}
    \caption{Plot of the QCD axion~\cite{PQ_mechanism, PQ_mechanism2, Weinberg_axion, Wilczek_axion, QCD_axion_mass, QCD_axion_mass_NNLO} parameter space, with mass $m_a$ and decay constant $f_a$. The blue patch shows the region where the superradiant population of the $211$ cloud can be large enough to trigger the growth of a $1s$-halo for stellar BHs with masses $2$--$100M_{\astrosun}$ and spin $\tilde a\approx1$. The yellow and crimson patches show these regions for BHs with the same masses and spins as Cyg X-1 and GRS 1915+105 respectively and black represents the exclusion region derived from spin measurements of these BHs. The teal and purple patches represent exclusion region from the GW231123 event~\cite{GW231123_SR} and the BH merger rate~\cite{Mergers_galactic_nuclei_SR}. In the plot we take $M_\text{Cyg X-1}=21.2\pm2.2M_{\astrosun}$, $M_\text{GRS 1915+105}=12.4^{+2.0}_{-1.0}M_{\astrosun}$ and $\tilde a_\text{Cyg X-1}=0.9985$, $\tilde a_\text{GRS 1915+105}=0.95$~\cite{Stellar_BH_spins}. Extreme spin values do not have a large impact on the plot.\\
    For completeness, we add the GW170817 exclusion~\cite{GW170817} and the nuclear clock sensitivity projection~\cite{MegaPlanck, QCD_axion_atomic_levels}.}
    \label{fig:Axion_fa}
\end{figure}

In fig. \ref{fig:Axion_fa} we show the regions in the $\mu=m_a$ vs. $f_a$ parameter space~\cite{AxionLimits} where this process might be relevant. We see that stellar BHs can cover particle masses $\mu\sim10^{-14}\div$$2\cdot10^{-12}\ \eV$. While practically all of the region where $1s$ can grow is excluded by bounds from MICROSCOPE~\cite{Microscope}, solar core~\cite{Solar_Core}, and white dwarfs~\cite{White_Dwarfs}, they rely on the coupling of the scalar to ordinary matter, which we do not assume.

The most interesting comparison is with the bound from stellar BH spins~\cite{Stellar_BH_spins}. We see that for the two BHs used in this bound there is a non-insignificant overlap, so the $211\rightarrow1s$ decay may play a role in their dynamics, which might warrant further discussion.

Another relevant comparison is with bounds inferred from gravitational wave signals~\cite{GW231123_SR,GW_spins_SR}. For these, all considered values of $\alpha$ lie above the region where the $211\rightarrow1s$ decay is relevant. A different approach is to consider the rate of BH mergers instead of specific signals~\cite{Mergers_galactic_nuclei_SR}. There the values of $\alpha$ overlap with the range of the decay, but the self-interaction constant $f_a\gtrsim10^{17}\ \GeV$ is too high for our process to be relevant.

At the fixed point the mass of the $211$ cloud is $M_{211}^\text{crit}$. If the $211$ cloud still experiences superradiant growth with $\dot N_{211}\supset\Gamma_{\text{SR}}N_{211}$, half of these particles will drop down to the $1s$-halo and half will be emitted to infinity in non-relativistic waves with frequency $\omega_\text{rad}=2\omega_{2p}-\omega_{1s}=\mu\alpha^2/4\ll\mu$ and velocity $v_\text{rad}=\alpha/\sqrt{2}$. These waves will be emitted asymmetrically due to the $211$ state's directionality, but we can estimate the density of this radiation far away from the BH by treating it as spherically symmetric:
\begin{equation}
    \rho_\text{rad}\sim\frac{\Gamma_{\text{SR}}M_{211}^\text{crit}}{8\pi r^2v_\text{rad}}
    =1.3\frac{\text{keV}}{\text{cm}^3}\ \tilde a\sqrt{\frac{1+\sqrt{1-\tilde a^2}}{2}}\alpha^{7.5}\left[\frac{f_a}{10^8\ \GeV}\right]^2\left[\frac{1\ \text{pc}}{r}\right]^2.
\end{equation}
Substituting $\tilde a=0.9$, $\alpha=0.015$ and $f_a=10^{15}\ \GeV$, we get $\rho_\text{rad}=2.1\frac{\text{keV}}{\text{cm}^3}\left[\frac{1\ \text{pc}}{r}\right]^2$. It is important to note that this is the radiation density from the fixed point. If the $211$-halo were to get larger than the fixed point, the luminosity achieved during relaxation to the fixed point could be much higher.

\section{Summary and outlook}
In this work, we extended the known dark matter (DM) halo formation mechanism shown in~\cite{gilads_halo_paper} to a Schwarzschild black hole (BH), and showed that in cases where the BH induces instability in the $1s$ ground state, DM can be captured into halos of excited states instead.

We showed that a $2p$-halo is generated in a similar fashion to the $1s$-halo, namely that it grows exponentially and then saturates when the decay into the ground state balances its growth. Notably, in some parts of the parameter space decay of the $2p$-halo into the $2s$ state prevents the formation of the $2p$-halo. Future work may include a continuation to higher $n$ bound states, necessary to describe capture above the $2p$-instability line and higher.

For the parameters where the $2p$-halo grows and stabilizes, the halo produces an overdensity of DM which can be as large as $\sim10^5$ of the initial background DM density. While they are very dense, these halos are not large in total mass compared to the BH. Even for a hypothetical extreme BH of  mass $M_{\rm extreme}=10^3\ M_{\astrosun}$, say formed by a sequence of mergers, the $2p$ cloud mass does not exceed $\sim3\cdot10^{-14} \,M_{\rm extreme}\,$,  while the $1s$ cloud mass does not exceed
$\sim  10^{-12}\, M_{\rm extreme}$\,.

The low cloud mass may be remedied by considering BHs in a background with a higher DM density, like the super massive BHs (SMBHs) in galactic centers or primordial BHs (PBHs). Both setups lead to significant changes to the treatment of the background. In SMBH environments our approximation for the scattering states distribution $f(\mathbf{k})$ is no longer valid. Additionally, interactions with passing stellar objects may induce transitions between different states which can be relevant even for superradiant clouds~\cite{Stars_quench_SR}. While for PBHs the $1s$-capture was recently considered in~\cite{Capture_PBH}, which may be supplemented by considering higher $n$ bound states.

Switching gears to the Kerr case, we also showed that the effects of the self-interaction-induced transitions are relevant for the dynamics of BH superradiance, due to the decay of the superradiant $211$ into the ground state, potentially stopping the superradiant growth of the state, and resulting in a smaller $211$-halo than the current lore. The decay may affect the bound on the axion couplings from stellar BH spins due to~\cite{Stellar_BH_spins} for $\mu\lesssim3\cdot10^{-13}\ \eV$.

During the process, the decay of the $211$ state will produce a flux of DM waves with frequency $\omega_\text{rad}=\mu\alpha^2/4$. While the equilibrium radiation density is quite low, the non-equilibrium properties might be of interest, as the radiation density during relaxation to the equilibrium could be large. This could happen for BH masses up to $\sim4\cdot10^6\ M_{\astrosun}$, is independent of the background DM parameters, and is worthy of further study.

The formalism we developed for the capturing of excited states can also be applied to the solar halo, to evaluate the $2p$ density around the Sun in the background of the large $1s$-halo. Unfortunately, the only region where the $2p$ cloud density is not negligible compared to the background is in the $\xi \lesssim1$ regime where our results are less applicable, and only happens for parameters where $r_b\ll 1\ \text{AU}$, which adds an additional exponential suppression for any Earth-based measurement of the densities, rendering the cloud undetectable. This last point is also important for the $1s$-halo, as even for parameters where it reaches a large overdensity compared to the background near the Sun, the exponential suppression can mean that no overdensity will be measurable on Earth. The treatment of halo formation for $\xi\lesssim1$, which would allow larger $r_b$, may be an area for future investigation. 

While preparing this manuscript for submission we became aware of~\cite{Rodd} which discusses potential modifications for the superradiance rates from gravitational interactions within the cloud. It is complementary to our results, as it focuses on a different region of the parameter space where the self-interactions are subdominant and, vice versa, we focus on the regime where the gravitational interactions within the cloud are becoming subdominant. 

\section*{Acknowledgements}
The authors are grateful to Dmitry Budker and Alessandro Lenoci for valuable discussions. The authors thank WEXAC HPC for providing access to the cluster. GP is supported by the Israel Science Foundation (ISF), Minerva, the NSF-BSF, and the European Research Council (ERC, DM-Dawn, Grant Agreement No. 101199868). O.T. is supported in part by the ISF grant No. 3533/24, the BSF grant No. 2024169., and by grant No. 2025785 from the US-Israel BSF/NSF.

The authors acknowledge use of Claude Code's Opus 5/5.5 for debugging the C++ code and implementing a function calculating the Coulomb wavefunction, which we cross-checked with Wolfram Mathematica.

\newpage
\appendix
\section{Derivation of diagrams}
\label{app:Derivation of diagrams}
In this Appendix we derive contributions of second-order diagrams to the occupation number of a state. We will use the expansion \eqref{eq:psi expansion} of the wavefunction $\psi$, expanding the amplitude $\tilde{c}_s=c^{(0)}_s+c^{(1)}_s+c^{(2)}_s+O\left(g^3\right)$ and the frequency $\tilde{\omega}_s=\omega^{(0)}_s+\omega^{(1)}_s+\omega^{(2)}_s+O\left(g^3\right)$ to second order:
\begin{equation}
    \begin{aligned}
        \psi=&\sumint_{s}\tilde{c}_{s}(t)e^{-i\tilde{\omega}_{s}t}\psi_{s}
        =\psi^{(0)}+\psi^{(1)}+\psi^{(2)}+O(g^3)\\
        =&\sumint_s c_s^{(0)}e^{-i\omega_s^{(0)}t}\psi_s
        +\sumint_s\left(c_s^{(1)}-c_s^{(0)}i\omega_s^{(1)}t\right)e^{-i\omega_s^{(0)}t}\psi_s\\
        +&\sumint_s\left(c_s^{(2)}-c_s^{(1)}i\omega_s^{(1)}t-c_s^{(0)}\left[i\omega_s^{(2)}t
        +{\omega_s^{(1)}}^2\frac{t^2}{2}\right]\right)e^{-i\omega_s^{(0)}t}\psi_s+O(g^3).
    \end{aligned}
\end{equation}
The corrections to the amplitude will necessarily change with time, as we want to discuss changes in the cloud's populations, while the frequency corrections can be taken to be time-independent.

The unperturbed solutions fulfill the unperturbed equation $(i\partial_t+\frac{\nabla^2}{2\mu}+\frac{\alpha}{r})e^{-i\omega_s^{(0)}t}\psi_s=0$, so the only terms in the order-by-order substitution into the system of Gross-Pitaevskii equations \eqref{eq:GP system} will be the ones where the time derivative acts on the coefficients outside of the zeroth-order wavefunctions:
\begin{equation}
    \begin{cases}
       \smallsumint_{\ s}i\partial_t\left(c_s^{(0)}\right)e^{-i\omega_s^{(0)}t}\psi_s=0,\\
       \smallsumint_{\ s}i\partial_t\left(c_s^{(1)}
       -c_s^{(0)}i\omega_s^{(1)}t\right)e^{-i\omega_s^{(0)}t}\psi_s
       =g{\psi^{(0)}}^2\overline{\psi^{(0)}},\\
       \smallsumint_{\ s}i\partial_t\left(c_s^{(2)}
       -c_s^{(1)}i\omega_s^{(1)}t-c_s^{(0)}\left[i\omega_s^{(2)}t
       +{\omega_s^{(1)}}^2\frac{t^2}{2}\right]\right)e^{-i\omega_s^{(0)}t}\psi_s
       =g\left({\psi^{(0)}}^2\overline{\psi^{(1)}}
       +2\psi^{(0)}\overline{\psi^{(0)}}\psi^{(1)}\right).
    \end{cases}
\end{equation}
To single out the corrections to a particular state, we can use the orthogonality of the hydrogen atom basis and integrate both sides with $\int d^3x\ \overline{\psi_s}$:
\begin{equation}
    \begin{cases}
       i\partial_t\left(c_s^{(0)}\right)=0,\\
       i\partial_t\left(c_s^{(1)}-c_s^{(0)}i\omega_s^{(1)}t\right)
       =ge^{i\omega_s^{(0)}t}\int d^3x\ {\psi^{(0)}}^2\overline{\psi^{(0)}\psi_s},\\
       i\partial_t\left(c_s^{(2)}
       -c_s^{(1)}i\omega_s^{(1)}t-c_s^{(0)}\left[i\omega_s^{(2)}t
       +{\omega_s^{(1)}}^2\frac{t^2}{2}\right]\right)
       =ge^{i\omega_s^{(0)}t}\int d^3x\left({\psi^{(0)}}^2\overline{\psi^{(1)}}
       +2\psi^{(0)}\overline{\psi^{(0)}}\psi^{(1)}\right)\overline{\psi_s}.
    \end{cases}
\end{equation}

As mentioned in the main text, the zeroth-order equation just yields $c_s^{(0)}=\text{const}$, in which from now on we will omit the superscript. To find $c_s^{(1)}$ we can take a time integral of the first-order equation, where we expand $\psi^{(0)}$ and denote $M_{s_1+s_2\rightarrow s_3+s}=\int d^3x\ \psi_{s_1}\psi_{s_2}\overline{\psi_{s_3}\psi_s}$:
\begin{equation}
    c_s^{(1)}(t)=c_si\omega_s^{(1)}t
            -ig\sumint_{s_i}c_{s_1}c_{s_2}\overline{c_{s_3}}
            M_{s_1+s_2\rightarrow s_3+s}
            \int_0^td\tau e^{-i\tau\left(\omega_{s_1}+\omega_{s_2}-\omega_{s_3}-\omega_s\right)}.
\end{equation}

Substituting the expression for $c_s^{(1)}$ and performing a similar time integral, we can find $c_s^{(2)}$:
\begin{equation}
    \begin{aligned}
        c_s^{(2)}&(t)
        =c_s\bigg(i\omega_s^{(2)}t-{\omega_s^{(1)}}^2\frac{t^2}{2}\bigg)
        +g\omega_{s}^{(1)}t\sumint_{s_{i}}c_{s_{1}}c_{s_{2}}\overline{c_{s_{3}}}
        M_{s_{1}+s_{2}\rightarrow s_{3}+s}\int_{0}^{t}d\tau 
        e^{-i\tau\left(\omega_{s_{1}}+\omega_{s_{2}}-\omega_{s_{3}}-\omega_{s}\right)}
        \\- 
        2&g^2\sumint_{s_i,q_i,d}
        c_{s_1}c_{s_2}c_{s_3}\overline{c_{q_2}c_{q_3}}M_{s_1+s_2\rightarrow q_3+d}M_{d+s_3\rightarrow q_2+s}
        \int_0^te^{it'\left(\omega_{s}+\omega_{q_2}-\omega_{s_3}-\omega_{d}\right)}dt'
        \int_0^{t'} e^{it''\left(\omega_{d}+\omega_{q_3}-\omega_{s_2}-\omega_{s_1}\right)}dt''
        \\+
        &g^2\sumint_{s_i,q_i,d}
        c_{s_1}c_{s_2}c_{s_3}\overline{c_{q_1}c_{q_2}}M_{s_1+s_2\rightarrow s+d}M_{d+s_3\rightarrow q_2+q_1}
        \int_0^te^{it'\left(\omega_{d}+\omega_{s}-\omega_{s_2}-\omega_{s_1}\right)}dt'
        \int_0^{t'} e^{it''\left(\omega_{q_1}+\omega_{q_2}-\omega_{s_3}-\omega_{d}\right)}dt''.
    \end{aligned}
\end{equation}

It is straightforward to check that the energy corrections cancel out of the expressions \eqref{eq:number of particles expansion} for $N_s$, so they do not change the number of particles in a state.

Now we can note that due to coherence time averaging, the sets of incoming and outgoing unbound states have to be the same. Similarly, but due to energy conservation, the sets of bound states also have to be similar, up to having different magnetic numbers $m$ in the $21m$ states.

This means that the energies of the incoming and outgoing states are the same, so we can denote:
\begin{equation}
    \Delta\omega=\omega_{s_1}+\omega_{s_2}-\omega_{q_3}-\omega_d
    =\omega_{q_1}+\omega_{q_2}-\omega_{s_3}-\omega_d,
\end{equation}
with $s_i$ the incoming, $q_i$ the outgoing and $d$ the intermediate states.

Finally, this gives us three types of diagrams: one coming from $\left|c_s^{(1)}\right|^2$ and two from the $-2g^2$ and $+g^2$ terms of $\overline{c_s}c_s^{(2)}+c_s\overline{c_s^{(2)}}$ coming from \eqref{eq:number of particles expansion}:

\begin{figure}[H]
    \centering
    \begin{subfigure}{.23\textwidth}
        \begin{tikzpicture}[scale=0.7, baseline=(current bounding box.center)]
            \coordinate (v1) at (0, 0);
            \coordinate (v2) at (1.5, -1.5);

            \draw[thick] (-1, 1) node[left] {$s_1$} -- (v1);
            \draw[thick] (-1, -1) node[left] {$s_2$} -- (v1);
            \draw[thick] (v1) -- (1, 1) node[right] {$q_3$};

            \draw[thick] (0.5, -2.5) node[left] {$s_3$} -- (v2);
            \draw[thick] (v2) -- (2.5, -0.5) node[right] {$q_2$};

            \draw[thick, dashed] (v1) -- node[above right] {$d$} (v2);
            \draw[thick] (v2) -- (2.5, -2.5) node[right] {$q_1$};

            \filldraw (v1) circle (1.5pt);
            \filldraw (v2) circle (1.5pt);
        \end{tikzpicture}
    \end{subfigure}
    \begin{subfigure}{.06\textwidth}
        \scalebox{1.5}{$\Rightarrow$}
    \end{subfigure}
    \begin{subfigure}{.65\textwidth}
        $\begin{aligned}
            N_d^{(2)}=(2-\delta^{s_1}_{s_2})(2-\delta^{q_1}_{q_2})&
            g^2\sumint_{\{\{s_1,s_2\},s_3\}}\sumint_{\{\{q_1,q_2\},q_3\}} 
            c_{s_i}\overline{c_{q_i}}\\
            \times M_{s_1+s_2\rightarrow q_3+d}&M_{d+s_3\rightarrow q_2+q_1}
            \int_0^tdt'\int_0^{t}dt''e^{i(t''-t')\Delta\omega},
        \end{aligned}$
    \end{subfigure}
    \hfill

    \centering
    \begin{subfigure}{.23\textwidth}
        \begin{tikzpicture}[scale=0.7, baseline=(current bounding box.center)]
            \coordinate (v1) at (0, 0);
            \coordinate (v2) at (1.5, -1.5);

            \draw[thick] (-1, 1) node[left] {$s_1$} -- (v1);
            \draw[thick] (-1, -1) node[left] {$s_2$} -- (v1);
            \draw[thick] (v1) -- (1, 1) node[right] {$q_3$};

            \draw[thick] (0.5, -2.5) node[left] {$s_3$} -- (v2);
            \draw[thick] (v2) -- (2.5, -0.5) node[right] {$q_2$};

            \draw[thick, dashed] (v1) -- node[above right] {$d$} (v2);
            \draw[thick, dashed] (v2) -- (2.5, -2.5) node[right] {$q_1$};

            \filldraw (v1) circle (1.5pt);
            \filldraw (v2) circle (1.5pt);
        \end{tikzpicture}
    \end{subfigure}
    \begin{subfigure}{.06\textwidth}
        \scalebox{1.5}{$\Rightarrow$}
    \end{subfigure}
    \begin{subfigure}{.65\textwidth}
        $\begin{aligned}
            N_{q_1}^{(2)}=&-2(2-\delta^{s_1}_{s_2})
            g^2\sumint_{\{\{s_1,s_2\},s_3\}}\sumint_{\{\{\underline{q_1},q_2\},q_3\}}
            \sumint_dc_{s_i}\overline{c_{q_i}}\\
            &\quad\times M_{s_1+s_2\rightarrow q_3+d}M_{d+s_3\rightarrow q_2+q_1}
            \int_0^tdt'\int_0^{t'}dt''e^{i(t'-t'')\Delta\omega}+\cc,
        \end{aligned}$
    \end{subfigure}
    \hfill

    \centering
    \begin{subfigure}{.23\textwidth}
        \begin{tikzpicture}[scale=0.7, baseline=(current bounding box.center)]
            \coordinate (v1) at (0, 0);
            \coordinate (v2) at (1.5, -1.5);

            \draw[thick] (-1, 1) node[left] {$s_1$} -- (v1);
            \draw[thick] (-1, -1) node[left] {$s_2$} -- (v1);
            \draw[thick, dashed] (v1) -- (1, 1) node[right] {$q_3$};

            \draw[thick] (0.5, -2.5) node[left] {$s_3$} -- (v2);
            \draw[thick] (v2) -- (2.5, -0.5) node[right] {$q_2$};

            \draw[thick, dashed] (v1) -- node[above right] {$d$} (v2);
            \draw[thick] (v2) -- (2.5, -2.5) node[right] {$q_1$};

            \filldraw (v1) circle (1.5pt);
            \filldraw (v2) circle (1.5pt);
        \end{tikzpicture}
    \end{subfigure}
    \begin{subfigure}{.06\textwidth}
        \scalebox{1.5}{$\Rightarrow$}
    \end{subfigure}
    \begin{subfigure}{.65\textwidth}
        $\begin{aligned}
            N_{q_3}^{(2)}=(2-\delta^{s_1}_{s_2})(2-\delta^{q_1}_{q_2})&
            g^2\sumint_{\{\{s_1,s_2\},s_3\}}\sumint_{\{\{q_1,q_2\},\underline{q_3}\}}
            \sumint_dc_{s_i}\overline{c_{q_i}}\\
            \times M_{s_1+s_2\rightarrow q_3+d}&M_{d+s_3\rightarrow q_2+q_1}
            \int_0^tdt'\int_0^{t'}dt''e^{i(t''-t')\Delta\omega}+\cc.
        \end{aligned}$
    \end{subfigure}
    \hfill
    \caption{All three types of second-order diagrams and their contributions to the number of particles in a particular state: $d$ for the first diagram, $q_1$ for the second diagram and $q_3$ for the third diagram. The notations of $\{\{s_1,s_2\},s_3\}$ and $\delta^{s_1}_{s_2}$ are explained below.}
    \label{fig:three second-order diagrams}
\end{figure}

In fig. \ref{fig:three second-order diagrams} the $\{\{s_1,s_2\},s_3\}$ denotes a triplet of incoming or outgoing states, with swaps of the first two states already accounted for, so the triplet  $\{\{s_2,s_1\},s_3\}$ does not appear in the sum. The underlined state in a triplet is the produced state, so we do not sum over it and consider triplets with that state fixed in place. 

The other notation $\delta_{s_2}^{s_1}$ is a Kronecker delta that will be $1$ when the states are the same and $0$ otherwise. Unbound states are usually counted as being the same, as the momentum integrals already include both permutations of the momenta. However, when considering two delta functions of the same pairs of unbound states, the 4 naive permutations are not fully compensated for by a factor of 2 from the momentum integral, so we define $(2-\delta^k_p)(2-\delta^k_p)=2$.  

These results will be used in Appendix \ref{app:Classification of diagrams} to classify all second-order diagrams.
\section{Conservation of angular momentum}\label{app:m conservation}

In this Appendix we show that the total magnetic quantum number of incoming and outgoing bound states is conserved for first- and second-order diagrams. We align our $z$ axis with $\mathbf{v}_\dm$, so that the unbound particle distribution $f(\mathbf{k})$ is axially symmetric. We will use spherical coordinates for all variables, denoting the azimuthal angle of each by $\varphi$, with a subscript denoting the variable. In what follows we only care about the azimuthal dependence of the terms, so $f(\mathbf{k})$ can be ignored, and $\psi_{nlm}(\mathbf{x})$ can be replaced by its azimuthal dependence $e^{im\varphi_x}$. Since conjugating $\psi_{nlm}$ is equivalent to taking $m\mapsto-m$, we can treat all wavefunctions as unconjugated without loss of generality.

In the case of a matrix element that only has bound wavefunctions $\psi_{m_i}$, we get:
\begin{equation}
    \int d\varphi_x\ e^{i\varphi_x\sum_im_i},
\end{equation}
which vanishes unless $\sum_i m_i=0$, meaning $\sum_\text{in}m_i=\sum_\text{out}m_i$.

Adding now unbound wavefunctions, the only azimuthal dependence on $\hat{k}$ in $\psi_\mathbf{k}(\mathbf{x})$ is through $\hat{x}\cdot\hat{k}\supset\cos(\varphi_k-\varphi_x)$. Therefore, for our purposes we can write $\psi_\mathbf{k}(\mathbf{x})$ schematically as $F[\cos(\varphi_k-\varphi_x)]$ (here as well, conjugating $\psi_\mathbf{k}(\mathbf{x})$ does not change this form, so we will ignore it).

For a first-order diagram with momenta $k_i$ and bound wavefunctions $\psi_{m_j}$, the azimuthal integrals are:
\begin{equation}
    \int \prod_id\varphi_{k_i}\int d\varphi_x\  e^{i\varphi_x\sum_j m_j}\prod_iF[\cos(\varphi_{k_i}-\varphi_x)]^2.
\end{equation}
For every $k_i$, we can switch the order of integration and shift variables $\varphi'_{k_i}= \varphi_{k_i}-\varphi_x$, so that there is no $\varphi_x$ dependence in $F$. This means that we get the same $\varphi_x$ integral as before, and still get $\sum_\text{in}m_i=\sum_\text{out}m_i$.

In the case of a second-order diagram, any pair of incoming-outgoing momenta in the same vertex can be treated the same way as the first-order diagram, so we will only look at momenta that are paired through different vertices (cross-contractions of App. \ref{app:Classification of diagrams}). Assuming bound wavefunctions $\psi_{m_j}$ in the first vertex and $\psi_{m_l}$ in the second, we get:

\begin{equation}
    \int \prod_id\varphi_{k_i}\int d\varphi_x\  e^{i\varphi_x\sum_j m_j}\prod_iF[\cos(\varphi_{k_i}-\varphi_x)]\int d\varphi_y\  e^{i\varphi_y\sum_l m_l}\prod_iF[\cos(\varphi_{k_i}-\varphi_y)].
\end{equation}

If we now again shift $\varphi'_{k_i}=\varphi_{k_i}-\varphi_x$, and then also shift $\varphi'_x=\varphi_x-\varphi_y$, we get that the only $\varphi_y$ dependence left in the integral is:
\begin{equation}
    \int d\varphi_y\ e^{i\varphi_y(\sum_jm_j+\sum_lm_l)},
\end{equation}
meaning that the diagram vanishes unless the sums of magnetic quantum numbers of all incoming and outgoing states are equal: $\sum_\text{in}m_i=\sum_\text{out}m_i$.
\section{Mixing between $2p$-states}
\label{app:Mixing}
In this Appendix we assume that we already have a large $2p$-halo, so that the effects of unbound or other bound states will be negligible. It will prove to be enough to consider first-order processes only involving the $2p$ states, and therefore we can use shorthand notation denoting the state $21m$ by $m$ alone. The following diagrams describe the relevant processes:

\begin{figure}[H]
    \centering
    $
    \displaystyle
    c^{(1)}_{0} = 
    2
    \begin{tikzpicture}[scale=0.55, baseline=(current bounding box.center)]
        \coordinate (v1) at (0, 0);

        \draw[thick] (-1, 1) node[left] {$\pm1$} -- (v1);
        \draw[thick] (-1, -1) node[left] {$0$} -- (v1);
        \draw[thick] (v1) -- (1, 1) node[right] {$\pm1$};
        \draw[thick, dashed] (v1) -- (1,-1) node[right] {$0$} ;
        \filldraw (v1) circle (1.5pt);
        
    \end{tikzpicture}
    +
        \begin{tikzpicture}[scale=0.55, baseline=(current bounding box.center)]
        \coordinate (v1) at (0, 0);

        \draw[thick] (-1, 1) node[left] {$0$} -- (v1);
        \draw[thick] (-1, -1) node[left] {$0$} -- (v1);
        \draw[thick] (v1) -- (1, 1) node[right] {$0$};
        \draw[thick, dashed] (v1) -- (1,-1) node[right] {$0$} ;
        \filldraw (v1) circle (1.5pt);
        
    \end{tikzpicture}
    +2
    \begin{tikzpicture}[scale=0.55, baseline=(current bounding box.center)]
        \coordinate (v1) at (0, 0);

        \draw[thick] (-1, 1) node[left] {$1$} -- (v1);
        \draw[thick] (-1, -1) node[left] {$-1$} -- (v1);
        \draw[thick] (v1) -- (1, 1) node[right] {$0$};
        \draw[thick, dashed] (v1) -- (1,-1) node[right] {$0$} ;
        \filldraw (v1) circle (1.5pt);
        
    \end{tikzpicture},
    $

    $
    \displaystyle
    c^{(1)}_{1} = 
    2
    \begin{tikzpicture}[scale=0.55, baseline=(current bounding box.center)]
        \coordinate (v1) at (0, 0);

        \draw[thick] (-1, 1) node[left] {$-1/0$} -- (v1);
        \draw[thick] (-1, -1) node[left] {$1$} -- (v1);
        \draw[thick] (v1) -- (1, 1) node[right] {$-1/0$};
        \draw[thick, dashed] (v1) -- (1,-1) node[right] {$1$} ;
        \filldraw (v1) circle (1.5pt);
        
    \end{tikzpicture}
    +
        \begin{tikzpicture}[scale=0.55, baseline=(current bounding box.center)]
        \coordinate (v1) at (0, 0);

        \draw[thick] (-1, 1) node[left] {$1$} -- (v1);
        \draw[thick] (-1, -1) node[left] {$1$} -- (v1);
        \draw[thick] (v1) -- (1, 1) node[right] {$1$};
        \draw[thick, dashed] (v1) -- (1,-1) node[right] {$1$} ;
        \filldraw (v1) circle (1.5pt);
        
    \end{tikzpicture}
    +
    \begin{tikzpicture}[scale=0.55, baseline=(current bounding box.center)]
        \coordinate (v1) at (0, 0);

        \draw[thick] (-1, 1) node[left] {$0$} -- (v1);
        \draw[thick] (-1, -1) node[left] {$0$} -- (v1);
        \draw[thick] (v1) -- (1, 1) node[right] {$-1$};
        \draw[thick, dashed] (v1) -- (1,-1) node[right] {$1$} ;
        \filldraw (v1) circle (1.5pt);
        
    \end{tikzpicture},
    $

    $
    \displaystyle
    c^{(1)}_{-1} = 
    2
    \begin{tikzpicture}[scale=0.55, baseline=(current bounding box.center)]
        \coordinate (v1) at (0, 0);

        \draw[thick] (-1, 1) node[left] {$1/0$} -- (v1);
        \draw[thick] (-1, -1) node[left] {$-1$} -- (v1);
        \draw[thick] (v1) -- (1, 1) node[right] {$1/0$};
        \draw[thick, dashed] (v1) -- (1,-1) node[right] {$-1$} ;
        \filldraw (v1) circle (1.5pt);
        
    \end{tikzpicture}
    +
        \begin{tikzpicture}[scale=0.55, baseline=(current bounding box.center)]
        \coordinate (v1) at (0, 0);

        \draw[thick] (-1, 1) node[left] {$-1$} -- (v1);
        \draw[thick] (-1, -1) node[left] {$-1$} -- (v1);
        \draw[thick] (v1) -- (1, 1) node[right] {$-1$};
        \draw[thick, dashed] (v1) -- (1,-1) node[right] {$-1$} ;
        \filldraw (v1) circle (1.5pt);
        
    \end{tikzpicture}
    +
    \begin{tikzpicture}[scale=0.55, baseline=(current bounding box.center)]
        \coordinate (v1) at (0, 0);

        \draw[thick] (-1, 1) node[left] {$0$} -- (v1);
        \draw[thick] (-1, -1) node[left] {$0$} -- (v1);
        \draw[thick] (v1) -- (1, 1) node[right] {$1$};
        \draw[thick, dashed] (v1) -- (1,-1) node[right] {$-1$} ;
        \filldraw (v1) circle (1.5pt);
        
    \end{tikzpicture}.
    $
    \caption{Diagrams contributing to occupation numbers of $2p$-states. $1/0/-\!1$ denotes the state $21m$. The factors of 2 in front of the diagrams are symmetry factors, as the diagrams $s_1+s_2\rightarrow\dots$ and $s_2+s_1\rightarrow\dots$ should be both counted when $s_1\neq s_2$. Note that these diagrams are for the amplitudes, not populations, so they are missing the factors of $c_s$ or $\overline{c_s}$ compared to equation \eqref{eq:1st correction to number of particles}.}
\end{figure}
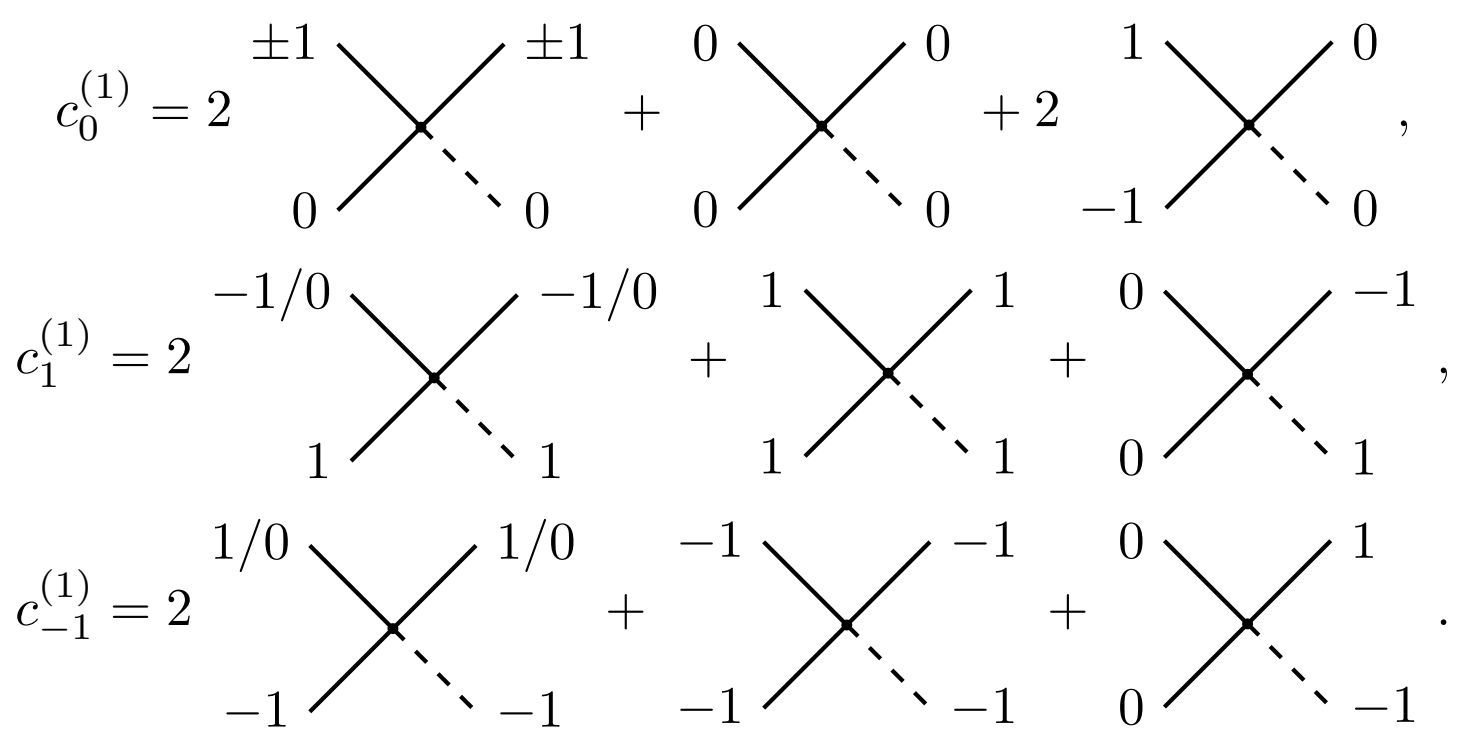

It is straightforward to evaluate all of the matrix elements involved. Denoting $\mathcal{M}=\frac{3}{2048\pi r_b^3}$, we get:
\begin{equation}
    \begin{aligned}
        &M_{0+0\leftrightarrow0+0}=3\mathcal{M},\quad M_{0+1\leftrightarrow 0+1}=M_{0+(-1)\leftrightarrow 0+(-1)}=\mathcal{M},\quad M_{0+0\leftrightarrow1+(-1)}=-\mathcal{M}\\
        &M_{1+1\leftrightarrow1+1}=M_{1+(-1)\leftrightarrow1+(-1)}=M_{(-1)+(-1)\leftrightarrow(-1)+(-1)}=2\mathcal{M}.
    \end{aligned}
\end{equation}
We get a system of three coupled differential equations for the amplitudes of the $2p$-states:
\begin{equation}
    \begin{cases}
        \dot{c}_{0}=-ig\mathcal{M}\left[\left(2|c_{1}|^{2}+2|c_{-1}|^{2}+3|c_{0}|^{2}\right)c_{0}-2c_{1}c_{-1}\overline{c_{0}}\right]\\
        \dot{c}_{1}=-ig\mathcal{M}\left[\left(2|c_{0}|^{2}+4|c_{-1}|^{2}+2|c_{1}|^{2}\right)c_{1}-c_{0}^{2}\overline{c_{-1}}\right]\\
        \dot{c}_{-1}=-ig\mathcal{M}\left[\left(2|c_{0}|^{2}+4|c_{1}|^{2}+2|c_{-1}|^{2}\right)c_{-1}-c_{0}^{2}\overline{c_{1}}\right]
    \end{cases}.
\end{equation}

One can verify that $|c_{1}|^{2}+|c_{-1}|^{2}+|c_{0}|^{2}\equiv 3N_{2p}=\const$, which has the meaning of conservation of the total number of particles. Another notable property is that $|c_1|^2-|c_{-1}|^2=\const$, so the first-order mixing cannot equalize populations of $21\pm\!1$ states.

Changing variables to the Cartesian basis $c_{0}=c_{z},\;c_{\pm1}=(\pm c_{x}+ic_{y})/\sqrt{2}$, the equations simplify:
\begin{equation}
    \label{eq:Cartesian mixing}
    \begin{cases}
        \dot{c}_{x}=-ig\mathcal{M}\left[6N_{2p}c_{x}+\left(c_{x}^{2}+c_{y}^{2}+c_{z}^{2}\right)\overline{c_{x}}\right]\\
        \dot{c}_{y}=-ig\mathcal{M}\left[6N_{2p}c_{y}+\left(c_{x}^{2}+c_{y}^{2}+c_{z}^{2}\right)\overline{c_{y}}\right]\\
        \dot{c}_{z}=-ig\mathcal{M}\left[6N_{2p}c_{z}+\left(c_{x}^{2}+c_{y}^{2}+c_{z}^{2}\right)\overline{c_{z}}\right].
    \end{cases}
\end{equation}
In equation \eqref{eq:Cartesian mixing} we see a manifest symmetry between the $x/y/z$ directions, which arises because we neglected the effects of the oncoming unbound DM, which was the only thing providing a preferred direction to the problem.

Combining the three equations of \eqref{eq:Cartesian mixing}, we find $c_x^2+c_y^2+c_z^2=Ae^{-18ig\mathcal{M}N_{2p}t}$. Without loss of generality, we will take $A$ to be real. Changing the variables to $c_n=e^{-9ig\mathcal{M}N_{2p}t}a_n,\ n\in\{x, y, z\}$, we get three simple decoupled differential equations:
\begin{equation}
    \dot{a}_n=-ig\mathcal{M}\left[-3N_{2p}a_n+A\overline{a_n}\right].
\end{equation}
Denoting $c_\pm=\sqrt{3N_{2p}\pm A}$ and $\omega=g\mathcal{M}c_+c_-$, we find solutions parametrized by real numbers $u_n$ and $v_n$:
\begin{equation}
    a_n(t)=u_n\left(\cos\omega t+i\frac{c_-}{c_+}\sin\omega t\right) + v_n\left(i\cos\omega t-\frac{c_+}{c_-}\sin\omega t\right).
\end{equation}

So we end up getting that the number of particles in a given $2p$-state $N_{21m}\sim|c_n|^2=|a_n|^2$ oscillates with frequency $2\omega$. Represented in more intuitive terms, that yields the mixing period:
\begin{equation}
    \label{eq:mixing time}
    T_{2p\rightarrow2p}=\frac{\pi}{\omega}=\frac{\pi}{g\mathcal{M}\sqrt{9N_{2p}^2-A^2}}\sim\frac{1}{g\mathcal{M}N_{2p}}\approx6\cdot10^4\years
    \left[\frac{f_a}{10^8 \GeV}\right]^2\left[\frac{\mu}{10^{-14}\eV}\right]\left[\frac{0.4\ \GeV/\cm^3}{\rho_{2p}}\right],
\end{equation}
where $\rho_{2p}$ is the maximal density of the $2p$-cloud. 

In \eqref{eq:mixing time} we assumed that $A$ is not too close to $3N_{2p}$. This requires the populations of the states in the $21m$ multiplet to be comparable and have different phases. The latter requirement is straightforward, as even if the phases were fine-tuned initially, their equality would be rapidly disrupted by interactions with unbound states, which would affect the phases of $c_x, c_y$ and $c_z$  differently.

The case for populations is more nuanced. As mentioned before, $2p$-mixing cannot equalize populations of $21\pm\!1$ states. Nevertheless, due to reflection symmetry their capture rates are the same and they are similar to the capture rate for the $210$ state, so their populations will be similar. If then one of the states were to be a few times larger than the others, the second-order decay processes would equalize it with the less populated states. However, if the halo is only produced in one state (e.g. superradiance), the $2p$-mixing can be small.
\section{Classification of diagrams}
\label{app:Classification of diagrams}
In this Appendix we will use expressions from fig. \ref{fig:three second-order diagrams} to classify all second-order diagrams and their contributions to occupation numbers. We will only consider diagrams that change the number of particles in $1s$, $2s$ and $2p$-states, with populations of all other bound states assumed to be negligible. Moreover, we will neglect the contributions that only redistribute particles between the $2p$-states, as the leading contribution to $2p$-mixing is given by the first-order diagrams considered in Appendix \ref{app:Mixing}.

\subsection{Energy-correcting diagrams}
A diagram is called energy-correcting if it has an energy-correcting vertex $a+b\rightarrow a+b$. In this subsection we will show that energy-correcting diagrams do not lead to changes in occupation numbers, except for a contribution to $2p$-mixing. From coherence time averaging, we know the sets of incoming and outgoing unbound states to be the same. From energy conservation, the sets of bound states have to be the same, when we count different $2p$-states to be equivalent. In particular, that means that:
\begin{equation}
    \omega_{s_1}+\omega_{s_2}-\omega_{q_3}-\omega_d=\omega_{q_1}+\omega_{q_2}-\omega_{s_3}-\omega_d=0,
\end{equation}
as one of the vertices is energy-correcting, so has $\Delta\omega=0$.

So the second vertex also has to conserve energy exactly. Due to coherence time averaging any incoming unbound state in the second vertex has to be paired with exactly the same outgoing one, so we either get that the second vertex is also energy-correcting or that it is a $2p$-mixing vertex.

Let us first consider the case where the second vertex is also energy-correcting $b+c\rightarrow b+c$. For conciseness, we take $a$, $b$, $c$ to be different states, though this assumption is not necessary. We can construct a family of 8 diagrams whose contributions will cancel:
\begin{figure}[H]
    \centering
    $\displaystyle
    \begin{tikzpicture}[scale=0.55, baseline=(current bounding box.center)]
        \coordinate (v1) at (0, 0);
        \coordinate (v2) at (1.5, -1.5);

        \draw[thick] (-1, 1) node[left] {$a$} -- (v1);
        \draw[thick] (-1, -1) node[left] {$b$} -- (v1);
        \draw[thick] (v1) -- (1, 1) node[right] {$a$};

        \draw[thick] (0.5, -2.5) node[left] {$c$} -- (v2);
        \draw[thick] (v2) -- (2.5, -0.5) node[right] {$b$};

        \draw[thick, dashed] (v1) -- node[above right] {$b$} (v2);
        \draw[thick] (v2) -- (2.5, -2.5) node[right] {$c$};

        \filldraw (v1) circle (1.5pt);
        \filldraw (v2) circle (1.5pt);
    \end{tikzpicture}
    +
    \begin{tikzpicture}[scale=0.55, baseline=(current bounding box.center)]
        \coordinate (v1) at (0, 0);
        \coordinate (v2) at (1.5, -1.5);

        \draw[thick] (-1, 1) node[left] {$a$} -- (v1);
        \draw[thick] (-1, -1) node[left] {$b$} -- (v1);
        \draw[thick, dashed] (v1) -- (1, 1) node[right] {$a$};

        \draw[thick] (0.5, -2.5) node[left] {$c$} -- (v2);
        \draw[thick] (v2) -- (2.5, -0.5) node[right] {$b$};

        \draw[thick, dashed] (v1) -- node[above right] {$b$} (v2);
        \draw[thick] (v2) -- (2.5, -2.5) node[right] {$c$};

        \filldraw (v1) circle (1.5pt);
        \filldraw (v2) circle (1.5pt);
    \end{tikzpicture}
    +
    \begin{tikzpicture}[scale=0.55, baseline=(current bounding box.center)]
        \coordinate (v1) at (0, 0);
        \coordinate (v2) at (1.5, -1.5);

        \draw[thick] (-1, 1) node[left] {$a$} -- (v1);
        \draw[thick] (-1, -1) node[left] {$b$} -- (v1);
        \draw[thick] (v1) -- (1, 1) node[right] {$a$};

        \draw[thick] (0.5, -2.5) node[left] {$c$} -- (v2);
        \draw[thick, dashed] (v2) -- (2.5, -0.5) node[right] {$b$};

        \draw[thick, dashed] (v1) -- node[above right] {$b$} (v2);
        \draw[thick] (v2) -- (2.5, -2.5) node[right] {$c$};

        \filldraw (v1) circle (1.5pt);
        \filldraw (v2) circle (1.5pt);
    \end{tikzpicture}
    +
    \begin{tikzpicture}[scale=0.55, baseline=(current bounding box.center)]
        \coordinate (v1) at (0, 0);
        \coordinate (v2) at (1.5, -1.5);

        \draw[thick] (-1, 1) node[left] {$a$} -- (v1);
        \draw[thick] (-1, -1) node[left] {$b$} -- (v1);
        \draw[thick] (v1) -- (1, 1) node[right] {$a$};

        \draw[thick] (0.5, -2.5) node[left] {$c$} -- (v2);
        \draw[thick] (v2) -- (2.5, -0.5) node[right] {$b$};

        \draw[thick, dashed] (v1) -- node[above right] {$b$} (v2);
        \draw[thick, dashed] (v2) -- (2.5, -2.5) node[right] {$c$};

        \filldraw (v1) circle (1.5pt);
        \filldraw (v2) circle (1.5pt);
    \end{tikzpicture}
    \quad+a\leftrightarrow c.
    $
\end{figure}

As all matrix elements and time integrals are real and the same between diagrams, we only need to make sure that the sum of symmetry factors comes out to 0. Indeed, for $N_a^{(2)}$ the $+4$ from the second diagram is canceled by the $-4$ coming from the fourth diagram with $a\leftrightarrow c$, with a similar result holding for $N_c^{(2)}$. For $N_b^{(2)}$ we get twice $+4$ coming from the first diagram and its partner and twice $-4$ coming from the third diagram and its partner. 

In the case where the second vertex is the $2p$-mixing vertex, we can use the fact that its matrix element is still real to get a similar cancellation:
\begin{figure}[H]
    \centering
    \begin{minipage}{.42\textwidth}
        $\displaystyle
        \begin{tikzpicture}[scale=0.55, baseline=(current bounding box.center)]
            \coordinate (v1) at (0, 0);
            \coordinate (v2) at (1.5, -1.5);

            \draw[thick] (-1, 1) node[left] {$a$} -- (v1);
            \draw[thick] (-1, -1) node[left] {$m$} -- (v1);
            \draw[thick, dashed] (v1) -- (1, 1) node[right] {$a$};

            \draw[thick] (0.5, -2.5) node[left] {$m_2$} -- (v2);
            \draw[thick] (v2) -- (2.5, -0.5) node[right] {$m_1$};

            \draw[thick, dashed] (v1) -- node[above right] {$m$} (v2);
            \draw[thick] (v2) -- (2.5, -2.5) node[right] {$m_3$};

            \filldraw (v1) circle (1.5pt);
            \filldraw (v2) circle (1.5pt);
        \end{tikzpicture}
        +
        \begin{tikzpicture}[scale=0.55, baseline=(current bounding box.center)]
            \coordinate (v1) at (0, 0);
            \coordinate (v2) at (1.5, -1.5);

            \draw[thick] (-1, 1) node[left] {$m_3$} -- (v1);
            \draw[thick] (-1, -1) node[left] {$m_1$} -- (v1);
            \draw[thick] (v1) -- (1, 1) node[right] {$m_2$};

            \draw[thick] (0.5, -2.5) node[left] {$a$} -- (v2);
            \draw[thick] (v2) -- (2.5, -0.5) node[right] {$m$};

            \draw[thick, dashed] (v1) -- node[above right] {$m$} (v2);
            \draw[thick, dashed] (v2) -- (2.5, -2.5) node[right] {$a$};

            \filldraw (v1) circle (1.5pt);
            \filldraw (v2) circle (1.5pt);
        \end{tikzpicture}
        $
    \end{minipage}
    \begin{minipage}{.03\textwidth}
        \scalebox{1.5}{$\Rightarrow$}
    \end{minipage}
    \begin{minipage}{.49\textwidth}
        $
        \begin{aligned}
            N_a^{(2)}\supset2\left(2-\delta^{m_1}_{m_3}\right)N_ac_mc_{m_2}\overline{c_{m_1}c_{m_3}}M_1M_2\frac{t^2}{2}+\cc&\\
            -2\left(2-\delta^{m_1}_{m_3}\right)N_ac_{m_1}c_{m_3}\overline{c_mc_{m_2}}M_1M_2\frac{t^2}{2}+\cc&=0,
        \end{aligned}
        $
    \end{minipage}
    \hfill
\end{figure}
where we take $a$ to be different from $m$ and get that these diagrams do not change the occupation number of the state $a$, so only contribute to $2p$-mixing. In the case where $a$ is the same as $m$, the whole diagram only has $2p$-states, so it cannot change occupation numbers of other states and we again only get a second-order contribution to $2p$-mixing.

\subsection{Cancellation of self-contractions}
We will call a pair of incoming and outgoing states $s_i$, $q_j$ a contraction if they are the same unbound state or the same $s$-state. A contraction is a cross-contraction if its states belong to different vertices and a self-contraction if they belong to the same vertex. In this subsection we will show that all diagrams with a self-contraction do not contribute to the occupation numbers of states, except for $2p$-mixing.

We will denote contracted states with the same letter $a/b/c$, the leftover $2p$-states will be denoted with $s_i$ and $q_i$. In the case where just one pair of $2p$-states is left, they will be the same due to magnetic number conservation. Then, all diagrams with a self-contraction belong to one of these classes:

\begin{figure}[H]
    \centering
    \begin{tikzpicture}[scale=0.7, baseline=(current bounding box.center)]
        \node[below] at (0.75, -2.75) {(1)};
        \coordinate (v1) at (0, 0);
        \coordinate (v2) at (1.5, -1.5);

        \draw[thick] (-1, 1) node[left] {$a$} -- (v1);
        \draw[thick] (-1, -1) node[left] {$s_2$} -- (v1);
        \draw[thick] (v1) -- (1, 1) node[right] {$a$};

        \draw[thick] (0.5, -2.5) node[left] {$s_3$} -- (v2);
        \draw[thick] (v2) -- (2.5, -0.5) node[right] {$q_2$};

        \draw[thick] (v1) -- node[above right] {$d$} (v2);
        \draw[thick] (v2) -- (2.5, -2.5) node[right] {$q_1$};

        \filldraw (v1) circle (1.5pt);
        \filldraw (v2) circle (1.5pt);
    \end{tikzpicture}
    \quad
    \begin{tikzpicture}[scale=0.7, baseline=(current bounding box.center)]
        \node[below] at (0.75, -2.75) {(2)};
        \coordinate (v1) at (0, 0);
        \coordinate (v2) at (1.5, -1.5);

        \draw[thick] (-1, 1) node[left] {$s_3$} -- (v1);
        \draw[thick] (-1, -1) node[left] {$s_2$} -- (v1);
        \draw[thick] (v1) -- (1, 1) node[right] {$q_1$};

        \draw[thick] (0.5, -2.5) node[left] {$a$} -- (v2);
        \draw[thick] (v2) -- (2.5, -0.5) node[right] {$q_2$};

        \draw[thick] (v1) -- node[above right] {$d$} (v2);
        \draw[thick] (v2) -- (2.5, -2.5) node[right] {$a$};

        \filldraw (v1) circle (1.5pt);
        \filldraw (v2) circle (1.5pt);
    \end{tikzpicture}
    \quad
    \begin{tikzpicture}[scale=0.7, baseline=(current bounding box.center)]
        \node[below] at (0.75, -2.75) {(3)};
        \coordinate (v1) at (0, 0);
        \coordinate (v2) at (1.5, -1.5);

        \draw[thick] (-1, 1) node[left] {$a$} -- (v1);
        \draw[thick] (-1, -1) node[left] {$s_2$} -- (v1);
        \draw[thick] (v1) -- (1, 1) node[right] {$a$};

        \draw[thick] (0.5, -2.5) node[left] {$b$} -- (v2);
        \draw[thick] (v2) -- (2.5, -0.5) node[right] {$s_2$};

        \draw[thick] (v1) -- node[above right] {$d$} (v2);
        \draw[thick] (v2) -- (2.5, -2.5) node[right] {$b$};

        \filldraw (v1) circle (1.5pt);
        \filldraw (v2) circle (1.5pt);
    \end{tikzpicture}
    
    \begin{tikzpicture}[scale=0.7, baseline=(current bounding box.center)]
        \node[below] at (0.75, -2.75) {(4)};
        \coordinate (v1) at (0, 0);
        \coordinate (v2) at (1.5, -1.5);

        \draw[thick] (-1, 1) node[left] {$b$} -- (v1);
        \draw[thick] (-1, -1) node[left] {$a$} -- (v1);
        \draw[thick] (v1) -- (1, 1) node[right] {$b$};

        \draw[thick] (0.5, -2.5) node[left] {$s_3$} -- (v2);
        \draw[thick] (v2) -- (2.5, -0.5) node[right] {$a$};

        \draw[thick] (v1) -- node[above right] {$d$} (v2);
        \draw[thick] (v2) -- (2.5, -2.5) node[right] {$s_3$};

        \filldraw (v1) circle (1.5pt);
        \filldraw (v2) circle (1.5pt);
    \end{tikzpicture}
    \quad
    \begin{tikzpicture}[scale=0.7, baseline=(current bounding box.center)]
        \node[below] at (0.75, -2.75) {(5)};
        \coordinate (v1) at (0, 0);
        \coordinate (v2) at (1.5, -1.5);

        \draw[thick] (-1, 1) node[left] {$s_3$} -- (v1);
        \draw[thick] (-1, -1) node[left] {$a$} -- (v1);
        \draw[thick] (v1) -- (1, 1) node[right] {$s_3$};

        \draw[thick] (0.5, -2.5) node[left] {$b$} -- (v2);
        \draw[thick] (v2) -- (2.5, -0.5) node[right] {$a$};

        \draw[thick] (v1) -- node[above right] {$d$} (v2);
        \draw[thick] (v2) -- (2.5, -2.5) node[right] {$b$};

        \filldraw (v1) circle (1.5pt);
        \filldraw (v2) circle (1.5pt);
    \end{tikzpicture}
    \quad
    \begin{tikzpicture}[scale=0.7, baseline=(current bounding box.center)]
        \node[below] at (0.75, -2.75) {(6)};
        \coordinate (v1) at (0, 0);
        \coordinate (v2) at (1.5, -1.5);

        \draw[thick] (-1, 1) node[left] {$b$} -- (v1);
        \draw[thick] (-1, -1) node[left] {$a$} -- (v1);
        \draw[thick] (v1) -- (1, 1) node[right] {$b$};

        \draw[thick] (0.5, -2.5) node[left] {$c$} -- (v2);
        \draw[thick] (v2) -- (2.5, -0.5) node[right] {$a$};

        \draw[thick] (v1) -- node[above right] {$d$} (v2);
        \draw[thick] (v2) -- (2.5, -2.5) node[right] {$c$};

        \filldraw (v1) circle (1.5pt);
        \filldraw (v2) circle (1.5pt);
    \end{tikzpicture}
    \caption{Classes of diagrams with a self-contraction, drawn without dashing of the lines. Same letters denote the same states, $a/b/c$ denote unbound or $s$-states, $s_i$ and $q_i$ denote $2p$-states.}
    \label{fig:templates with a self-contraction}
\end{figure}
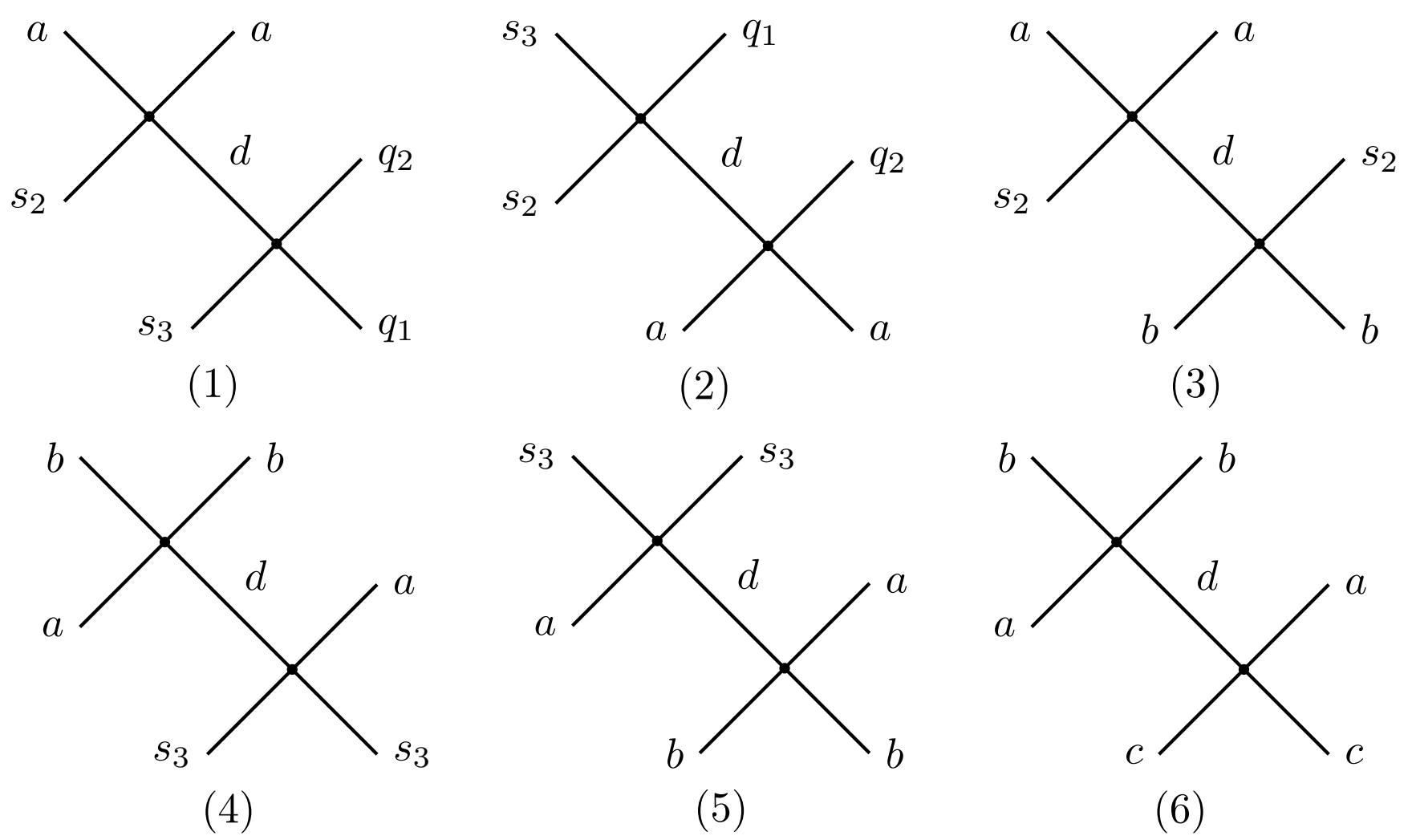

\paragraph{Classes 1, 2 \& 3}\hfill

Due to energy conservation the intermediate state $d$ has to be a $2p$-state. The state $a$ is either an $s$-state, so spherically symmetric, or unbound, making it axially symmetric after taking the integral over the momentum. Then, for the matrix element of the $a+s_2\rightarrow a+d$ ($a+d\rightarrow a+q_2$) to be non-zero, we need to choose $d=s_2$ ($d=q_2$), which makes the diagrams energy-correcting.

\paragraph{Classes 4, 5 \& 6}\hfill

These classes give rise to contributions that cancel between themselves. From energy conservation, we get that either $d=a$, making the diagram energy-correcting, or both $a$ and $d$ are unbound with momenta of the same magnitude. There are four diagrams contributing to occupation numbers of bound states:

\begin{figure}[H]
    \centering
    $\displaystyle
    \begin{tikzpicture}[scale=0.55, baseline=(current bounding box.center)]
        \coordinate (v1) at (0, 0);
        \coordinate (v2) at (1.5, -1.5);

        \draw[thick] (-1, 1) node[left] {$b$} -- (v1);
        \draw[thick] (-1, -1) node[left] {$k$} -- (v1);
        \draw[thick] (v1) -- (1, 1) node[right] {$b$};

        \draw[thick] (0.5, -2.5) node[left] {$s_1$} -- (v2);
        \draw[thick] (v2) -- (2.5, -0.5) node[right] {$k$};

        \draw[thick, dashed] (v1) -- node[above right] {$p$} (v2);
        \draw[thick, dashed] (v2) -- (2.5, -2.5) node[right] {$s_1$};

        \filldraw (v1) circle (1.5pt);
        \filldraw (v2) circle (1.5pt);
    \end{tikzpicture}
    \quad
    \begin{tikzpicture}[scale=0.55, baseline=(current bounding box.center)]
        \coordinate (v1) at (0, 0);
        \coordinate (v2) at (1.5, -1.5);

        \draw[thick] (-1, 1) node[left] {$s_1$} -- (v1);
        \draw[thick] (-1, -1) node[left] {$k$} -- (v1);
        \draw[thick, dashed] (v1) -- (1, 1) node[right] {$s_1$};

        \draw[thick] (0.5, -2.5) node[left] {$b$} -- (v2);
        \draw[thick] (v2) -- (2.5, -0.5) node[right] {$k$};

        \draw[thick, dashed] (v1) -- node[above right] {$p$} (v2);
        \draw[thick] (v2) -- (2.5, -2.5) node[right] {$b$};

        \filldraw (v1) circle (1.5pt);
        \filldraw (v2) circle (1.5pt);
    \end{tikzpicture}
    \quad\quad\quad
        \begin{tikzpicture}[scale=0.55, baseline=(current bounding box.center)]
        \coordinate (v1) at (0, 0);
        \coordinate (v2) at (1.5, -1.5);

        \draw[thick] (-1, 1) node[left] {$b$} -- (v1);
        \draw[thick] (-1, -1) node[left] {$k$} -- (v1);
        \draw[thick, dashed] (v1) -- (1, 1) node[right] {$b$};

        \draw[thick] (0.5, -2.5) node[left] {$s_1$} -- (v2);
        \draw[thick] (v2) -- (2.5, -0.5) node[right] {$k$};

        \draw[thick, dashed] (v1) -- node[above right] {$p$} (v2);
        \draw[thick] (v2) -- (2.5, -2.5) node[right] {$s_1$};

        \filldraw (v1) circle (1.5pt);
        \filldraw (v2) circle (1.5pt);
    \end{tikzpicture}
    \quad
    \begin{tikzpicture}[scale=0.55, baseline=(current bounding box.center)]
        \coordinate (v1) at (0, 0);
        \coordinate (v2) at (1.5, -1.5);

        \draw[thick] (-1, 1) node[left] {$s_1$} -- (v1);
        \draw[thick] (-1, -1) node[left] {$k$} -- (v1);
        \draw[thick,] (v1) -- (1, 1) node[right] {$s_1$};

        \draw[thick] (0.5, -2.5) node[left] {$b$} -- (v2);
        \draw[thick] (v2) -- (2.5, -0.5) node[right] {$k$};

        \draw[thick, dashed] (v1) -- node[above right] {$p$} (v2);
        \draw[thick, dashed] (v2) -- (2.5, -2.5) node[right] {$b$};

        \filldraw (v1) circle (1.5pt);
        \filldraw (v2) circle (1.5pt);
    \end{tikzpicture}
    $
\end{figure}

The first pair gives canceling contributions to $N_{s_1}$ and the second to $N_b$. For the 6th class replace $s_1$ with $c$ to get the same result, and when $b=c$ every diagram is counted twice, which doesn't change the answer.

\subsection{Diagrams contributing to $N^{(2)}_{2p}$}
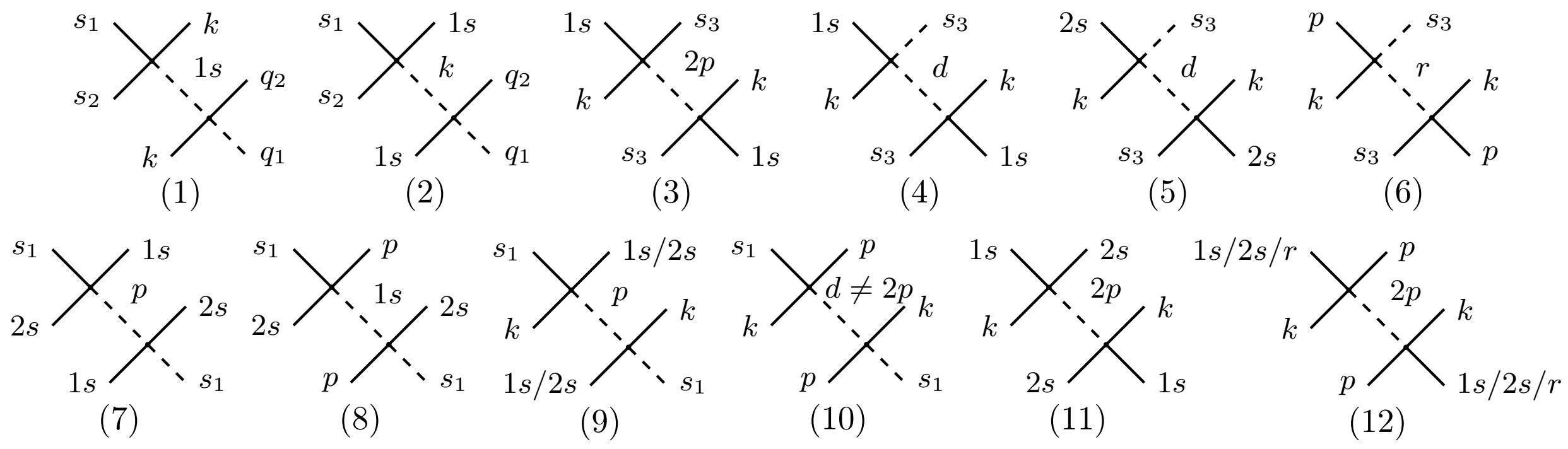
\begin{figure}[H]
    \centering
    $\displaystyle
    \begin{tikzpicture}[scale=0.4, baseline=(current bounding box.center)]
        \coordinate (v1) at (0, 0);
        \coordinate (v2) at (1.5, -1.5);

        \draw[thick] (-1, 1) node[left] {\small $s_1$} -- (v1);
        \draw[thick] (-1, -1) node[left] {\small $s_2$} -- (v1);
        \draw[thick] (v1) -- (1, 1) node[right] {\small $k$};

        \draw[thick] (0.5, -2.5) node[left] {\small $k$} -- (v2);
        \draw[thick] (v2) -- (2.5, -0.5) node[right] {\small $q_2$};

        \draw[thick, dashed] (v1) -- node[above right] {\small $1s$} (v2);
        \draw[thick, dashed] (v2) -- (2.5, -2.5) node[right] {\small $q_1$};

        \filldraw (v1) circle (1.5pt);
        \filldraw (v2) circle (1.5pt);
        
        \node at (0.75, -3.5) {(1)};
    \end{tikzpicture}
    \:
    \begin{tikzpicture}[scale=0.4, baseline=(current bounding box.center)]
        \coordinate (v1) at (0, 0);
        \coordinate (v2) at (1.5, -1.5);

        \draw[thick] (-1, 1) node[left] {\small $s_1$} -- (v1);
        \draw[thick] (-1, -1) node[left] {\small $s_2$} -- (v1);
        \draw[thick] (v1) -- (1, 1) node[right] {\small $1s$};

        \draw[thick] (0.5, -2.5) node[left] {\small $1s$} -- (v2);
        \draw[thick] (v2) -- (2.5, -0.5) node[right] {\small $q_2$};

        \draw[thick, dashed] (v1) -- node[above right] {\small $k$} (v2);
        \draw[thick, dashed] (v2) -- (2.5, -2.5) node[right] {\small $q_1$};

        \filldraw (v1) circle (1.5pt);
        \filldraw (v2) circle (1.5pt);
        
        \node at (0.75, -3.5) {(2)};
    \end{tikzpicture}
    \:
    \begin{tikzpicture}[scale=0.4, baseline=(current bounding box.center)]
        \coordinate (v1) at (0, 0);
        \coordinate (v2) at (1.5, -1.5);

        \draw[thick] (-1, 1) node[left] {\small $1s$} -- (v1);
        \draw[thick] (-1, -1) node[left] {\small $k$} -- (v1);
        \draw[thick] (v1) -- (1, 1) node[right] {\small $s_3$};

        \draw[thick] (0.5, -2.5) node[left] {\small $s_3$} -- (v2);
        \draw[thick] (v2) -- (2.5, -0.5) node[right] {\small $k$};

        \draw[thick, dashed] (v1) -- node[above right] {\small $2p$} (v2);
        \draw[thick] (v2) -- (2.5, -2.5) node[right] {\small $1s$};

        \filldraw (v1) circle (1.5pt);
        \filldraw (v2) circle (1.5pt);
        
        \node at (0.75, -3.5) {(3)};
    \end{tikzpicture}
    \:
    \begin{tikzpicture}[scale=0.4, baseline=(current bounding box.center)]
        \coordinate (v1) at (0, 0);
        \coordinate (v2) at (1.5, -1.5);

        \draw[thick] (-1, 1) node[left] {\small $1s$} -- (v1);
        \draw[thick] (-1, -1) node[left] {\small $k$} -- (v1);
        \draw[thick, dashed] (v1) -- (1, 1) node[right] {\small $s_3$};

        \draw[thick] (0.5, -2.5) node[left] {\small $s_3$} -- (v2);
        \draw[thick] (v2) -- (2.5, -0.5) node[right] {\small $k$};

        \draw[thick, dashed] (v1) -- node[above right] {\small $d$} (v2);
        \draw[thick] (v2) -- (2.5, -2.5) node[right] {\small $1s$};

        \filldraw (v1) circle (1.5pt);
        \filldraw (v2) circle (1.5pt);
        
        \node at (0.75, -3.5) {(4)};
    \end{tikzpicture}
    \:
    \begin{tikzpicture}[scale=0.4, baseline=(current bounding box.center)]
        \coordinate (v1) at (0, 0);
        \coordinate (v2) at (1.5, -1.5);

        \draw[thick] (-1, 1) node[left] {\small $2s$} -- (v1);
        \draw[thick] (-1, -1) node[left] {\small $k$} -- (v1);
        \draw[thick, dashed] (v1) -- (1, 1) node[right] {\small $s_3$};

        \draw[thick] (0.5, -2.5) node[left] {\small $s_3$} -- (v2);
        \draw[thick] (v2) -- (2.5, -0.5) node[right] {\small $k$};

        \draw[thick, dashed] (v1) -- node[above right] {\small $d$} (v2);
        \draw[thick] (v2) -- (2.5, -2.5) node[right] {\small $2s$};

        \filldraw (v1) circle (1.5pt);
        \filldraw (v2) circle (1.5pt);
        
        \node at (0.75, -3.5) {(5)};
    \end{tikzpicture}
    \:
    \begin{tikzpicture}[scale=0.4, baseline=(current bounding box.center)]
        \coordinate (v1) at (0, 0);
        \coordinate (v2) at (1.5, -1.5);

        \draw[thick] (-1, 1) node[left] {\small $p$} -- (v1);
        \draw[thick] (-1, -1) node[left] {\small $k$} -- (v1);
        \draw[thick, dashed] (v1) -- (1, 1) node[right] {\small $s_3$};

        \draw[thick] (0.5, -2.5) node[left] {\small $s_3$} -- (v2);
        \draw[thick] (v2) -- (2.5, -0.5) node[right] {\small $k$};

        \draw[thick, dashed] (v1) -- node[above right] {\small $r$} (v2);
        \draw[thick] (v2) -- (2.5, -2.5) node[right] {\small $p$};

        \filldraw (v1) circle (1.5pt);
        \filldraw (v2) circle (1.5pt);
        
        \node at (0.75, -3.5) {(6)};
    \end{tikzpicture}
    $
    \vspace{1.5em}
    $
    \displaystyle
    \begin{tikzpicture}[scale=0.4, baseline=(current bounding box.center)]
        \coordinate (v1) at (0, 0);
        \coordinate (v2) at (1.5, -1.5);

        \draw[thick] (-1, 1) node[left] {\small $s_1$} -- (v1);
        \draw[thick] (-1, -1) node[left] {\small $2s$} -- (v1);
        \draw[thick] (v1) -- (1, 1) node[right] {\small $1s$};

        \draw[thick] (0.5, -2.5) node[left] {\small $1s$} -- (v2);
        \draw[thick] (v2) -- (2.5, -0.5) node[right] {\small $2s$};

        \draw[thick, dashed] (v1) -- node[above right] {\small $p$} (v2);
        \draw[thick, dashed] (v2) -- (2.5, -2.5) node[right] {\small $s_1$};

        \filldraw (v1) circle (1.5pt);
        \filldraw (v2) circle (1.5pt);
        
        \node at (0.75, -3.5) {(7)};
    \end{tikzpicture}
    \:
    \begin{tikzpicture}[scale=0.4, baseline=(current bounding box.center)]
        \coordinate (v1) at (0, 0);
        \coordinate (v2) at (1.5, -1.5);

        \draw[thick] (-1, 1) node[left] {\small $s_1$} -- (v1);
        \draw[thick] (-1, -1) node[left] {\small $2s$} -- (v1);
        \draw[thick] (v1) -- (1, 1) node[right] {\small $p$};

        \draw[thick] (0.5, -2.5) node[left] {\small $p$} -- (v2);
        \draw[thick] (v2) -- (2.5, -0.5) node[right] {\small $2s$};

        \draw[thick, dashed] (v1) -- node[above right] {\small $1s$} (v2);
        \draw[thick, dashed] (v2) -- (2.5, -2.5) node[right] {\small $s_1$};

        \filldraw (v1) circle (1.5pt);
        \filldraw (v2) circle (1.5pt);
        
        \node at (0.75, -3.5) {(8)};
    \end{tikzpicture}
    \:
    \begin{tikzpicture}[scale=0.4, baseline=(current bounding box.center)]
        \coordinate (v1) at (0, 0);
        \coordinate (v2) at (1.5, -1.5);

        \draw[thick] (-1, 1) node[left] {\small $s_1$} -- (v1);
        \draw[thick] (-1, -1) node[left] {\small $k$} -- (v1);
        \draw[thick] (v1) -- (1, 1) node[right] {\small $1s/2s$};

        \draw[thick] (0.5, -2.5) node[left] {\small $1s/2s$} -- (v2);
        \draw[thick] (v2) -- (2.5, -0.5) node[right] {\small $k$};

        \draw[thick, dashed] (v1) -- node[above right] {\small $p$} (v2);
        \draw[thick, dashed] (v2) -- (2.5, -2.5) node[right] {\small $s_1$};

        \filldraw (v1) circle (1.5pt);
        \filldraw (v2) circle (1.5pt);
        
        \node at (0.75, -3.5) {(9)};
    \end{tikzpicture}
    \:
    \begin{tikzpicture}[scale=0.4, baseline=(current bounding box.center)]
        \coordinate (v1) at (0, 0);
        \coordinate (v2) at (1.5, -1.5);

        \draw[thick] (-1, 1) node[left] {\small $s_1$} -- (v1);
        \draw[thick] (-1, -1) node[left] {\small $k$} -- (v1);
        \draw[thick] (v1) -- (1, 1) node[right] {\small $p$};

        \draw[thick] (0.5, -2.5) node[left] {\small $p$} -- (v2);
        \draw[thick] (v2) -- (2.5, -0.5) node[right] {\small $k$};

        \draw[thick, dashed] (v1) -- node[above] {\small $\qquad d\neq 2p$} (v2);
        \draw[thick, dashed] (v2) -- (2.5, -2.5) node[right] {\small $s_1$};

        \filldraw (v1) circle (1.5pt);
        \filldraw (v2) circle (1.5pt);
        
        \node at (0.75, -3.5) {(10)};
    \end{tikzpicture}
    \:
    \begin{tikzpicture}[scale=0.4, baseline=(current bounding box.center)]
        \coordinate (v1) at (0, 0);
        \coordinate (v2) at (1.5, -1.5);

        \draw[thick] (-1, 1) node[left] {\small $1s$} -- (v1);
        \draw[thick] (-1, -1) node[left] {\small $k$} -- (v1);
        \draw[thick] (v1) -- (1, 1) node[right] {\small $2s$};

        \draw[thick] (0.5, -2.5) node[left] {\small $2s$} -- (v2);
        \draw[thick] (v2) -- (2.5, -0.5) node[right] {\small $k$};

        \draw[thick, dashed] (v1) -- node[above right] {\small $2p$} (v2);
        \draw[thick] (v2) -- (2.5, -2.5) node[right] {\small $1s$};

        \filldraw (v1) circle (1.5pt);
        \filldraw (v2) circle (1.5pt);
        
        \node at (0.75, -3.5) {(11)};
    \end{tikzpicture}
    \!\!\!
    \begin{tikzpicture}[scale=0.4, baseline=(current bounding box.center)]
        \coordinate (v1) at (0, 0);
        \coordinate (v2) at (1.5, -1.5);

        \draw[thick] (-1, 1) node[left] {\small $1s/2s/r$} -- (v1);
        \draw[thick] (-1, -1) node[left] {\small $k$} -- (v1);
        \draw[thick] (v1) -- (1, 1) node[right] {\small $p$};

        \draw[thick] (0.5, -2.5) node[left] {\small $p$} -- (v2);
        \draw[thick] (v2) -- (2.5, -0.5) node[right] {\small $k$};

        \draw[thick, dashed] (v1) -- node[above right] {\small $2p$} (v2);
        \draw[thick] (v2) -- (2.5, -2.5) node[right] {\small $1s/2s/r$};

        \filldraw (v1) circle (1.5pt);
        \filldraw (v2) circle (1.5pt);
        
        \node at (0.75, -3.5) {(12)};
    \end{tikzpicture}
    $
    \caption{All contributions to $N_{2p}^{(2)}$, where $s_i, q_i$ represent $2p$-states and $k,p,r$ represent unbound states. Intermediate state $d$ can be replaced with any state fulfilling energy conservation.}
    \label{fig:All_2p_diagrams}
\end{figure}

For diagrams 4, 5 and 10 there are infinitely many possible bound intermediate states. For completeness' sake we always take $d=1s,2s,2p,k$ (with $k$ denoting an unbound state) when possible, but we also try to look at other states that can give an effective contribution. For diagrams 4 and 10 we simply take all $3lm$ states as a "next to leading order" contribution (as higher $n$ states will have smaller and smaller matrix elements). For diagram 5, energy conservation gives $k^2=\mu^2\alpha^4+2\mu\omega_d$, so $d$ is either an unbound state or a very high $n$ bound state (with $n>1/\alpha$). For high $\alpha$ this can possibly give a large contribution, since $n$ is not too large, and the $k^2$ can be fine-tuned to be of order $\mu^2v_\dm^2$. We evaluated some of these points, and they do not appear to give a meaningful contribution to the dynamics.

\subsection{Diagrams contributing to $N^{(2)}_{1s/2s}$}
\begin{figure}[H]
    \centering
    
    $\displaystyle
    \begin{tikzpicture}[scale=0.4, baseline=(current bounding box.center)]
        \coordinate (v1) at (0, 0);
        \coordinate (v2) at (1.5, -1.5);

        \draw[thick] (-1, 1) node[left] {\small $s_1$} -- (v1);
        \draw[thick] (-1, -1) node[left] {\small $s_2$} -- (v1);
        \draw[thick] (v1) -- (1, 1) node[right] {\small $k$};

        \draw[thick] (0.5, -2.5) node[left] {\small $k$} -- (v2);
        \draw[thick] (v2) -- (2.5, -0.5) node[right] {\small $q_2$};

        \draw[thick, dashed] (v1) -- node[above right] {\small $1s$} (v2);
        \draw[thick] (v2) -- (2.5, -2.5) node[right] {\small $q_1$};

        \filldraw (v1) circle (1.5pt);
        \filldraw (v2) circle (1.5pt);
        
        \node at (0.75, -3.5) {\color{red}(1)};
    \end{tikzpicture}
    \:
    \begin{tikzpicture}[scale=0.4, baseline=(current bounding box.center)]
        \coordinate (v1) at (0, 0);
        \coordinate (v2) at (1.5, -1.5);

        \draw[thick] (-1, 1) node[left] {\small $s_1$} -- (v1);
        \draw[thick] (-1, -1) node[left] {\small $s_2$} -- (v1);
        \draw[thick, dashed] (v1) -- (1, 1) node[right] {\small $1s$};

        \draw[thick] (0.5, -2.5) node[left] {\small $1s$} -- (v2);
        \draw[thick] (v2) -- (2.5, -0.5) node[right] {\small $q_2$};

        \draw[thick, dashed] (v1) -- node[above right] {\small $k$} (v2);
        \draw[thick] (v2) -- (2.5, -2.5) node[right] {\small $q_1$};

        \filldraw (v1) circle (1.5pt);
        \filldraw (v2) circle (1.5pt);
        
        \node at (0.75, -3.5) {\color{red}(2)};
    \end{tikzpicture}
    \:
    \begin{tikzpicture}[scale=0.4, baseline=(current bounding box.center)]
        \coordinate (v1) at (0, 0);
        \coordinate (v2) at (1.5, -1.5);

        \draw[thick] (-1, 1) node[left] {\small $1s$} -- (v1);
        \draw[thick] (-1, -1) node[left] {\small $k$} -- (v1);
        \draw[thick] (v1) -- (1, 1) node[right] {\small $s_3$};

        \draw[thick] (0.5, -2.5) node[left] {\small $s_3$} -- (v2);
        \draw[thick] (v2) -- (2.5, -0.5) node[right] {\small $k$};

        \draw[thick, dashed] (v1) -- node[above right] {\small $2s$} (v2);
        \draw[thick] (v2) -- (2.5, -2.5) node[right] {\small $1s$};

        \filldraw (v1) circle (1.5pt);
        \filldraw (v2) circle (1.5pt);
        
        \node at (0.75, -3.5) {\color{blue}(3)};
    \end{tikzpicture}
    \:
    \begin{tikzpicture}[scale=0.4, baseline=(current bounding box.center)]
        \coordinate (v1) at (0, 0);
        \coordinate (v2) at (1.5, -1.5);

        \draw[thick] (-1, 1) node[left] {\small $1s$} -- (v1);
        \draw[thick] (-1, -1) node[left] {\small $k$} -- (v1);
        \draw[thick] (v1) -- (1, 1) node[right] {\small $s_3$};

        \draw[thick] (0.5, -2.5) node[left] {\small $s_3$} -- (v2);
        \draw[thick] (v2) -- (2.5, -0.5) node[right] {\small $k$};

        \draw[thick, dashed] (v1) -- node[above right] {\small $d$} (v2);
        \draw[thick, dashed] (v2) -- (2.5, -2.5) node[right] {\small $1s$};

        \filldraw (v1) circle (1.5pt);
        \filldraw (v2) circle (1.5pt);
        
        \node at (0.75, -3.5) {\color{red}(4)};
    \end{tikzpicture}
    \:
    \begin{tikzpicture}[scale=0.4, baseline=(current bounding box.center)]
        \coordinate (v1) at (0, 0);
        \coordinate (v2) at (1.5, -1.5);

        \draw[thick] (-1, 1) node[left] {\small $2s$} -- (v1);
        \draw[thick] (-1, -1) node[left] {\small $k$} -- (v1);
        \draw[thick] (v1) -- (1, 1) node[right] {\small $s_3$};

        \draw[thick] (0.5, -2.5) node[left] {\small $s_3$} -- (v2);
        \draw[thick] (v2) -- (2.5, -0.5) node[right] {\small $k$};

        \draw[thick, dashed] (v1) -- node[above right] {\small $d$} (v2);
        \draw[thick, dashed] (v2) -- (2.5, -2.5) node[right] {\small $2s$};

        \filldraw (v1) circle (1.5pt);
        \filldraw (v2) circle (1.5pt);
        
        \node at (0.75, -3.5) {\color{blue}(5)};
    \end{tikzpicture}
    \:
    \begin{tikzpicture}[scale=0.4, baseline=(current bounding box.center)]
        \coordinate (v1) at (0, 0);
        \coordinate (v2) at (1.5, -1.5);

        \draw[thick] (-1, 1) node[left] {\small $s_1$} -- (v1);
        \draw[thick] (-1, -1) node[left] {\small $2s$} -- (v1);
        \draw[thick, dashed] (v1) -- (1, 1) node[right] {\small $1s$};

        \draw[thick] (0.5, -2.5) node[left] {\small $1s$} -- (v2);
        \draw[thick] (v2) -- (2.5, -0.5) node[right] {\small $2s$};

        \draw[thick, dashed] (v1) -- node[above right] {\small $p$} (v2);
        \draw[thick] (v2) -- (2.5, -2.5) node[right] {\small $s_1$};

        \filldraw (v1) circle (1.5pt);
        \filldraw (v2) circle (1.5pt);
        
        \node at (0.75, -3.5) {\color{red}(6)};
    \end{tikzpicture}
    $
    \vspace{1em}
    $
    \displaystyle
    \begin{tikzpicture}[scale=0.4, baseline=(current bounding box.center)]
        \coordinate (v1) at (0, 0);
        \coordinate (v2) at (1.5, -1.5);

        \draw[thick] (-1, 1) node[left] {\small $s_1$} -- (v1);
        \draw[thick] (-1, -1) node[left] {\small $2s$} -- (v1);
        \draw[thick] (v1) -- (1, 1) node[right] {\small $1s$};

        \draw[thick] (0.5, -2.5) node[left] {\small $1s$} -- (v2);
        \draw[thick, dashed] (v2) -- (2.5, -0.5) node[right] {\small $2s$};

        \draw[thick, dashed] (v1) -- node[above right] {\small $p$} (v2);
        \draw[thick] (v2) -- (2.5, -2.5) node[right] {\small $s_1$};

        \filldraw (v1) circle (1.5pt);
        \filldraw (v2) circle (1.5pt);
        
        \node at (0.75, -3.5) {\color{blue}(7)};
    \end{tikzpicture}
    \:
    \begin{tikzpicture}[scale=0.4, baseline=(current bounding box.center)]
        \coordinate (v1) at (0, 0);
        \coordinate (v2) at (1.5, -1.5);

        \draw[thick] (-1, 1) node[left] {\small $s_1$} -- (v1);
        \draw[thick] (-1, -1) node[left] {\small $2s$} -- (v1);
        \draw[thick] (v1) -- (1, 1) node[right] {\small $p$};

        \draw[thick] (0.5, -2.5) node[left] {\small $p$} -- (v2);
        \draw[thick] (v2) -- (2.5, -0.5) node[right] {\small $2s$};

        \draw[thick, dashed] (v1) -- node[above right] {\small $1s$} (v2);
        \draw[thick] (v2) -- (2.5, -2.5) node[right] {\small $s_1$};

        \filldraw (v1) circle (1.5pt);
        \filldraw (v2) circle (1.5pt);
        
        \node at (0.75, -3.5) {\color{red}(8)};
    \end{tikzpicture}
    \:
    \begin{tikzpicture}[scale=0.4, baseline=(current bounding box.center)]
        \coordinate (v1) at (0, 0);
        \coordinate (v2) at (1.5, -1.5);

        \draw[thick] (-1, 1) node[left] {\small $s_1$} -- (v1);
        \draw[thick] (-1, -1) node[left] {\small $2s$} -- (v1);
        \draw[thick] (v1) -- (1, 1) node[right] {\small $p$};

        \draw[thick] (0.5, -2.5) node[left] {\small $p$} -- (v2);
        \draw[thick, dashed] (v2) -- (2.5, -0.5) node[right] {\small $2s$};

        \draw[thick, dashed] (v1) -- node[above right] {\small $1s$} (v2);
        \draw[thick] (v2) -- (2.5, -2.5) node[right] {\small $s_1$};

        \filldraw (v1) circle (1.5pt);
        \filldraw (v2) circle (1.5pt);
        
        \node at (0.75, -3.5) {\color{blue}(9)};
    \end{tikzpicture}
    \:
    \begin{tikzpicture}[scale=0.4, baseline=(current bounding box.center)]
        \coordinate (v1) at (0, 0);
        \coordinate (v2) at (1.5, -1.5);

        \draw[thick] (-1, 1) node[left] {\small $s_1$} -- (v1);
        \draw[thick] (-1, -1) node[left] {\small $k$} -- (v1);
        \draw[thick, dashed] (v1) -- (1, 1) node[right] {\small $1s/2s$};

        \draw[thick] (0.5, -2.5) node[left] {\small $1s/2s$} -- (v2);
        \draw[thick] (v2) -- (2.5, -0.5) node[right] {\small $k$};

        \draw[thick, dashed] (v1) -- node[above right] {\small $p$} (v2);
        \draw[thick] (v2) -- (2.5, -2.5) node[right] {\small $s_1$};

        \filldraw (v1) circle (1.5pt);
        \filldraw (v2) circle (1.5pt);
        
        \node at (0.75, -3.5) {\color{green}(10)};
    \end{tikzpicture}
    \:
    \begin{tikzpicture}[scale=0.4, baseline=(current bounding box.center)]
        \coordinate (v1) at (0, 0);
        \coordinate (v2) at (1.5, -1.5);

        \draw[thick] (-1, 1) node[left] {\small $s_1$} -- (v1);
        \draw[thick] (-1, -1) node[left] {\small $k$} -- (v1);
        \draw[thick] (v1) -- (1, 1) node[right] {\small $p$};

        \draw[thick] (0.5, -2.5) node[left] {\small $p$} -- (v2);
        \draw[thick] (v2) -- (2.5, -0.5) node[right] {\small $k$};

        \draw[thick, dashed] (v1) -- node[above right] {\small $\!\!1s/2s$} (v2);
        \draw[thick] (v2) -- (2.5, -2.5) node[right] {\small $s_1$};

        \filldraw (v1) circle (1.5pt);
        \filldraw (v2) circle (1.5pt);
        
        \node at (0.75, -3.5) {\color{green}(11)};
    \end{tikzpicture}
    \:
    \begin{tikzpicture}[scale=0.4, baseline=(current bounding box.center)]
        \coordinate (v1) at (0, 0);
        \coordinate (v2) at (1.5, -1.5);

        \draw[thick] (-1, 1) node[left] {\small $2s$} -- (v1);
        \draw[thick] (-1, -1) node[left] {\small $2s$} -- (v1);
        \draw[thick, dashed] (v1) -- (1, 1) node[right] {\small $1s$};

        \draw[thick] (0.5, -2.5) node[left] {\small $1s$} -- (v2);
        \draw[thick] (v2) -- (2.5, -0.5) node[right] {\small $2s$};

        \draw[thick, dashed] (v1) -- node[above right] {\small $p$} (v2);
        \draw[thick] (v2) -- (2.5, -2.5) node[right] {\small $2s$};

        \filldraw (v1) circle (1.5pt);
        \filldraw (v2) circle (1.5pt);
        
        \node at (0.75, -3.5) {\color{red}(12)};
    \end{tikzpicture}
    $
    \vspace{1em}
    $\displaystyle
    \begin{tikzpicture}[scale=0.4, baseline=(current bounding box.center)]
        \coordinate (v1) at (0, 0);
        \coordinate (v2) at (1.5, -1.5);

        \draw[thick] (-1, 1) node[left] {\small $2s$} -- (v1);
        \draw[thick] (-1, -1) node[left] {\small $2s$} -- (v1);
        \draw[thick] (v1) -- (1, 1) node[right] {\small $1s$};

        \draw[thick] (0.5, -2.5) node[left] {\small $1s$} -- (v2);
        \draw[thick] (v2) -- (2.5, -0.5) node[right] {\small $2s$};

        \draw[thick, dashed] (v1) -- node[above right] {\small $p$} (v2);
        \draw[thick, dashed] (v2) -- (2.5, -2.5) node[right] {\small $2s$};

        \filldraw (v1) circle (1.5pt);
        \filldraw (v2) circle (1.5pt);
        
        \node at (0.75, -3.5) {\color{blue}(13)};
    \end{tikzpicture}
    \:
    \begin{tikzpicture}[scale=0.4, baseline=(current bounding box.center)]
        \coordinate (v1) at (0, 0);
        \coordinate (v2) at (1.5, -1.5);

        \draw[thick] (-1, 1) node[left] {\small $2s$} -- (v1);
        \draw[thick] (-1, -1) node[left] {\small $2s$} -- (v1);
        \draw[thick] (v1) -- (1, 1) node[right] {\small $p$};

        \draw[thick] (0.5, -2.5) node[left] {\small $p$} -- (v2);
        \draw[thick] (v2) -- (2.5, -0.5) node[right] {\small $2s$};

        \draw[thick, dashed] (v1) -- node[above right] {\small $1s$} (v2);
        \draw[thick, dashed] (v2) -- (2.5, -2.5) node[right] {\small $2s$};

        \filldraw (v1) circle (1.5pt);
        \filldraw (v2) circle (1.5pt);
        
        \node at (0.75, -3.5) {\color{blue}(14)};
    \end{tikzpicture}
    \:
    \begin{tikzpicture}[scale=0.4, baseline=(current bounding box.center)]
        \coordinate (v1) at (0, 0);
        \coordinate (v2) at (1.5, -1.5);

        \draw[thick] (-1, 1) node[left] {\small $2s$} -- (v1);
        \draw[thick] (-1, -1) node[left] {\small $2s$} -- (v1);
        \draw[thick] (v1) -- (1, 1) node[right] {\small $p$};

        \draw[thick] (0.5, -2.5) node[left] {\small $p$} -- (v2);
        \draw[thick] (v2) -- (2.5, -0.5) node[right] {\small $2s$};

        \draw[thick, dashed] (v1) -- node[above right] {\small $1s$} (v2);
        \draw[thick] (v2) -- (2.5, -2.5) node[right] {\small $2s$};

        \filldraw (v1) circle (1.5pt);
        \filldraw (v2) circle (1.5pt);
        
        \node at (0.75, -3.5) {\color{red}(15)};
    \end{tikzpicture}
    \:
    \begin{tikzpicture}[scale=0.4, baseline=(current bounding box.center)]
        \coordinate (v1) at (0, 0);
        \coordinate (v2) at (1.5, -1.5);

        \draw[thick] (-1, 1) node[left] {\small $p$} -- (v1);
        \draw[thick] (-1, -1) node[left] {\small $k$} -- (v1);
        \draw[thick] (v1) -- (1, 1) node[right] {\small $q$};

        \draw[thick] (0.5, -2.5) node[left] {\small $q$} -- (v2);
        \draw[thick] (v2) -- (2.5, -0.5) node[right] {\small $k$};

        \draw[thick, dashed] (v1) -- node[above right] {\small $\!\!1s/2s$} (v2);
        \draw[thick] (v2) -- (2.5, -2.5) node[right] {\small $p$};

        \filldraw (v1) circle (1.5pt);
        \filldraw (v2) circle (1.5pt);
        
        \node at (0.75, -3.5) {\color{green}(16)};
    \end{tikzpicture}
    \:
    \begin{tikzpicture}[scale=0.4, baseline=(current bounding box.center)]
        \coordinate (v1) at (0, 0);
        \coordinate (v2) at (1.5, -1.5);

        \draw[thick] (-1, 1) node[left] {\small $1s$} -- (v1);
        \draw[thick] (-1, -1) node[left] {\small $k$} -- (v1);
        \draw[thick] (v1) -- (1, 1) node[right] {\small $2s$};

        \draw[thick] (0.5, -2.5) node[left] {\small $2s$} -- (v2);
        \draw[thick] (v2) -- (2.5, -0.5) node[right] {\small $k$};

        \draw[thick, dashed] (v1) -- node[above right] {\small $2s$} (v2);
        \draw[thick] (v2) -- (2.5, -2.5) node[right] {\small $1s$};

        \filldraw (v1) circle (1.5pt);
        \filldraw (v2) circle (1.5pt);
        
        \node at (0.75, -3.5) {\color{blue}(17)};
    \end{tikzpicture}
    \:
    \begin{tikzpicture}[scale=0.4, baseline=(current bounding box.center)]
        \coordinate (v1) at (0, 0);
        \coordinate (v2) at (1.5, -1.5);

        \draw[thick] (-1, 1) node[left] {\small $1s$} -- (v1);
        \draw[thick] (-1, -1) node[left] {\small $k$} -- (v1);
        \draw[thick, dashed] (v1) -- (1, 1) node[right] {\small $2s$};

        \draw[thick] (0.5, -2.5) node[left] {\small $2s$} -- (v2);
        \draw[thick] (v2) -- (2.5, -0.5) node[right] {\small $k$};

        \draw[thick, dashed] (v1) -- node[above right] {\small $d$} (v2);
        \draw[thick] (v2) -- (2.5, -2.5) node[right] {\small $1s$};

        \filldraw (v1) circle (1.5pt);
        \filldraw (v2) circle (1.5pt);
        
        \node at (0.75, -3.5) {\color{blue}(18)};
    \end{tikzpicture}
    $
    \vspace{1em}
    $
    \displaystyle
    \begin{tikzpicture}[scale=0.4, baseline=(current bounding box.center)]
        \coordinate (v1) at (0, 0);
        \coordinate (v2) at (1.5, -1.5);

        \draw[thick] (-1, 1) node[left] {\small $1s$} -- (v1);
        \draw[thick] (-1, -1) node[left] {\small $k$} -- (v1);
        \draw[thick] (v1) -- (1, 1) node[right] {\small $2s$};

        \draw[thick] (0.5, -2.5) node[left] {\small $2s$} -- (v2);
        \draw[thick] (v2) -- (2.5, -0.5) node[right] {\small $k$};

        \draw[thick, dashed] (v1) -- node[above right] {\small $d$} (v2);
        \draw[thick, dashed] (v2) -- (2.5, -2.5) node[right] {\small $1s$};

        \filldraw (v1) circle (1.5pt);
        \filldraw (v2) circle (1.5pt);
        
        \node at (0.75, -3.5) {\color{red}(19)};
    \end{tikzpicture}
    \:
    \begin{tikzpicture}[scale=0.4, baseline=(current bounding box.center)]
        \coordinate (v1) at (0, 0);
        \coordinate (v2) at (1.5, -1.5);

        \draw[thick] (-1, 1) node[left] {\small $2s$} -- (v1);
        \draw[thick] (-1, -1) node[left] {\small $k$} -- (v1);
        \draw[thick, dashed] (v1) -- (1, 1) node[right] {\small $1s$};

        \draw[thick] (0.5, -2.5) node[left] {\small $1s$} -- (v2);
        \draw[thick] (v2) -- (2.5, -0.5) node[right] {\small $k$};

        \draw[thick, dashed] (v1) -- node[above right] {\small $p$} (v2);
        \draw[thick] (v2) -- (2.5, -2.5) node[right] {\small $2s$};

        \filldraw (v1) circle (1.5pt);
        \filldraw (v2) circle (1.5pt);
        
        \node at (0.75, -3.5) {\color{red}(20)};
    \end{tikzpicture}
    \:
    \begin{tikzpicture}[scale=0.4, baseline=(current bounding box.center)]
        \coordinate (v1) at (0, 0);
        \coordinate (v2) at (1.5, -1.5);

        \draw[thick] (-1, 1) node[left] {\small $2s$} -- (v1);
        \draw[thick] (-1, -1) node[left] {\small $k$} -- (v1);
        \draw[thick] (v1) -- (1, 1) node[right] {\small $1s$};

        \draw[thick] (0.5, -2.5) node[left] {\small $1s$} -- (v2);
        \draw[thick] (v2) -- (2.5, -0.5) node[right] {\small $k$};

        \draw[thick, dashed] (v1) -- node[above right] {\small $p$} (v2);
        \draw[thick, dashed] (v2) -- (2.5, -2.5) node[right] {\small $2s$};

        \filldraw (v1) circle (1.5pt);
        \filldraw (v2) circle (1.5pt);
        
        \node at (0.75, -3.5) {\color{blue}(21)};
    \end{tikzpicture}
    \:
    \begin{tikzpicture}[scale=0.4, baseline=(current bounding box.center)]
        \coordinate (v1) at (0, 0);
        \coordinate (v2) at (1.5, -1.5);

        \draw[thick] (-1, 1) node[left] {\small $q$} -- (v1);
        \draw[thick] (-1, -1) node[left] {\small $k$} -- (v1);
        \draw[thick, dashed] (v1) -- (1, 1) node[right] {\small $1s/2s$};

        \draw[thick] (0.5, -2.5) node[left] {\small $1s/2s$} -- (v2);
        \draw[thick] (v2) -- (2.5, -0.5) node[right] {\small $k$};

        \draw[thick, dashed] (v1) -- node[above right] {\small $p$} (v2);
        \draw[thick] (v2) -- (2.5, -2.5) node[right] {\small $q$};

        \filldraw (v1) circle (1.5pt);
        \filldraw (v2) circle (1.5pt);
        
        \node at (0.75, -3.5) {\color{green}(22)};
    \end{tikzpicture}
    \:
    \begin{tikzpicture}[scale=0.4, baseline=(current bounding box.center)]
        \coordinate (v1) at (0, 0);
        \coordinate (v2) at (1.5, -1.5);

        \draw[thick] (-1, 1) node[left] {\small $1s/2s$} -- (v1);
        \draw[thick] (-1, -1) node[left] {\small $k$} -- (v1);
        \draw[thick] (v1) -- (1, 1) node[right] {\small $p$};

        \draw[thick] (0.5, -2.5) node[left] {\small $p$} -- (v2);
        \draw[thick] (v2) -- (2.5, -0.5) node[right] {\small $k$};

        \draw[thick, dashed] (v1) -- node[above right] {\small $d$} (v2);
        \draw[thick, dashed] (v2) -- (2.5, -2.5) node[right] {\small $1s/2s$};

        \filldraw (v1) circle (1.5pt);
        \filldraw (v2) circle (1.5pt);
        
        \node at (0.75, -3.5) {\color{green}(23)};
    \end{tikzpicture}
    \!\!\!\!\!
    \begin{tikzpicture}[scale=0.4, baseline=(current bounding box.center)]
        \coordinate (v1) at (0, 0);
        \coordinate (v2) at (1.5, -1.5);

        \draw[thick] (-1, 1) node[left] {\small $1s/2s$} -- (v1);
        \draw[thick] (-1, -1) node[left] {\small $k$} -- (v1);
        \draw[thick] (v1) -- (1, 1) node[right] {\small $p$};

        \draw[thick] (0.5, -2.5) node[left] {\small $p$} -- (v2);
        \draw[thick] (v2) -- (2.5, -0.5) node[right] {\small $k$};

        \draw[thick, dashed] (v1) -- node[above right] {\small $\!\!2s/1s$} (v2);
        \draw[thick] (v2) -- (2.5, -2.5) node[right] {\small $1s/2s$};

        \filldraw (v1) circle (1.5pt);
        \filldraw (v2) circle (1.5pt);

        \node at (0.75, -3.5) {\color{green}(24)};
    \end{tikzpicture}
    $
    \caption{All contributions to $N_{1s/2s}^{(2)}$, where $s_i, q_i$ represent $2p$-states and $k,p,r$ represent unbound states. Intermediate state $d$ can be replaced with any state fulfilling energy conservation. Colors are by the state which they contribute to: R,B,G for 1s, 2s, or both.}
    \label{fig:All_s_diagrams}
\end{figure}
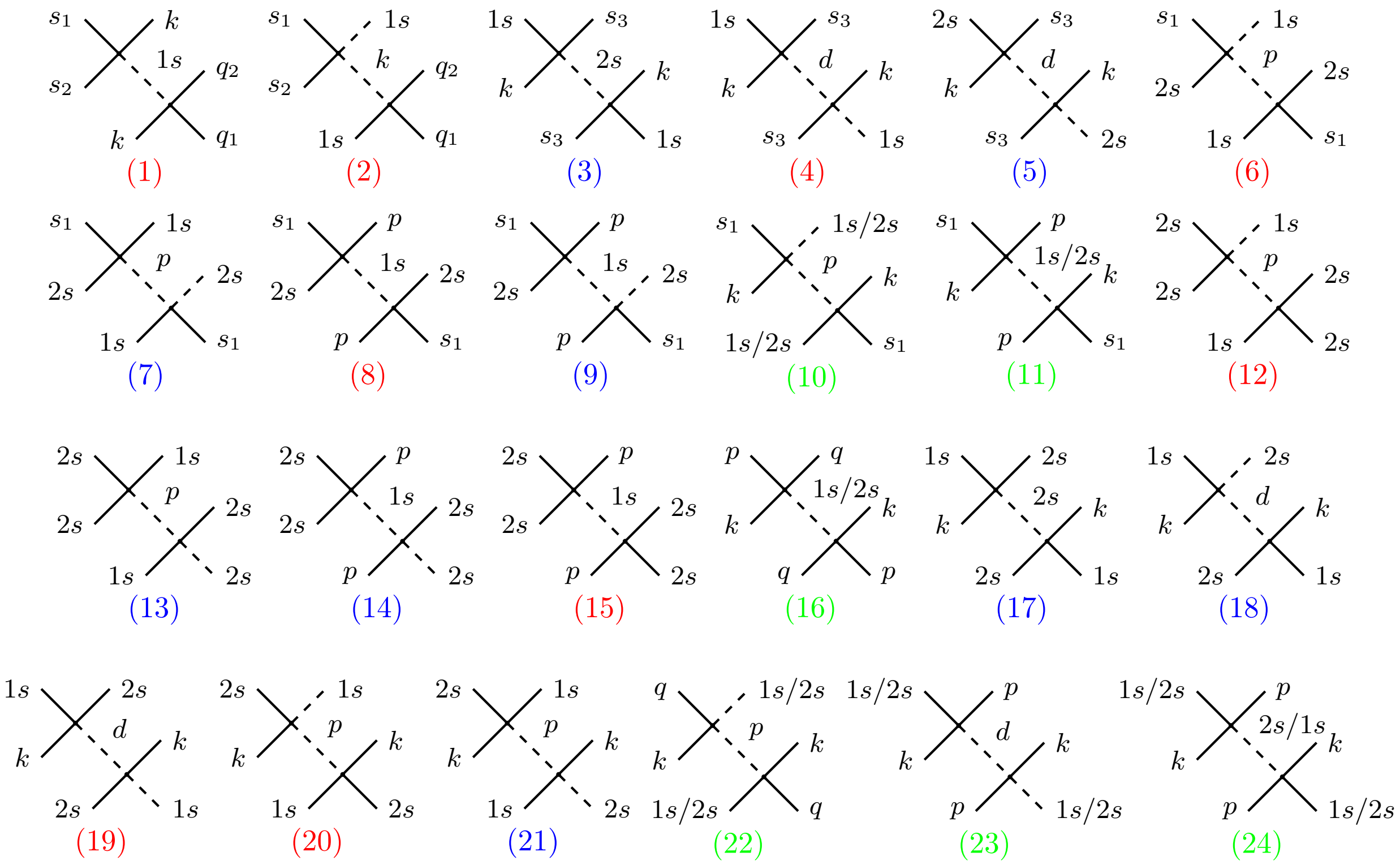

We again have an infinite number of possible intermediate states for diagrams 4,5,18,19,23. Here we also always take $d=1s,2s,2p,k$ when possible, and for diagrams 4, 18, 19, 23 we also evaluate them with $d=3lm$ as well. Diagram 5 is the same as diagram 5 in fig. \ref{fig:All_2p_diagrams}.
\section{Numerical evaluation of the matrix element integral}
\label{app:Numerical matrix element}
In this Appendix we describe the numerical methods for taking integrals $\mathcal{I}$ required to find rates of processes, like ones in equations \eqref{eq:growth rate}, \eqref{eq:2p->2s decay rate}, \eqref{eq:2p->1s decay rate}, \eqref{eq:2p->1s mixing rate}. All of the integrals are taken using Monte Carlo integration with importance sampling. First we discuss evaluation of $\hat M_{a+b\rightarrow c+d}=\int d^3\tilde x\ \tilde\psi_a\tilde\psi_b\overline{\tilde\psi_c\tilde\psi_d}$, then we go on to talk about the integrals over the momenta of unbound states. For a C++ implementation see~\cite{code}.

Throughout this Appendix we are using $\tilde k=\frac{\mathbf k}{\mu v_\dm}$, $\tilde\psi_{nlm}=r_b^{3/2}\psi_{nlm}$, $\tilde\psi_k=\sqrt{kr_b}\psi_k$, $\tilde x=\frac{x}{r_b}$.

We consider second-order processes that change occupation numbers of bound states, so the process involves at least one bound state at both its vertices. This allows us to calculate matrix elements using Monte-Carlo integration with importance sampling utilizing the lowest bound state. Consider a matrix element $a+b\rightarrow c+d$ and let its lowest state's principal quantum number be $n$, then: 
\begin{equation}
    \hat M_{a+b\rightarrow c+d}
    =\int d^3\tilde x\ \tilde\psi_a\tilde\psi_b\overline{\tilde\psi_c\tilde\psi_d}
    =4\pi n\left\langle|\tilde x|^2e^{|\tilde x|/n}
    \tilde\psi_a\tilde\psi_b\overline{\tilde\psi_c\tilde\psi_d}
    \right\rangle_{\hat x\in U(S^2)}^{|\tilde x|\in \text{Exp}(n)},
\end{equation}
with $|\tilde x|$ sampled from $\text{Exp}(n)$ with a p.d.f. $f_n(x)=e^{-x/n}/n$ for $x>0$, while $\hat x$ denotes the direction $\tilde x/|\tilde x|$ and is sampled uniformly over a sphere. 

The expression inside the averaging does not grow exponentially, as the explicit $e^{|\tilde x|/n}$ is canceled by the inverse factor from the wavefunction of the lowest bound state. Therefore, we are still averaging a well-behaved function, but with a significantly reduced variance.

From now on we treat $\hat M$ as a known quantity and move to the momentum integrals. 

We can use the energy-conserving $\delta$-function to get rid of the integral with the highest momentum. As stated before, every vertex has a bound state, so there is an unbound state coming with the lowest bound state (in case of a draw we are free to choose either) which we denote $\tilde k_1$ and choose to be the one whose magnitude is set by the $\delta$-function.

This choice means that for any realization of all other momenta, we have $|\tilde k_1|\ge|\tilde k_j|\ge0$, so we can always choose $|\tilde k_1|$ to fulfill energy conservation, and we are left with the integral over $\hat k_1=\tilde k_1/|\tilde k_1|\in S^2$.

We use index $j$ to denote all other unbound states and introduce $I_1$, $I_j$ which is $0$ if the state is the intermediate one and $1$ otherwise. The general form of the integral is:
\begin{equation}
    \mathcal{I} = \int d\hat k_1 e^{-I_1(\tilde{k}_1-\hat{z})^2}\prod_j\int d^3\tilde{k}_{j}
    \frac{e^{-I_j(\tilde{k}_j-\hat{z})^2}}{\tilde{k}_{j}}
    \cdot\hat M_1 \hat M_2(\tilde k_1, \tilde k_j),
\end{equation}
where we take $\mathbf{v}_\dm$ to point in the $\hat z$ direction.

The integral over $\hat k_1$ is taken by sampling uniformly over a sphere. The $\tilde k_j$ integral involves a $1/\tilde k_j$ factor, which would hurt its numerical convergence. However, this can be mitigated by changing the coordinate system. In the case of $I_j=1$, we will change to a cylindrical coordinate system:
\begin{equation}
    \int d^3\tilde{k}\frac{e^{-(\tilde{k}-\hat z)^2}}{\tilde{k}}
    =\pi^2\int_{-\infty}^\infty\frac{e^{-(\tilde{k}_z-1)^2}}{\sqrt\pi}d\tilde{k}_z
    \int_0^\infty\frac{e^{-\tilde{k}_{xy}^2}}{\sqrt\pi/2}d\tilde{k}_{xy}
    \int_0^{2\pi}\frac{d\varphi_{xy}}{2\pi}\frac{\tilde{k}_{xy}}{\sqrt{\tilde{k}_{xy}^2+\tilde{k}_z^2}}
    =\pi^2\left\langle\frac{\tilde{k}_{xy}}{\sqrt{\tilde{k}_{xy}^2+\tilde{k}_z^2}}\right\rangle,
\end{equation}
where the angle brackets denote average sampled with $\tilde{k}_z\in\mathcal N\left(1, \frac{1}{2}\right)$, $\tilde{k}_{xy}\in\left|\mathcal N\left(0, \frac{1}{2}\right)\right|$ and $\varphi_{xy}\in U[0,2\pi]$. 

In the case of $I_j=0$ we change to spherical coordinates within a sphere of radius $\tilde{k}_{max}$:
\begin{equation}
    \int d^3\tilde{k}\frac{1}{\tilde{k}}\approx\int_0^{\tilde{k}_\text{max}} |\tilde k|d|\tilde k|\int dS^2=
    4\pi\tilde k_{\text{max}}\left\langle |\tilde{k}|\right\rangle
    _{|\tilde k|\in U[0, \tilde k_\text{max}],\ \hat k\in U(S^2)}.
\end{equation}
As $I_j=0$, the intermediate state is not the highest momentum unbound state, as we have $|\tilde k_1|\ge|\tilde k_j|$. This means that the part of the integral we neglected has $|\tilde k_1|\ge \tilde k_\text{max}$, for a non-intermediate state that comes with the occupation number factor $\sim e^{-(\tilde k_1-\hat z)^2}$. We choose $\tilde k_\text{max}=10$, so that all neglected points are suppressed by at least a factor of $e^{-81}$, which is more than enough for our precision.

For convenience, let us restate the final expression:
\begin{equation}
    \mathcal I = \Bigg\langle 4\pi e^{-I_1(\tilde k_1-\hat{z})^2}
    \Big|_{{\hat k}_1\in U(S^2)}
    \prod_{j}
    \begin{cases}
        \pi^2\frac{\tilde k_{xy}}{\sqrt{\tilde k_{xy}^2+\tilde k_z^2}}
        \Big|_{\varphi_{xy}\in U[0,2\pi]}
        ^{\tilde{k}_z\in\mathcal N\left(1, \frac{1}{2}\right),\ 
        \tilde{k}_{xy}\in\left|\mathcal N\left(0, \frac{1}{2}\right)\right|},\ I_j=1\\
        4\pi\tilde k_\text{max} |\tilde k|\Big|_{|\tilde k|\in U[0, \tilde k_\text{max}],\ \hat k\in U(S^2)},\ I_j=0
    \end{cases}
    \!\!\!\!\times\hat M_1\hat M_2\left(\tilde k_1, \tilde k_j\right)
    \Bigg\rangle.
\end{equation}
The average here is taken on the outermost brackets with distributions denoted inside the expression.

In our realization we generate order of $10^6$ sets of unbound states $\tilde k_1, \tilde k_j$ and for every one of them take order of $10^3$ points inside each of the $\hat M$ integrals. For smaller $\xi$ the unbound wavefunctions oscillate many times on the scale of the Bohr radius $r_b$, increasing the variance and requiring more points for the integral to converge. The error is estimated by taking the sample variance over all unbound state realizations.
\section{The maximal size of the DM cloud}\label{app:cloud size}
In this Appendix we define the size of the cloud where our perturbative treatment fails, and the relate the cloud's density to its occupation number.

In order to treat the self-interaction perturbatively its energy has to be smaller than the gravitational potential. The energy density of the system is~\cite{gilads_halo_paper}:
\begin{equation}
    \epsilon=\frac{|\nabla \psi|^2}{2\mu}+\mu \Phi |\psi|^2+\frac{1}{2}g|\psi|^4,
\end{equation}
with $\Phi=-GM/r$ the gravitational potential (the self-gravity and general relativistic corrections are neglected). Thus, we want $\mu|\Phi||\psi|^2\gg \frac{1}{2}|g||\psi|^4$ for our perturbative description to hold. Defining the density $\rho(r)\equiv\mu|\psi(r)|^2$ allows us to rewrite the condition as $\rho(r)\ll 2|\Phi(r)|\mu^2/|g|\equiv\rho_\text{crit}(r)$. This inequality should hold everywhere, but we can evaluate $\Phi$ at $r_b$ to give us a benchmark value (accounting for the spatial dependence of $\Phi$ doesn't change the answer by more than an order of magnitude, although it does mean that different states have slightly different critical densities):
\begin{equation}
    \rho_\text{crit}\equiv \frac{2|\Phi(r_b)|\mu^2}{|g|}=\frac{2\alpha^2\mu^2}{|g|}
    \cong 3\cdot 10^4 \rho_\dm \left[\frac{f_{a}}{10^{8}\ \GeV}\right]^{2}\left[\frac{\mu}{10^{-14}\ \eV}\right]^{4}\left[\frac{M}{M_{\astrosun}}\right]^{2}\left[\frac{0.4\ \GeV/\cm^{3}}{\rho_{\dm}}\right].
\end{equation}

For the $1s$ state, we can just evaluate it at its maximal point ($r=0$) to find the critical occupation number\footnote{This result is the same as that appearing in \cite{gilads_halo_paper}, except for a missing factor of 2 in the original.}:
\begin{multline}
    \rho_{1s}(0)=\mu N_{1s}^{\max}|\psi_{1s}(0)|^2=\frac{\mu}{\pi r_b^3}N_{1s}^{\max}=\frac{2\alpha^2\mu^2}{|g|}\\
    \Rightarrow N_{1s}^{\max}=\frac{2\pi}{|g|\alpha\mu^2}
     \cong7\cdot 10^{67}\left[\frac{f_{a}}{10^{8}\,\text{GeV}}\right]^{2}\left[\frac{10^{-14}\,\text{eV}}{\mu}\right]^{3}\left[\frac{M_{\astrosun}}{M}\right].
\end{multline}

For $2s$ the procedure is similar. For the $2p$ states, assuming that all three states are equally populated with $N_{2p}^{\max}$ particles:
\begin{equation}
    |\psi_{2p}|^2=|\tilde{c}_{210}\psi_{210}+\tilde{c}_{211}\psi_{211}+\tilde{c}_{21-\!1}\psi_{21-\!1}|^2\approx N_{2p}^{\max}(|\psi_{210}|^2+|\psi_{211}|^2+|\psi_{21-\!1}|^2),
\end{equation}
where we took an angular average over the states, which removes the $\overline{\psi}_{21m}\psi_{21m'}+\cc$ terms. Under this averaging there is no angular dependence and we can write:
\begin{equation}
    |\psi_{2p}(r)|^2=\frac{1}{32\pi r_b^3}\left(\frac{r}{r_b}\right)^2e^{-r/r_b},
\end{equation}
which is maximal at $2r_b$. The results are summarized in table \ref{tab:critical occupation numbers}, along with the density profile (which is just the hydrogen atom probability distributions).

\begin{table}[H]
    \centering
    \begin{tabular}{|c|c|c|c|}
        \hline
        state & $N^{\max}$ & $\rho (r_b)$ & Density profile $\rho(r)$\\
        \hline\hline
         $1s$ & $\frac{2\pi}{|g|\alpha\mu^2} \cong 7\cdot 10^{67}$ & $\frac{\mu N_{1s}}{\pi e^2 r_b^3}\cong0.1 \rho_\text{crit}\left[\frac{N_{1s}}{N_{1s}^{\max}}\right]$ & $\frac{\mu N_{1s}}{\pi r_b^3}e^{-2r/r_b}$\\
         \hline
         $2s$ & $\frac{16 \pi}{|g|\alpha\mu^2}\cong5\cdot 10^{68}$ & $\frac{\mu N_{2s}}{32\pi e r_b^3}\cong0.1\rho_\text{crit}\left[\frac{N_{2s}}{N_{2s}^{\max}}\right]$ & $\frac{\mu N_{2s}}{32\pi r_b^3}(2-\frac{r}{r_b})^2e^{-r/r_b}$\\
         \hline
         $2p$ & $\frac{16 e^2\pi}{|g|\alpha\mu^2}\cong 4\cdot10^{69}$ & $\frac{\mu N_{2p}}{32\pi e r_b^3}\cong0.7 \rho_\text{crit}\left[\frac{N_{2p}}{N_{2p}^{\max}}\right]$ & $\frac{\mu N_{2p}}{32\pi r_b^3}\left(\frac{r}{r_b}\right)^2e^{-r/r_b}$\\
         \hline
    \end{tabular}
    \caption{Maximal occupation number, density at $r_b$, and density profile for the $1s,2s,2p$ states. For the $2p$ states $N_{2p}$ is the average occupation number of each of the three $21m$ states, and an angular average was taken over the density. Expressions are evaluated for the benchmark parameters $f_a=10^8\ \GeV$, $\mu=10^{-14}\ \eV$, $M=M_{\astrosun}$, $\rho_\dm=0.4\ \GeV/\cm^3$.}
    \label{tab:critical occupation numbers}
\end{table}

\section{Fixed point of the $2p$-halo occupation number and its $\xi$ dependence}
\label{app:static point}
In this Appendix we write the approximate fixed point of the $2p$-halo, applicable for both the dilute and dense $2p$-halo regimes. We write a simplified set of differential equations with the dominant processes:
\begin{equation}
    \begin{cases}
\dot{N}_{2p}=\Gamma_{\rel}\gamma_{\infty\rightarrow2p}N_{2p}-\Gamma_{\rel}\xi^4\left(\gamma_{2p \rightarrow1s}^{k}+\gamma_{2p \rightarrow1s}^{1s}\frac{N_{1s}}{N_{k}}\right)\frac{N_{2p}^{2}}{N_{k}}\\
\dot{N}_{1s}=\frac{3}{2}\Gamma_{\rel}\xi^4\gamma_{2p \rightarrow1s}^{k}\frac{N_{2p}^{2}}{N_{k}}+\left(\frac{3}{2}\Gamma_{\rel}\xi^4\gamma_{2p \rightarrow1s}^{1s}\frac{N_{2p}^{2}}{N_{k}^2}-\Gamma_{1s\rightarrow\BH}\right)N_{1s}
\end{cases},
\end{equation}
with $\gamma_{\infty\rightarrow 2p}=\Gamma_{\infty\rightarrow2p}/\Gamma_{\text{rel}}$ (whose value appears in fig. \ref{fig:capture vs decay}), and $\gamma_{2p \rightarrow1s}^{1s},\gamma_{2p \rightarrow1s}^{k}$ defined in eqs. \eqref{eq:2p->1s decay high xi}, \eqref{eq:2p->1s decay low xi}, respectively. These equations have a stable fixed point, which gives for $N_{2p}$:
\begin{equation}\label{eq:fixed point}
    \frac{N_{2p}}{N_k}=\frac{\sqrt{6\Gamma(\gamma_{\infty\rightarrow2p})^{2}\gamma_{2p \rightarrow1s}^{1s}+\xi^{4}\Gamma^{2}(\gamma_{2p \rightarrow1s}^{k})^{2}}-\xi^{2}\Gamma\gamma_{2p \rightarrow1s}^{k}}{3\xi^{2}\gamma_{\infty\rightarrow2p}\gamma_{2p \rightarrow1s}^{1s}},\quad \Gamma=\frac{\Gamma_{1s\rightarrow\BH}}{\Gamma_\rel}.
\end{equation}
This result should be viewed as an order of magnitude approximation, which holds as long as $\Gamma_\rel$ is large enough to allow the halo to form on astrophysical timescales.

This expression has a nontrivial $\xi$ dependence due to the $\gamma^k_{2p\rightarrow1s}\sim e^{-\xi^2/2}$ process, but in our allowed parameter space the second term in the square root (the $\xi^4\Gamma^2(\gamma^k_{2p\rightarrow1s})^2$ term) is not much larger than the first, so we can neglect it in favor of a clearer scaling (and for $\xi\gtrsim10$ it is completely negligible). This gives us:
\begin{equation}
    N_{2p}\sim \frac{\sqrt{6\Gamma(\gamma_{\infty\rightarrow2p})^{2}\gamma_{2p \rightarrow1s}^{1s}}}{3\xi^{2}\gamma_{\infty\rightarrow2p}\gamma_{2p \rightarrow1s}^{1s}}N_k \sim \frac{\sqrt{\Gamma}}{\xi^6}\sim \frac{f_a^2}{\xi^{3/2}},
\end{equation}
so the scaling of the cloud mass and density are:
\begin{equation}
   M_{2p}\sim\mu N_{2p}\sim\frac{f_a^2}{\sqrt{\xi}},\quad \rho_{2p}\sim \frac{M_{2p}}{r_b^3}\sim \xi^{11/2}f_a^2,
\end{equation}
Since our allowed parameter space (fig. \ref{fig:parameter space}) requires to decrease $f_a$ as $\xi$ increases, we see that the total cloud mass in the high $\xi$ regime will be smaller (as can be seen in the numerical results of fig. \ref{fig:mass parameter space}), while the strong scaling of the density with $\xi$  means that at high $\xi$ it will be larger (as can be seen in fig. \ref{fig:density parameter space}).
\section{Numerical results for the cloud mass and their scaling} 
\label{app:mass plot and scaling}
\begin{figure}[H]
    \centering
    \includegraphics[width=0.6\textwidth]{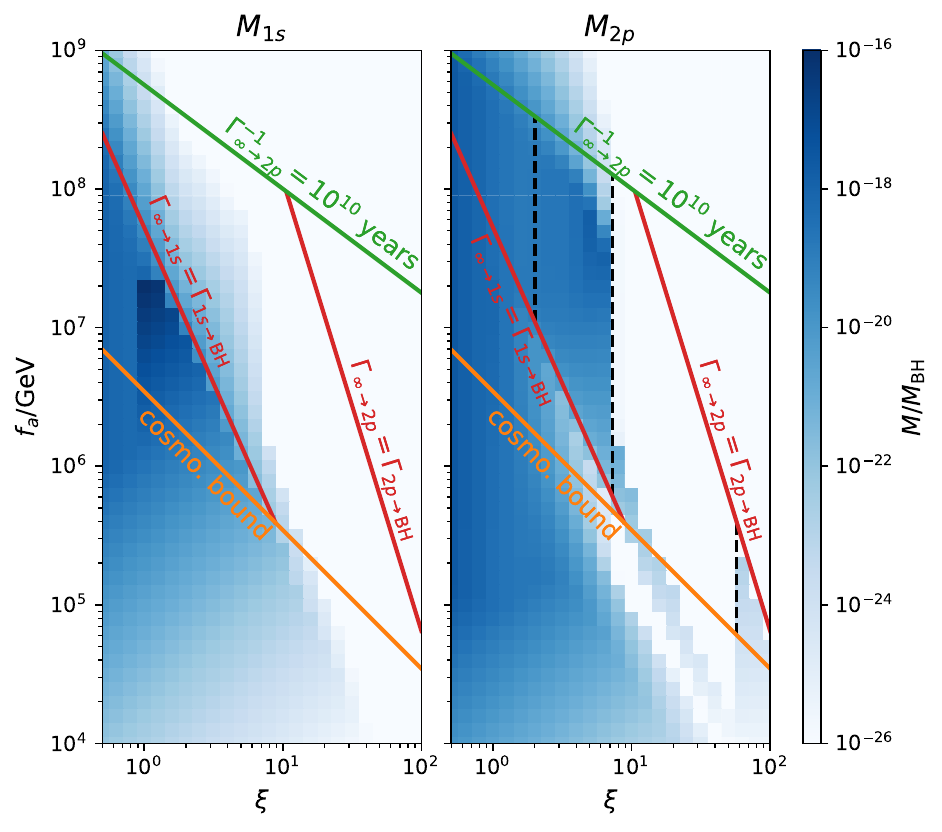}
    \caption{Halo masses after $1.5\cdot10^{10}$ years for a Schwarzschild BH of mass $M=10 M_{\astrosun}$ at 1 kpc from the galactic center (meaning $\rho_\dm\approx6\ \GeV/\cm^3$).}
    \label{fig:mass parameter space}
\end{figure}

In this Appendix we show the numerical results for the masses of the clouds (fig. \ref{fig:mass parameter space}), and discuss the scaling of the numerical results with $M,\rho_\dm$. To see the scaling, we need to rewrite any $\mu$ dependence we have in our parameters in terms of $\xi$. When normalizing the masses to $\mu N_k\propto\rho_\dm M^3/(\xi^3 v_\dm^6)$ and the time to $\tau_\rel$, the numerical results depend only on $\xi$ and the ratios $\Gamma_{nlm\rightarrow \BH}/\Gamma_\rel\propto f_a^4\xi^{9+4l}v_\dm^{11+4l}/(\rho_\dm^2M^4)$.

Scaling $\rho_\dm \mapsto a\rho_\dm,\ f_a\mapsto\sqrt{a}f_a$ leaves $\Gamma_{nlm\rightarrow \BH}/\Gamma_\rel$ constant, so the simulation results will remain the same. Thus, the only effect on the stable point is through changing $\mu N_k\mapsto a\mu N_k$  (since $N_k\propto\rho_\dm$), so $\rho_{nlm},M_{nlm}$ scale by $a$. This means that $\rho_{nlm}/\rho_\dm$ remains unchanged while the cloud mass $M_{nlm}/M$ will increase by a factor of $a$. The only other difference is that the cosmological bound will appear to have shifted down by a factor of $\sqrt{a}$, since we are now considering higher values of $f_a$ (see eq. \eqref{eq:cosmo bound}).

In a similar fashion, changing $M\mapsto aM$, $f_a\mapsto af_a$ still leaves  $\Gamma_{nlm\rightarrow \BH}/\Gamma_\rel$ constant, so again the only change will be through $\mu N_k\mapsto a^3\mu N_k$, and $\rho_{nlm}\propto\mu N_k/r_b^3\mapsto \rho_{nlm}$ (since $r_b\propto M/(\xi^2v^2)\mapsto a r_b$). Thus $\rho_{nlm}/\rho_\dm$ remains constant, while $M_{nlm}/M\mapsto a^2 M_{nlm}/M$. The cosmological bound does not change, as $\mu\mapsto a^{-1}\mu$ to keep $\xi$ constant, so $\mu f_a\mapsto\mu f_a$. On the other hand, the formation-time bound will change, since while $\Gamma_{nlm\rightarrow \BH}/\Gamma_\rel$ remained unchanged, $\Gamma_\rel\mapsto a^{-1}\Gamma_\rel$, so for a larger region in the parameter space the halo formation will be too slow.

\begin{figure}[H]
    \centering
    \includegraphics[width=0.6\textwidth]{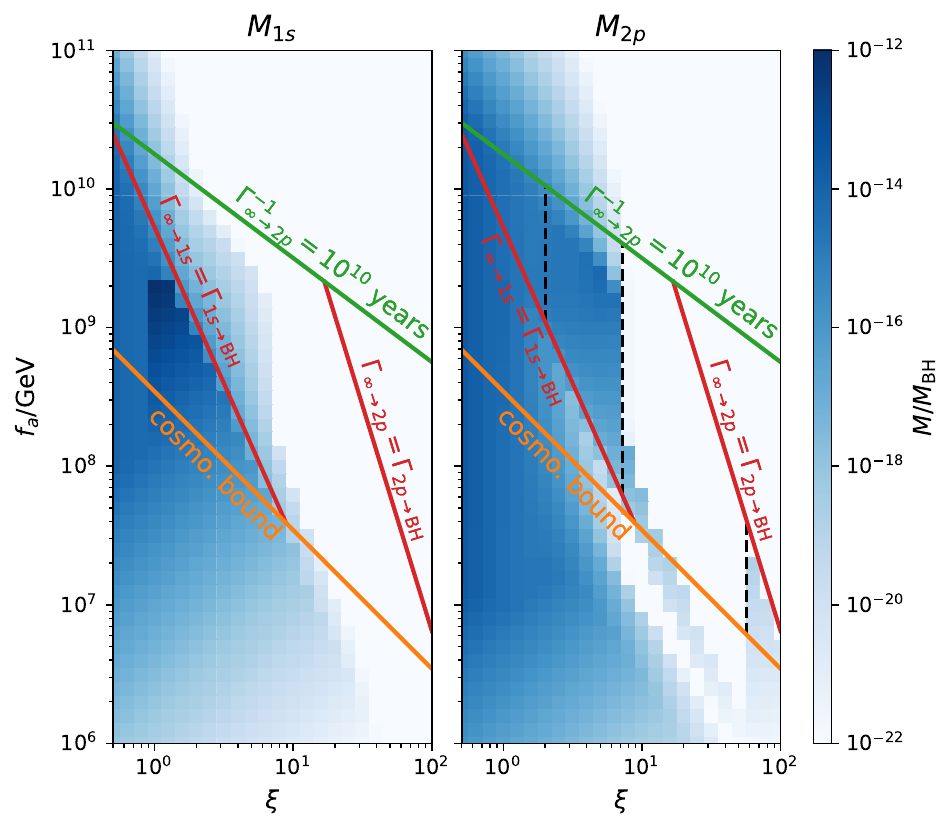}
    \caption{Halo masses after $1.5\cdot10^{10}$ years for a Schwarzschild BH of mass $M=10^3\ M_{\astrosun}$ at  1 kpc from the galactic center ($\rho_\dm\approx6\ \GeV/\cm^3$).}
    \label{fig:mass parameter space SMBH}
\end{figure}

For example, in fig. \ref{fig:mass parameter space} the maximal $M_{2p}$ we get (for parameters that do not violate the cosmological bound) is $2.7\cdot 10^{-18} M_{\BH}$, and is achieved for $(\xi,\ f_a)=(0.5,\ 8\cdot 10^{6}\ \GeV)$. The numerical results for an intermediate-mass BH with mass $10^3 M_{\astrosun}$ are shown in fig. \ref{fig:mass parameter space SMBH}, where we can see that after rescaling $f_a\mapsto 10^{2}f_a$ the bound from the halo lifetime indeed moved down relative to the plot by half an order. Looking at the data, the maximal $M_{2p}$ we achieve is $2.7\cdot10^{-14} M_\BH$, and it is achieved at $(\xi,\ f_a)=(0.5,\ 8\cdot 10^{8}\ \GeV)$, meaning that indeed $M_{2p}/M\mapsto10^4 M_{2p}/M$.

\
\bibliography{bibliography}
\end{document}